\documentclass{aastex631} %Para arxiv

\usepackage{amsmath}

\usepackage{hyperref}

\usepackage{xcolor} % or \usepackage{color}

\begin{document}

%\title{Absolute colors of trans-Neptunian objects}
\title{Historical Surveys to Rubin First Look: \\ Absolute Colors of trans-Neptunian objects}

\author[0000-0001-6082-2477]{Milagros Colazo}
\affiliation{Astronomical Observatory Institute, Faculty of Physics and Astronomy\\
A. Mickiewicz University\\
Słoneczna 36, 60-286 Poznań, Poland}

\author[0000-0002-5045-9675]{Alvaro Alvarez-Candal}
\affiliation{Instituto de Astrof\'isica de Andaluc\'ia, CSIC\\ 
Apt 3004, E18080 Granada, Spain}

%\collaboration{20}{(AAS Journals Data Editors)}

%\author{F.X Timmes}
%\affiliation{Arizona State University}
%\affiliation{AAS Journals Associate Editor-in-Chief}

%\altaffiliation{AASTeX v6+ programmer}

%\affiliation{AAS Director of Publishing}
%\affiliation{American Astronomical Society \\
%1667 K Street NW, Suite 800 \\



\begin{abstract}

We present a comprehensive photometric study of trans-Neptunian objects (TNOs) by combining data from SDSS, Col-OSSOS, DES, and the recent Rubin First Look (RFL) data. Our database comprises 43\,677 measurements in the $u$, $g$, $r$, $i$, and $z$ filters, from which we derived 2\,193 phase curves for 781 unique objects. From these data, we computed 2\,542 absolute color measurements for 633 objects, 
%and derived spectral slope differences for % 
%437 objects, 
allowing a statistical characterization of phase coloring effects. Our results show correlations between colors at opposition and their variation with phase angle, indicating that redder (bluer) objects tend to become redder (bluer) as the phase angle increases. With a larger sample and the application of phase corrections, the colors show no strong bimodality nor correlation with orbital parameters.

Notably, our dataset includes the first photometric measurements from Rubin Observatory during RFL, covering eight objects—five newly discovered TNOs and three previously known. These early LSST observations occupy sparsely sampled regions of parameter space, particularly at faint magnitudes, highlighting the discovery and characterization potential of the full survey.

%The correlations between the change in the spectral slope with the phase angle increase with wavelength, except for the larger wavelength color, $H_i-H_z$, which tends to neutralize, consistent with the spectral flattening previously reported in visible wavelengths.

%{\bf the change of the spectral slope with the phase angle}$\Delta(dS'/d\alpha)$ and $\alpha$ strengthen with increasing $\Delta\lambda$, 

\end{abstract}



\section{Introduction} \label{sec:intro}

The trans-Neptunian objects, or TNOs, are relics of the Solar System's formation. Their orbital distribution is the result of the dynamical evolution suffered by the planetary system, being transported out of their original location by the migration of the planets, especially after the gas dissipated from the disk \cite[for instance see][]{tsiganis2005NiceModel,nesvornymorbidelli2012}. 
      
Therefore, their physical properties may reflect this evolution. In particular, in this work, we focus on photometric data of TNOs.

The colors of the TNOs show a wide distribution, ranging from solar-like (neutral) or even slightly bluish, up to extremely red \cite[see][and references therein]{luujewitt1996AJ,perna2010,Fraser2023,bernardinelli2025AJ}. Early works discussed the existence of a global bimodality of the color distribution (for example \citealt{telgerroma1998Natur} in favor, or \citealt{barucci2000AJ} against). The argument in favor of the possible bimodality was the precision needed to detect it \citep{telgerroma1998Natur}. Later studies reduced this possibility to only Centaurs and, possibly, small TNOs \citep{peixinho2012,peixinho2025cent}. 

Recent large observing surveys, which will be described in more detail below, such as Col-OSSOS \citep{Fraser2023,marsset2023PSJ} and DES \citep{bernardinelli2025AJ}, found that the colors are split into two populations. Nevertheless, one aspect that has not been considered in these works is the effect of phase coloring. Note that \cite{Alvarez-Candal2019} did not find any evidence of a bi-modal color distribution using the difference of absolute magnitudes $H_V-H_R$ in a sample of $\approx 100$ objects.

\smallskip
Phase coloring is an effect that changes the colors of small bodies (and not-so-small ones) of the Solar System as the phase angle (the angular distance between the Sun and the observer as seen from the object) changes. Phase coloring is detected via multi-filter phase curves. The phase curve shows the change in the reduced magnitude, $M(\alpha)$, with the phase angle $\alpha$. The observational phase curves are fitted using a suitable photometric model. In the case of TNOs, the most commonly used model is a simple linear model:
\begin{equation} \label{Eq:1}
M(\alpha) = H + \alpha\beta,
\end{equation}
where $\beta$ is the phase coefficient and $H$ is the absolute magnitude. The reduced magnitude is related to the apparent magnitude, $m$, through
\begin{equation}\label{Eq:2}
    M(\alpha) = m - 5\log{r\Delta}.
\end{equation}
In Eq. \ref{Eq:2}, $r$ is the Sun-object distance and $\Delta$ is the observer-object distance (both in astronomical units). The linear model is well-suited for TNOs because, although we are close to the opposition-effect regime, the changes in the phase curves are well-described by a linear behavior and significant departures tend to happen at very low-$\alpha$ \citep[$\lesssim0.1$ deg,][]{verbiscer2022PSJ}. 

\smallskip
Multi-filter phase curves were analyzed for a few objects: \cite{rabino2007AJ} studied the phase curves of a sample of TNOs in the VBI filters detecting a large range of variation of the phase coefficients (between slightly negative up to 0.3 - 0.4 mag deg$^{-1}$) and a possible wavelength dependence on the phase coefficients of some objects, but they did not explore the multi-filter nature of their results population-wise in depth. Similarly, \cite{ofek2012ApJ} searched for TNOs in the Sloan Digital Sky Survey (SDSS) images, presenting the phase curves in the $g$ and $r$ filters for 13 objects, but without going further into the analysis. \cite{ayalaloera2018MNRAS} made the first systematic analyses of colors obtained from phase curves in two filters: $ H_V-H_R$, dubbed as absolute color. The authors looked for correlations with several physical-chemical information, finding only one strong correlation: $ H_V-H_R$ and $\Delta\beta=\beta_V-\beta_R$, indicating that bluer colors at opposition become bluer with increasing $\alpha$, while redder objects become even redder. Interestingly, this behavior was confirmed for asteroids in the visible and the near-infrared ranges \citep{alvarezcandal2022,alvarezcandal2025}\footnote{In these works, the statement is wrongly put as blue (red) at opposition becomes redder (bluer) with increasing phase angle. Note, however, that the results are not affected by the statement itself because these were data-driven.}.

\smallskip
To study color distributions of the TNO population, but corrected by the effect of phase coloring with a much larger database than in \cite{ayalaloera2018MNRAS} and \cite{Alvarez-Candal2019}, we compiled data from the Col-OSSOS, the DES, and the SDSS surveys to reach a critical number. We present the data and the methodology used in the following two sections. The results are presented in Section \ref{subsec:results} and are discussed in the last section. This work also aims to serve as a first step towards understanding the multi-filter nature of the Legacy Survey for Space and Time \citep[LSST,][]{ivezic2019ApJ-LSST}, which will dramatically increase the number of known small bodies, particularly those in the outer Solar System. The estimates are to observe between 1\,200 and 2\,000 centaurs, up to the limiting magnitude of $m_r\approx24.7$, in the ten years of duration of the survey \citep{murtagh2025AJ} and over 30\,000 new TNOs, with high-quality colors for about 50\% of them \citep{kurlander2025AJ} and with some objects being observed hundreds of times, which will allow to create high-quality phase curves of thousands of outer Solar System objects. In this work, we present a first look into Rubin's data from the \textit{Rubin First Look} (RFL) observations, consisting of 1$\,$966 observations in the $g$, $r$, and $i$ filters for 8 objects as described in Sect. \ref{Rubin}.

\section{Data} \label{sec:data}
Our data is composed from three catalogs observed with different facilities: the SDSS, the Col-OSSOS,  and the DES. These surveys will be described below, along with the transformations between them to convert all magnitudes to the SDSS system.

\subsection{SDSS}

We used the small body data extracted from the \textit{Sloan Digital Sky Survey} (SDSS), specifically the re-analysis performed by \citet{Sergeyev2021}, who released more than one million observations of nearly 380$\,$000 small bodies—tripling the number of objects compared to the last release of the Moving Objects Catalog \citep{Ivezic2001, Juric2002}. We include TNOs (considered as objects with $a > 30$~AU) and Centaurs. 

The \citet{Sergeyev2021} catalog provides point-spread function (PSF) magnitudes in the $ugriz$ filter system, along with their respective uncertainties. It also includes heliocentric and topocentric distances, the phase angle, and other relevant parameters. In total, we have 18$\,$320 observations (including all filters) of 1$\,$144 objects. Possible trailing of moving objects is not expected to significantly affect our analysis: the catalog photometry already includes quality checks, and our method assumes relatively large uncertainties due to rotational light-curve variability, which mitigates potential small flux losses in PSF photometry.

We aimed to bring all surveys onto a standard photometric system. Since the filters used in Col-OSSOS and DES are based on the SDSS system, and this catalog already provides all necessary parameters, the only computation we performed for this dataset was the reduced magnitude (Eq. \ref{Eq:2}).

\subsection{Col-OSSOS}

The \textit{Outer Solar System Origins Survey} (OSSOS), conducted between 2013 and 2017 using the \textit{Canada--France--Hawaii Telescope} (CFHT), was a wide-field imaging program that systematically mapped 155~deg$^2$ of the sky to depths ranging from $m_r$ = 24.1--25.2. Observations were obtained with the \textit{MegaCam imager} \citep{Boulade2003} at the CFHT on Maunakea, Hawai‘i, and the survey design and methodology were detailed by \citet{Bannister2016a}. OSSOS provided a precisely characterized sample of trans-Neptunian objects, forming the foundation for subsequent compositional and color studies. Building upon this dataset, the \textit{Colours of the Outer Solar System Origins Survey} (Col-OSSOS; \citealt{Schwamb2019}) utilized both the \textit{Gemini North Telescope} and the CFHT to obtain high-quality, near-simultaneous visible and near-infrared (NIR) photometry of a magnitude-limited subset of TNOs discovered by OSSOS \citep{Bannister2016b, Bannister2018}. The Col-OSSOS data set consists primarily of observations in the Sloan $u$, $g$, and $r$ filters and the Maunakea $J$ filter, which were acquired using the \textit{Gemini Multi-Object Spectrograph} (GMOS; \citealt{Hook2004}) and \textit{Near Infrared Imager} (NIRI; \citealt{Hodapp2003}) at Gemini-North, as well as the \textit{MegaPrime imager} \citep{Boulade2003} at CFHT. Designed to achieve a homogeneous signal-to-noise ratio across all filters and targets, Col-OSSOS aimed to enable robust taxonomic classification of Kuiper Belt Objects and to address the long-standing question of how many distinct compositional classes exist within the outer Solar System \citep{Fraser2023}.

We obtained the catalog from \citet{Fraser2023}. The original catalog did not provide $\alpha$, $r$, or $\Delta$, parameters required to calculate the reduced magnitude (Eq. \ref{Eq:2}). These columns were added using \texttt{astroquery.jplhorizons.Horizons}. For Col-OSSOS, we have 
971 observations across  
$u$, $g$, $r$, $i$, and $z$ filters for 103 objects.
The Col-OSSOS data in the visible filters were not modified and assumed to be identical to the SDSS. We do not include in this work an analysis of the $J$ data because of its low numbers compared to the visible filters.

\subsection{DES}\label{DES}

The \textit{Dark Energy Survey} (DES; \citealt{DES2016}) was allocated 575 observing nights on the 4~m \textit{Blanco Telescope} at Cerro Tololo between 2013 and 2019, with the primary goal of studying the accelerated expansion of the universe and mapping the spatial distribution of dark matter. Beyond its cosmological objectives, DES has enabled the discovery of hundreds of outer Solar System objects \citep{Bernardinelli2020, Bernardinelli2022}. The 3~deg$^2$, 520~Mpix \textit{Dark Energy Camera} (DECam; \citealt{Flaugher2015}) was specifically built for DES, allowing the wide survey component to image a contiguous 5000~deg$^2$ area of the southern sky approximately ten times in each of the $g$, $r$, $i$, $z$, and $Y$ bands over six years.

We extracted the catalog from \citet{Bernardinelli2023}. In this dataset, we applied a filter to remove cometary objects by excluding any designations starting with "C/*".
Phase angles are provided in radians, so we converted them to degrees. The observations are reported as mean fluxes in each band, normalized to 30~au from both the observer and the Sun, with a zero point defined as 30~mag. These fluxes were converted to magnitudes using  
\begin{equation}
    m = -2.5 \log_{10}(\text{flux}_{30}) + 30.
\end{equation}

 To ensure consistency with the other datasets, we transformed the magnitudes into the SDSS photometric system using the following relations \citep{Abbott2021} and assuming solar colors:
\begin{equation}
\begin{aligned}
g_\mathrm{SDSS} &= g_{\rm DES} + 0.060\,(g-i)_{\rm DES} - 0.005,\\
r_\mathrm{SDSS} &= r_{\rm DES} + 0.150\,(r-i)_{\rm DES} + 0.014,\\
i_\mathrm{SDSS} &= i_{\rm DES} + 0.167\,(r-i)_{\rm DES} - 0.027,\\
z_\mathrm{SDSS} &= z_{\rm DES} + 0.054\,(r-i)_{\rm DES} - 0.024.
\end{aligned}
\end{equation}

For DES, we have 22$\,$420 observations across all filters for 813 objects. In this work, we do not use the $Y$ filter data because it was already used in \cite{alvarezcandal2025}.

%\subsection{Rubin First Look Data}\label{Rubin}
\subsection{RFL}\label{Rubin}

The {\it Rubin First Look} dataset consists of early imaging obtained with the LSST Camera during an engineering test campaign conducted between April 21 and May 5, 2025. Covering roughly 24 deg$^2$ near the Virgo Cluster, these observations were designed to validate Rubin Observatory’s capabilities in image coaddition, difference imaging, and moving-object detection. The data include densely sampled sequences collected over nine nights, enabling both light curve analysis and the identification of new Solar System objects \citep{Greenstreet2026}. Within this dataset \citep{koumjian2025}, five trans-Neptunian objects (TNOs) were discovered: 2025 ML58, 2025 MM66, 2025 MP35, 2025 MV13, and 2025 MW47. In addition, three previously known TNOs (2014 WN510, 2014 WV580, and 1998 BU48) were recovered in the RFL data. We further searched for TNOs and Centaurs in Data Preview 1 (DP1), obtained with the commissioning camera, but found no flux measurements for these populations.

As for Col-OSSOS and DES, all magnitudes should be converted to the SDSS system. While published transformations exist from DES to LSST magnitudes\footnote{\url{https://rtn-099.lsst.io/}}, we require the inverse (LSST to DES) to proceed. Given that the original color-dependent corrections are very small, we assumed that LSST colors might closely approximate DES colors, allowing a first-order inversion of the transformations. The resulting approximate equations are:
\begin{align}
g_\mathrm{DES} &\approx g_\mathrm{LSST} - 0.016 \,(g-i)_\mathrm{LSST} + 0.003 \,(g-i)_\mathrm{LSST}^2 - 0.006,\\
r_\mathrm{DES} &\approx r_\mathrm{LSST} - 0.185 \,(r-i)_\mathrm{LSST} + 0.015 \,(r-i)_\mathrm{LSST}^2 - 0.010,\\
i_\mathrm{DES} &\approx i_\mathrm{LSST} - 0.150 \,(r-i)_\mathrm{LSST} + 0.003 \,(r-i)_\mathrm{LSST}^2 + 0.009,\\
z_\mathrm{DES} &\approx z_\mathrm{LSST} - 0.270 \,(i-z)_\mathrm{LSST} - 0.036 \,(i-z)_\mathrm{LSST}^2 + 0.003.
\end{align}

We used solar colors as a reference. We emphasize that RFL observations were not combined with other surveys to compute phase curves because object matches were generally lacking (except for a single asteroid, 1998 BU48). The color corrections were applied primarily to enable a consistent comparison of derived parameters in the final plots. Any errors introduced by this approximation are likely smaller than the typical uncertainties in the absolute magnitudes $H$ and thus may not significantly affect the positions of points in the plots. Finally, the resulting DES magnitudes were converted to SDSS as described in Section~\ref{DES}, again using solar colors as a reference. The phase curves for all RFL asteroids are presented in Appendix~\ref{LSST-phasecurves}.

\subsection{Combined database}

We homogenized the four databases to include the following columns: \texttt{MPC}, \texttt{BAND}, \texttt{red\_MAG}, \texttt{MAG}, \texttt{e\_MAG}, \texttt{PHASE}, $\Delta$, $r$, and \texttt{SURVEY}, so that merging could be performed easily using the \texttt{MPC} column. All database management was carried out using database logic implemented in \texttt{pandas}. The overlap between the surveys is as follows: SDSS $\cap$ Col-OSSOS: 49 objects, SDSS $\cap$ DES: 93 objects, SDSS $\cap$ RFL: 1 object, Col-OSSOS $\cap$ DES: 11 objects, and SDSS $\cap$ Col-OSSOS $\cap$ DES: 9. In total, the combined catalog comprises 2$\,$035 TNOs with 43$\,$878 observations (Summary in Table \ref{overlaps}).
\begin{table}[h!]
\centering
\caption{Summary of survey overlaps and total catalog content.}
\label{overlaps}
\begin{tabular}{lc}
\hline
\textbf{Survey combination} & \textbf{Number of common objects}\\
\hline
SDSS $\cap$ Col-OSSOS & 49  \\
SDSS $\cap$ DES & 93   \\
SDSS $\cap$ RFL & 1  \\
Col-OSSOS $\cap$ DES & 11  \\
SDSS $\cap$ Col-OSSOS $\cap$ DES & 9  \\
\hline
\end{tabular}
\end{table}

\begin{figure}[ht]
    \centering
    \includegraphics[width=0.45\linewidth]{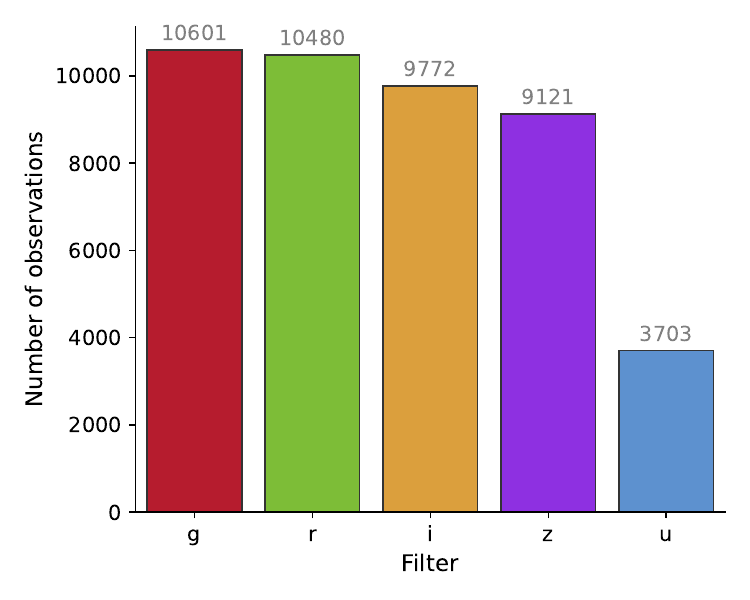}
    \caption{Number of observations per photometric filter in the combined TNO catalog. The filters $u$, $g$, $r$, $i$, and $z$  are shown with distinct colors. The numeric labels above each bar indicate the exact number of observations for that filter.}
    %\caption{Number of observations per photometric filter in the combined TNO catalog. The filters $u$, $g$, $r$, $i$, $z$, and $J$ are shown with distinct colors. The numeric labels above each bar indicate the exact number of observations for that filter.}
    \label{obs_per_filter}
\end{figure}

In Figure~\ref{obs_per_filter}, we show the distribution of observations per photometric filter across our combined TNO catalog. The plot highlights how the different filters are represented in the dataset, enabling a visual comparison of coverage across bands.
The $g$ band has the largest number of observations, while the $u$ band has significantly fewer, as expected, because of the usual difficulties in observing TNOs in the bluer wavelengths due to a decrease in reflectivity and overall lower efficiency of the detectors, joint with the fact that DES did not use this filter. Among the $g$, $r$, $i$, and $z$ bands, the number of observations is roughly comparable, reflecting a relatively uniform coverage across these filters.

\begin{figure}[ht]
    \centering
    \includegraphics[width=0.45\linewidth]{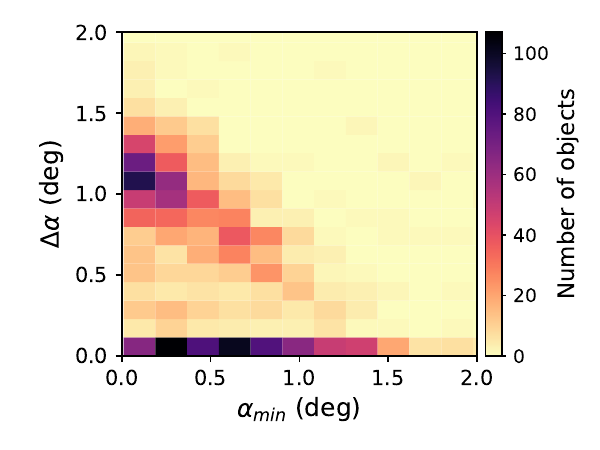}
    \caption{Phase angle coverage for the TNOs in the combined catalog. The x-axis shows the minimum observed phase angle, $\alpha_\mathrm{min}$, and the y-axis shows the phase angle range, $\Delta \alpha = \alpha_\mathrm{max} - \alpha_\mathrm{min}$. The color scale indicates the number of TNOs in each bin.}
    \label{phase_coverage}
\end{figure}

Figure~\ref{phase_coverage} presents the distribution of minimum phase angles, $\alpha_\mathrm{min}$, versus the range of phase angles, $\Delta \alpha$, for the objects in our merged catalog. Most TNOs have minimum phase angles below 1$^\circ$, and the range of phase angles spans up to approximately 2$^\circ$. Note that the limits in the plot were set to enhance visibility because Centaurs reach up to $\alpha_{min}=14.3$ deg and $\Delta\alpha=7.3$ deg, which would otherwise dilute the graphical information.
 
\section{Method} \label{sec:method}
%\subsection{Objects with three or more observations per filter}
The analysis is performed independently for each filter, and we only use those filters with three or more observations per object in the phase-curve fitting. This criterion leaves us with  
1$\,$139 TNOs, corresponding to about 50\% of the sample.
 
The fit of the phase curve follows methods similar to those described in \cite{Alvarez-Candal2016, Alvarez-Candal2019}. We provide here a brief description and refer the reader to those works for additional details.

Since our observations were obtained on different nights, we account for possible rotational variability by modeling the expected lightcurve amplitude. For this purpose, we used amplitudes reported in The Lightcurve Database (LCDB; \citealt{Warner2009}). When an amplitude from LCDB was available for an object, we used it as $\Delta m$. For each object and filter, we generated 2$\,$000 magnitudes as:
\begin{equation}
m_i^{\mathrm{sim}} = m_i + \epsilon_i, \quad \epsilon_i \in \left[-\frac{\Delta m}{2}, +\frac{\Delta m}{2}\right],
\end{equation}
where $\epsilon_i$ is drawn from a uniform distribution. For each iteration, we performed a linear fit of the simulated magnitudes versus phase angle following Equation \ref{Eq:1}, to obtain the slope $\beta$ (the phase coefficient) and the intercept $H$ (the absolute magnitude).

% \begin{equation}
% m_i^{\mathrm{sim}} = H + \beta \, \alpha_i,
% \end{equation}

\begin{figure}[ht]
    \centering
    \includegraphics[width=0.45\linewidth]{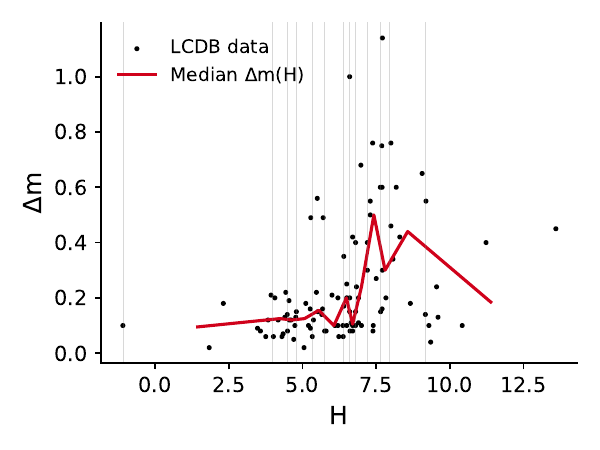}
    \caption{Median rotational amplitude as a function of absolute magnitude. Black points show individual TNO amplitudes reported in the LCDB, while the red line represents the median $\Delta m(H)$ curve obtained by binning the data into 13 bins containing equal numbers of objects; horizontal segments indicate the H-range spanned by each bin.}
    \label{mediancurve}
\end{figure}

For objects without a reported amplitude in LCDB, we applied a two-step estimation. First, we calculated the standard deviation of the observed reduced magnitudes. We then generated a perturbed set of magnitudes by adding random Gaussian noise with this standard deviation. A linear fit of these perturbed magnitudes versus phase angle was performed to obtain a preliminary estimate of the absolute magnitude $H$. Then, using the $H$ values and amplitudes from the LCDB, we constructed a median curve in the $\Delta m$-$H$ plane (red curve in Fig. \ref{mediancurve}) by binning the data into 13 bins containing equal numbers of objects, spanning the full range of $H$ values. 
 
For objects with no known rotational amplitude, we used their preliminary $H$ estimate to determine the appropriate bin and assigned the median $\Delta m$ value of that bin. If an object had an $H$ value outside the LCDB range (i.e., fainter or brighter than the catalog limits), we assigned the amplitude corresponding to the first or last bin, respectively. We then re-ran the Monte Carlo procedure with 2$\,$000 iterations following the same steps as for objects with known amplitude.

\begin{figure}[ht]
    \centering
    % Primera fila
    % \includegraphics[width=0.4\linewidth]{2006QF181_phasecurve_pdf_u.png}
    \includegraphics[width=0.6\linewidth]{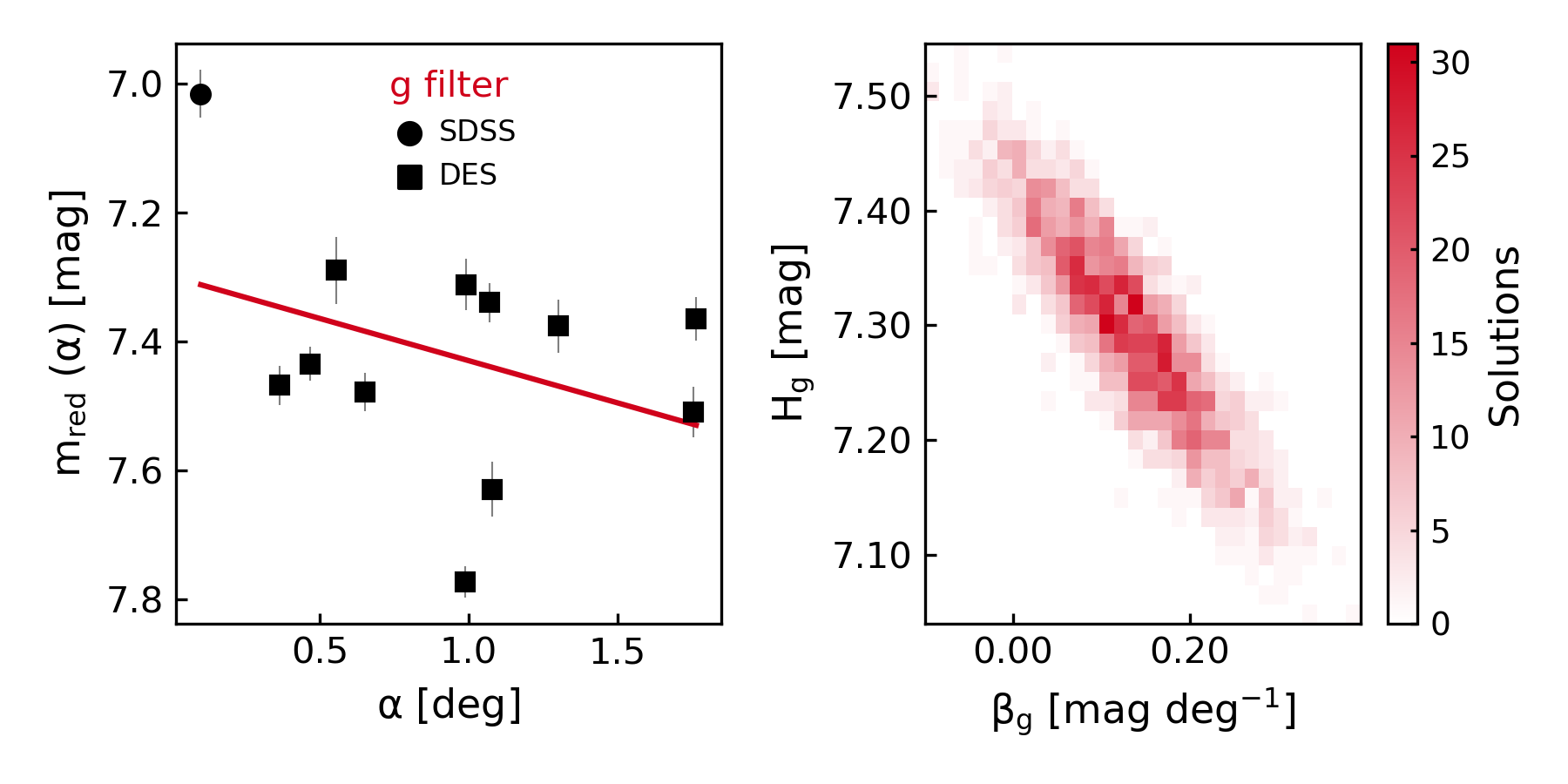}\\[3pt]
    % Segunda fila
    \includegraphics[width=0.6\linewidth]{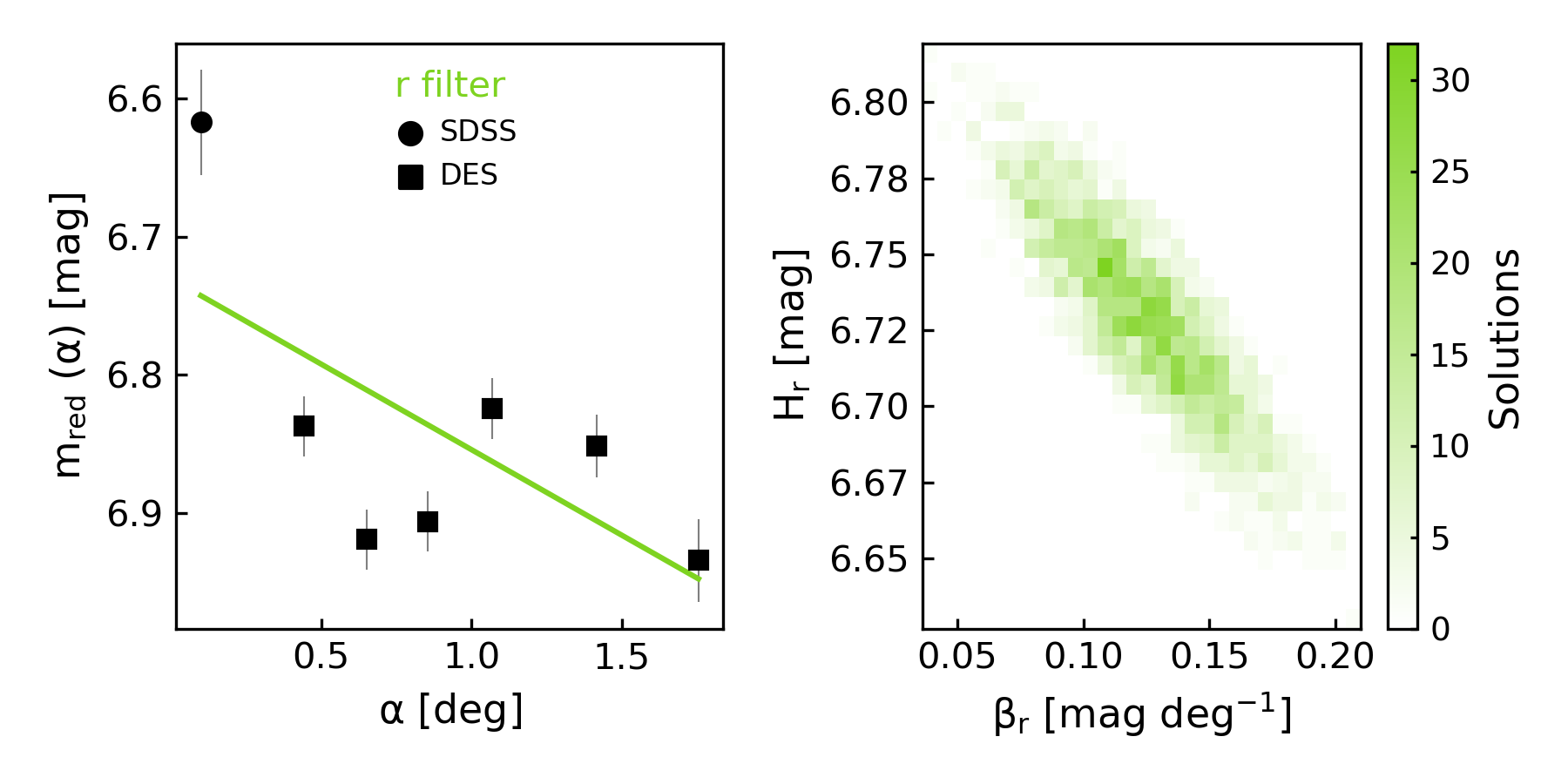}
    \includegraphics[width=0.6\linewidth]{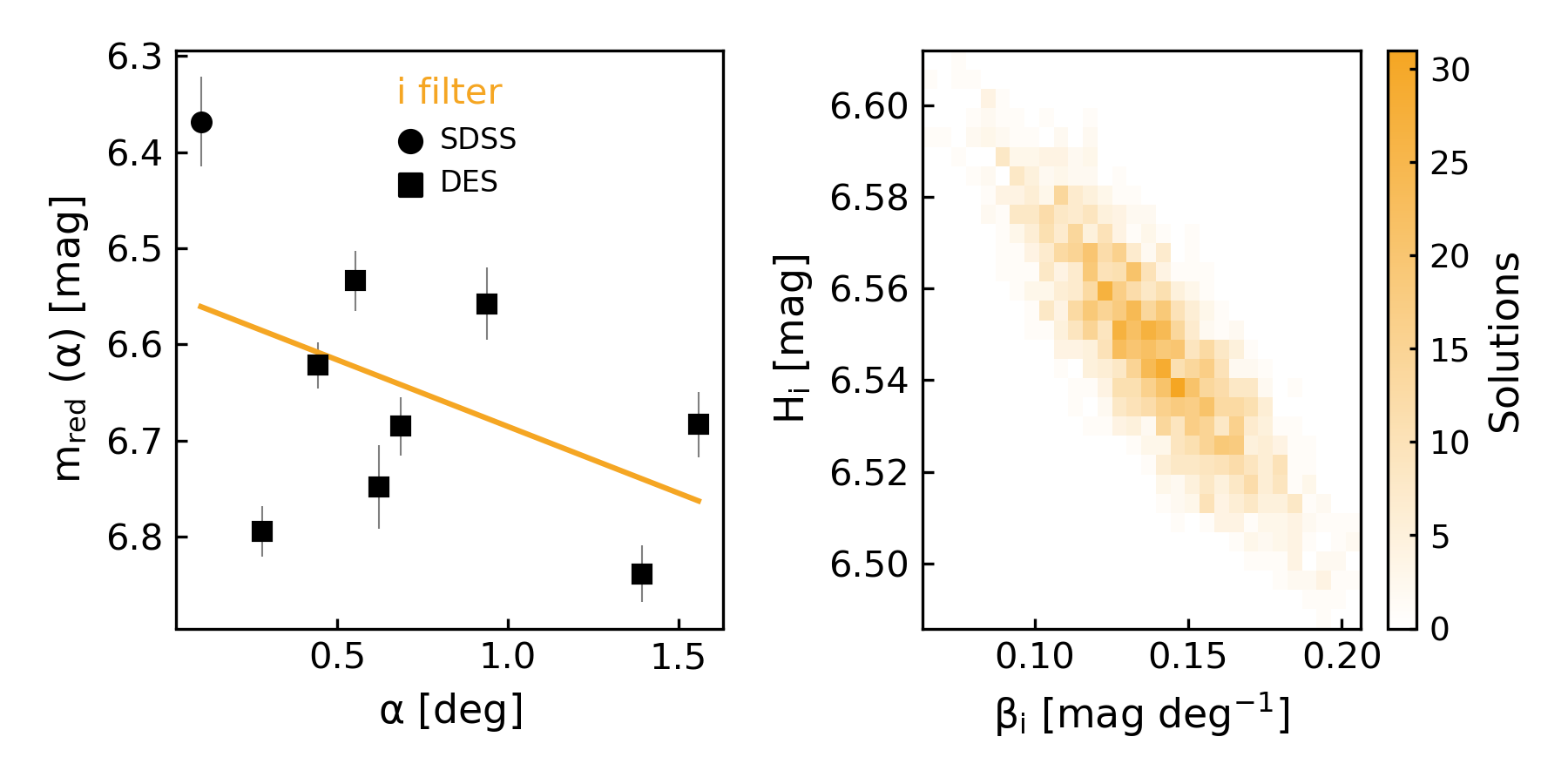}\\[3pt]
    % Última figura centrada (z)
    \includegraphics[width=0.6\linewidth]{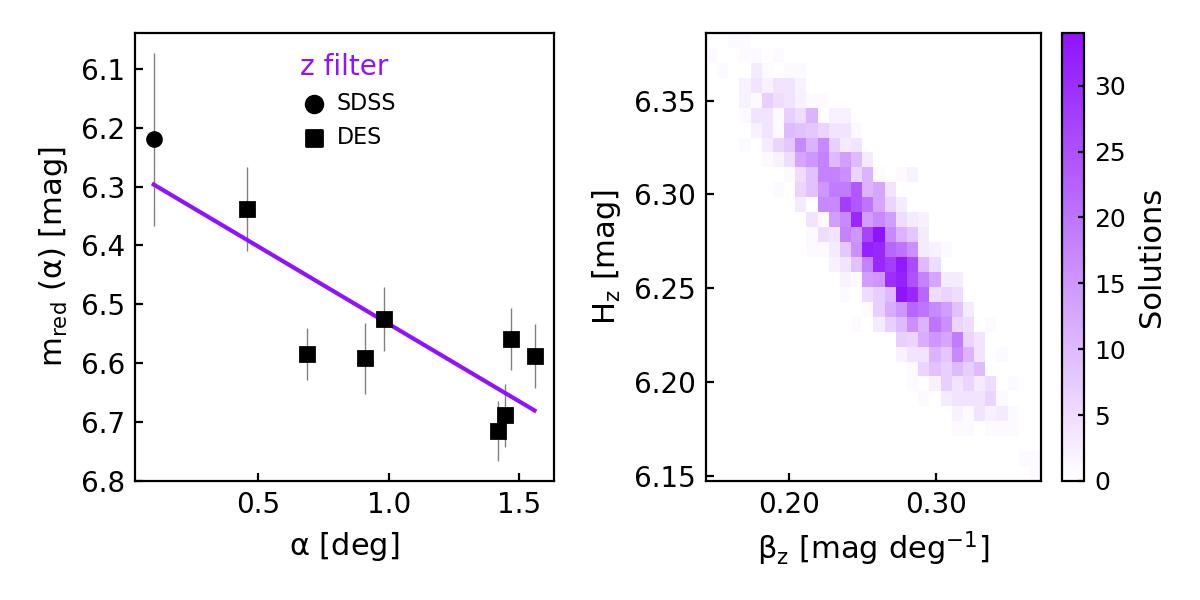}
    \caption{
        Example phase-curve fitting for a single TNO 2007 RW10 across different filters. Left panels: Observed reduced magnitudes versus phase angle for different surveys (markers indicate survey origin), with the median linear fit overplotted in a solid line. Right panels: Two-dimensional histogram of 2000 Monte Carlo iterations of the linear fit,
        showing the distribution of phase slope ($\beta$) versus absolute magnitude ($H$).
    }
    \label{example_phasecurve}
\end{figure}

Individual results for each object and filter were saved as separate CSV files. Additionally, a summary catalog containing the median $\beta$, median $H$, their standard deviations ($\sigma$), the minimum and range of observed phase angles, and the number of observations was produced for further analysis. One example of the processing is shown in Fig. \ref{example_phasecurve} for 2007 RW10.

\section{Results} \label{subsec:results}

\subsection{Phase curve parameters}

We have derived a total of 2\,333 phase curves in $u$, $g$, $r$, $i$, and $z$ filters. Approximately 68\% of the objects lie within the range $-0.31 < \beta < 0.54$ mag deg$^{-1}$,
and 95\% between $-1.75 < \beta < 1.62$ mag deg$^{-1}$. However, the 99\% interval extends to much more extreme values ($-6.78 < \beta < 6.37$ mag deg$^{-1}$), indicating the presence of a small number of outliers with exceptionally steep or inverted slopes. A quick inspection of these outliers reveals that most are objects with very poor phase-angle coverage, typically showing $\Delta\alpha$ values close to zero. The median $\Delta\alpha$ for the outlier subset is 0.0035 deg, compared to 0.62 deg for objects within the range $|\beta| < 1.5$ mag deg$^{-1}$. We will call the sample $|\beta| < 1.5$ mag deg$^{-1}$ as filtered, noting that large values of $|\beta|$ may not necessarily imply unphysical solutions because these may be driven by undetected phenomena, such as satellites, binaries, ring systems, etc.

We also analyzed the distribution of absolute magnitudes for the entire sample. For the unfiltered dataset, the mean absolute magnitude is $\langle H \rangle = 6.27$~mag, with 68\% of objects falling in the range 4.95-7.88~mag, 95\% within 3.18-11.04~mag, and 99\% spanning from -1.83 to 16.79~mag.

Considering only the filtered sample, we are left with phase curves for 781 objects, with 60, 578, 559, 532, and 464 phase-curve fits were obtained for the $u$, $g$, $r$, $i$, and $z$ filters, respectively.
% The mean $\beta$ for 
% % 
% our sample is 
% % 
% % 
% $\langle \beta \rangle = 0.10~\mathrm{mag~deg^{-1}}$.
%In the following analysis, we restrict our study to this final sample.

\begin{figure}
    \centering
    \includegraphics[width=1\linewidth]{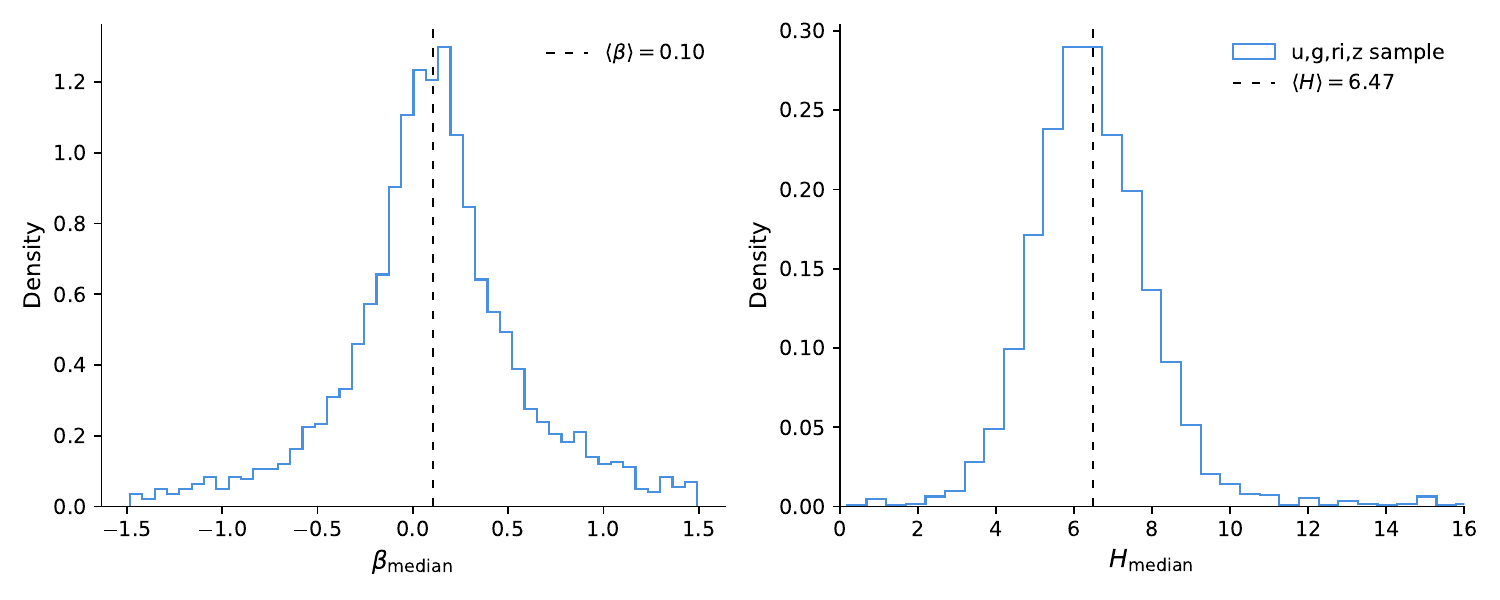}
    \caption{
     Distributions of the median $\beta$ (left) and $H$ (right). The black dashed vertical lines indicate the mean values, $\langle \beta \rangle$ and $\langle H \rangle$, computed from the filtered sample.}
    \label{H-beta}
\end{figure}

%For this filtered sample, the mean phase slope parameter and absolute magnitude values per filter are summarized in Table~\ref{tab:beta_H}. The distributions of both parameters are shown in Figure~\ref{H-beta}. As seen in Figure~\ref{H-beta}, the overall distribution of $\beta$ values shows a slight excess toward negative slopes, resulting in a negative mean. This asymmetry is mainly driven by the subset of objects with fewer than three observations ($N < 3$). When excluding these poorly sampled cases, the mean phase slope parameter shifts to $\langle \beta \rangle = 0.10~\mathrm{mag~deg^{-1}}$, consistent with the average value of $\beta = 0.09~\mathrm{mag~deg^{-1}}$ reported by \citet{ayalaloera2018MNRAS} for TNOs. The $\beta$ values reported correspond to this cleaned sample, excluding the $N<3$ cases. We report all data for the $N<3$ objects, but the user is cautioned as to the actual accuracy of these $\beta$. $H$ is less sensitive to the actual processing shown above because we are working close to the opposition, for the most part.
The mean phase slope parameter and absolute magnitude values per filter for our final sample are summarized in Table~\ref{tab:beta_H}. The distributions of both parameters are shown in Figure~\ref{H-beta}. The distribution of beta values is approximately symmetric, with the mean and median in close agreement. We find a mean phase slope parameter of $\langle \beta \rangle = 0.10~\mathrm{mag~deg^{-1}}$, consistent with the average value of $\beta = 0.09~\mathrm{mag~deg^{-1}}$ reported by \citet{ayalaloera2018MNRAS} for Trans-Neptunian Objects.

\begin{table}[ht]
\centering
\caption{Mean phase slope parameter $\langle \beta \rangle$ and absolute magnitude $\langle H \rangle$ per filter for our sample.}
\label{tab:beta_H}
\begin{tabular}{lcc}
\hline
\textbf{Filter} & $\langle \beta \rangle$ [mag~deg$^{-1}$] & $\langle H \rangle$ [mag] \\
\hline
$u$ &  $0.11$ &  $8.23$ \\
$g$ &  $0.12$ &  $7.19$ \\
$r$ & $0.09$ &  $6.44$ \\
$i$ &  $0.10$ &  $6.14$ \\
$z$ &  $0.10$ &  $5.77$ \\
%$J$ & \hlc{$0.46$} &  \hlc{$5.12$} \\
\hline
\end{tabular}
\end{table}

Within the RFL database, several LSST objects were identified as outliers because the median of their distributions fell outside the $\beta$ constraints: -1.5 $\leq$ $\beta$ $\leq$ 1.5 mag deg$^{-1}$. These cases correspond to 1998 BU48 $(g, r, i)$, 2025 MM66 $(r)$, 2025 MP35 $(g)$, and 2025 MW47 $(g, r)$. For these objects, we did not adopt the median values directly. Instead, we inspected the full two-dimensional $H-\beta$ probability distributions (Figure \ref{LSST-phasecurves}) and selected solutions at the boundary of the accepted $\beta$ range. Specifically, when the median $\beta$ lay outside the limits, we adopted $\beta = \pm$ 1.5 mag deg$^{-1}$ and assigned the corresponding $H$ value from the distribution at that $\beta$. This approach allows us to retain these sources in subsequent analyses while enforcing physically motivated $\beta$ constraints and maintaining internal consistency across the sample.

The adopted corrections are as follows: for 2025~MM66 ($r$), $\beta = 1.5$ mag deg$^{-1}$ and $H = 14.15$ (8.5\% of solutions within the range); for 2025~MP35 ($g$), $\beta = -1.5$ mag deg$^{-1}$ and $H = 17.75$ (9.75\% with $\beta \leq -1.5$ mag deg$^{-1}$); for 2025~MW47, $\beta = -1.5$ mag deg$^{-1}$ with $H$ of 16.67 ($g$, 11.55\%) and 16.11 ($r$, 5.20\%); and 1998~BU48, which remains an outlier under these $\beta$ constraints and was therefore not adjusted. This object would require further observations to better understand the origin of its anomalous behavior.

\subsection{Colors}
We split the color analyses into three parts. In the first we show the color-color diagrams, while in the second we discuss the evidence of a two-color population in our dataset. Last, we explore the relations between magnitude and color and color vs. orbital elements.

\subsubsection{Color-color plots}
\begin{figure}[ht]
    \centering
    \includegraphics[width=1\textwidth]{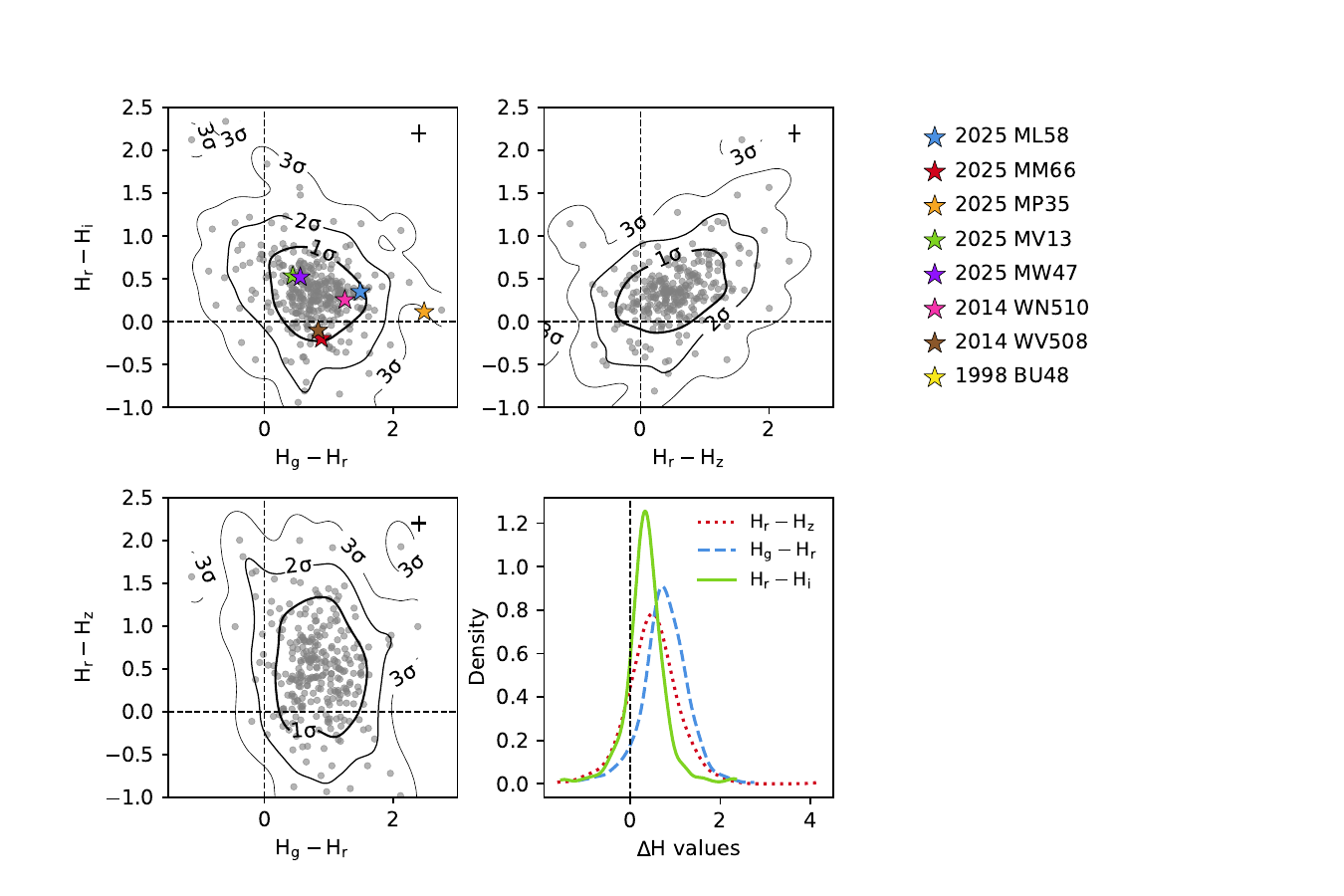}
    \caption{Two-dimensional and one-dimensional distributions of absolute magnitude differences between filters (absolute colors). Top and bottom-left panels: Scatter plots with 2D KDE contours showing the correlations between $\Delta$H values for different filter combinations. Contours indicate the 1$\sigma$, 2$\sigma$, and 3$\sigma$ levels, and the cross represents the typical mean absolute deviation of the measurements. Colored stars highlight objects observed by Rubin, including both recent discoveries and previously known targets. Bottom-right panel: One-dimensional KDE distributions for $\Delta$H values in all relevant filter combinations.}
    \label{abs_colors}
\end{figure}

From the nominal absolute magnitudes ($H_{\mathrm{median}}$) derived for each object and filter, we computed the corresponding absolute colors. The number of objects for which each color could be determined is as follows (Summary in Table \ref{table_colors}): 57 for $(u-g)$, 59 for $(u-r)$, 59 for $(u-i)$, 54 for $(u-z)$, 436 for $(g-r)$, 410 for $(g-i)$, 353 for $(g-z)$, 429 for $(r-i)$, 344 for $(r-z)$, and 341 for $(i-z)$. These colors were computed using our final sample, which we recall is defined as objects with $|\beta| < 1.5$ mag deg$^{-1}$ and $N\geq3$ in the $u$, $g$, $r$, $i$, and $z$ filters. Figure \ref{abs_colors} shows the two-dimensional and one-dimensional distributions of absolute colors between filters. The top-left, top-right, and bottom-left panels display scatter plots with 2D kernel density estimation (KDE) contours, which provide a smoothed representation of the point density and illustrate the correlations in color across different filter combinations. Contours correspond to the 1$\sigma$, 2$\sigma$, and 3$\sigma$ confidence levels, and the cross indicates the typical mean absolute deviation of the measurements. The bottom-right panel shows the one-dimensional KDEs of the different colors. Across all color--color diagrams analyzed, the 1$\sigma$ contours, which enclose roughly 68\% of the sample, are consistently located in the quadrant where both color indices are positive indicating overall red colors \citep{ayalaloera2018MNRAS, Schwamb2019}. 

Objects observed within the RFL (including discoveries) are highlighted in the plot containing measurements in $g$, $r$, and $i$. These objects generally exhibit moderately red colors (with $H_g-H_i>0$ and $H_r-H_i>0$), consistent with the bulk of the population. Exceptions are 2014~WV508 and 2025~MM66, which show $H_g - H_r > 0$ but $H_r - H_i < 0$, indicating comparatively neutral or slightly bluish colors. All RFL objects fall within the $1\sigma$ contour, except for 2025~MP35, which lies near the edge of the $3\sigma$ ,region and displays a relatively high $H_g - H_r \approx 2$. Although 1998 BU48 falls outside our adopted $\beta$ thresholds, its absolute colors still yield reasonable values, clustering near the main concentration of objects in color-color space. 

\subsubsection{On the two color populations}
\citet{Fraser2023} and \citet{bernardinelli2025AJ} used similar color-color diagrams, although not strictly phase-corrected, and detected two types of surfaces: faintIR and brightIR, the former, and NIRB and NIRF, the latter. We inspected our color diagrams, without including $(H_u-H_g)$ because of its low number of objects (only 57), and we do not find a clear evidence of multi-population, neither in the 2D plots, nor in histograms of single color. 

\citet{Fraser2023} identifies two populations through a Gaussian mixture model, but explicitly notes that these mixture curves are not direct fits to the observed data, but rather representative models. Likewise, in \citet{bernardinelli2025AJ}, the two populations are clearly distinguished in the model contours, but the observed data points plotted over them—particularly when their associated uncertainties are considered—show a less pronounced separation. These differences likely arise because our analysis is based on absolute colors, corrected for phase effects, whereas the previous studies rely on apparent colors, sometimes the average or median of multiple observations.

\begin{figure}
    \centering
    \includegraphics[width=\textwidth]{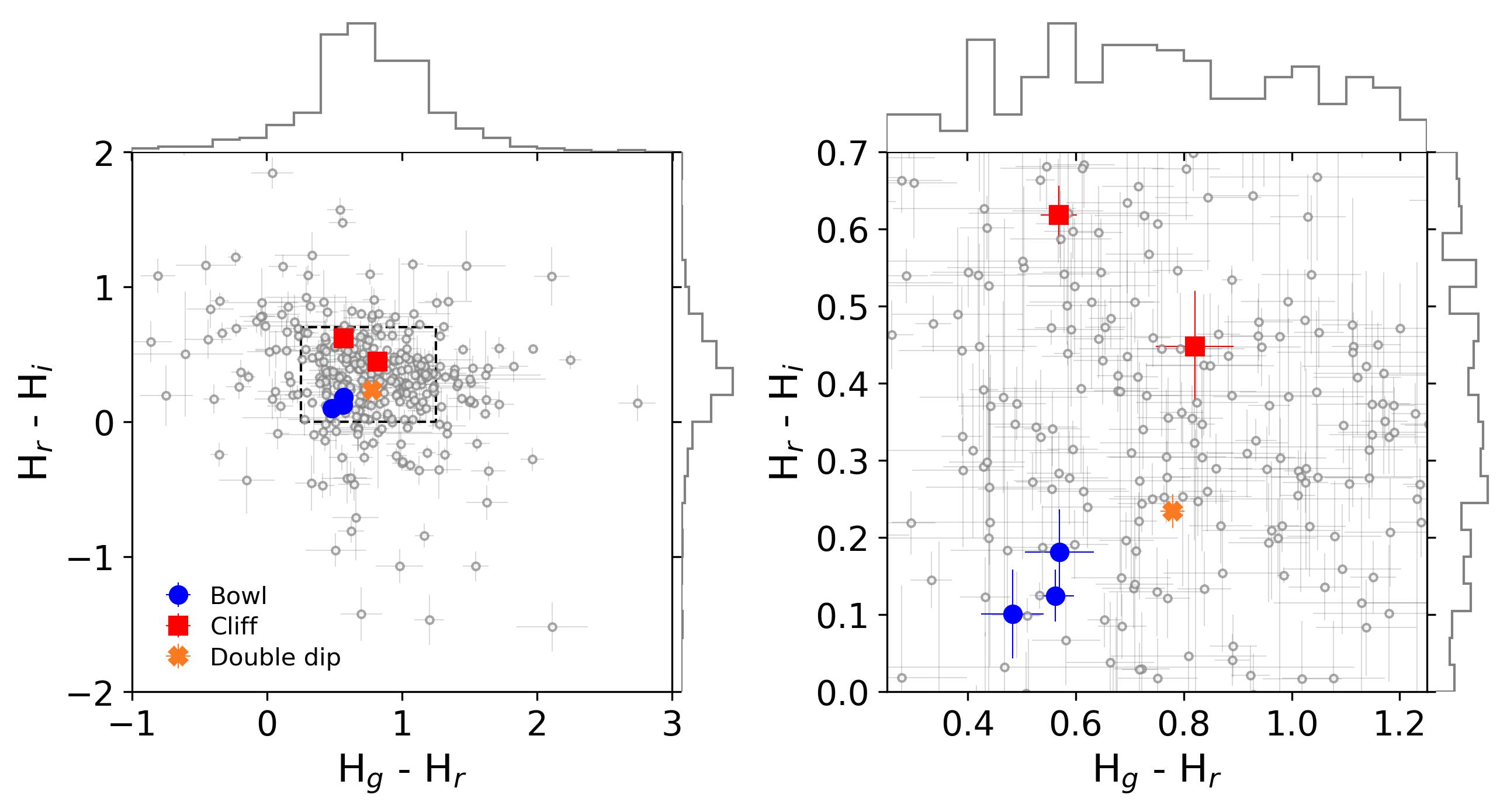}
    \caption{Color-color diagram for objects with N$\geq$8 measurements per band.   
     The left panel shows the full distribution, while the right panel presents a zoomed-in view of the region highlighted by the dashed rectangle. Black points represent the full dataset, with error bars reflecting photometric uncertainties. Objects in common with \cite{pinillaalonso2025NatAs} are highlighted with colored markers according to their spectral classification: Water-rich (Bowl, blue circles), CO2-rich (Double dip, orange crosses), and Organic-rich (Cliff, red squares).}
    \label{appendix:bestsample}
\end{figure}

To further investigate the apparent lack of statistically significant bimodality in our dataset, we re-examined the color-color distributions using a restricted subsample of objects with $N \geq 8$ measurements per band, comparable to the selection adopted in \citep[for example,][]{Bernardinelli2023}. The resulting color-color diagram, shown in Figure~\ref{appendix:bestsample}, confirms that the same global trends are recovered, with reduced scatter as expected, but sill no evident bi-modal distribution.

\begin{figure}
\centering

\textbf{Ofek (2012)}\\[2pt]
\includegraphics[width=0.32\textwidth]{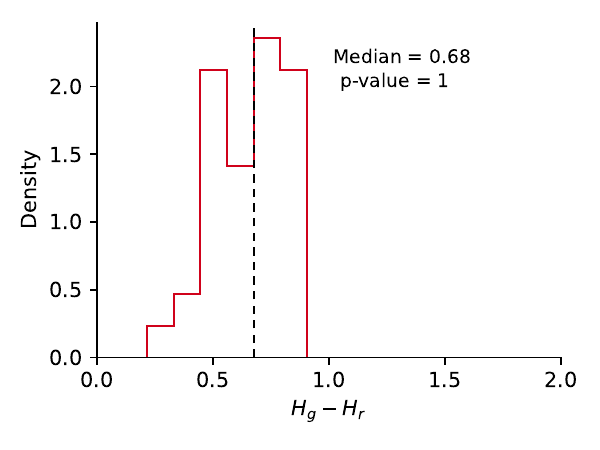}
\includegraphics[width=0.32\textwidth]{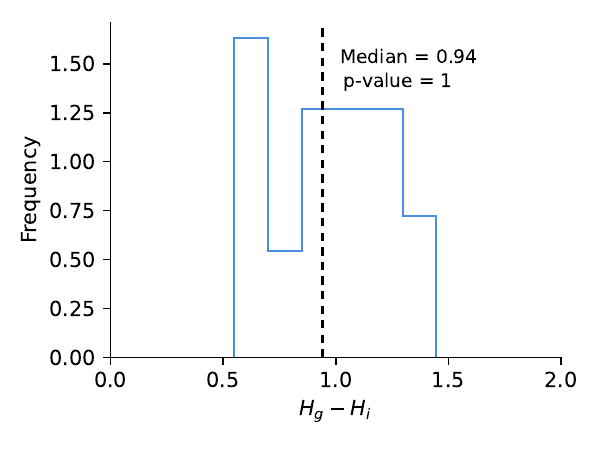}
\includegraphics[width=0.32\textwidth]{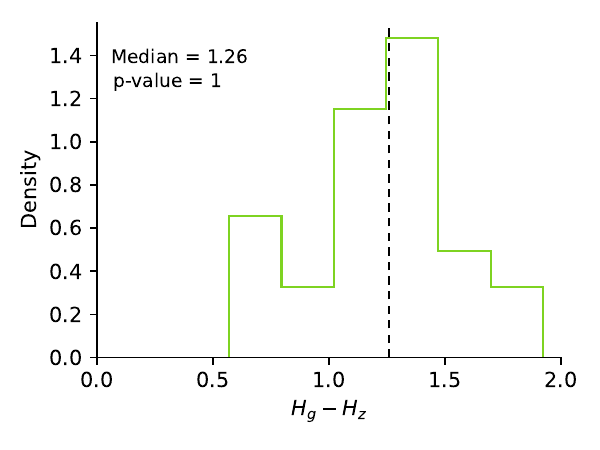}

\vspace{6pt}

\textbf{Ferreira et al. (2025)}\\[2pt]
\includegraphics[width=0.32\textwidth]{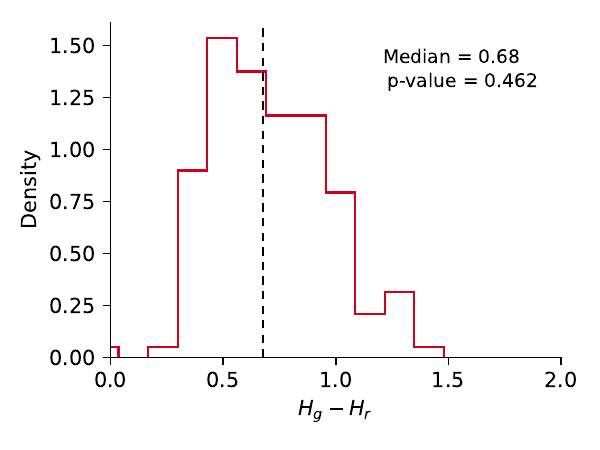}
\includegraphics[width=0.32\textwidth]{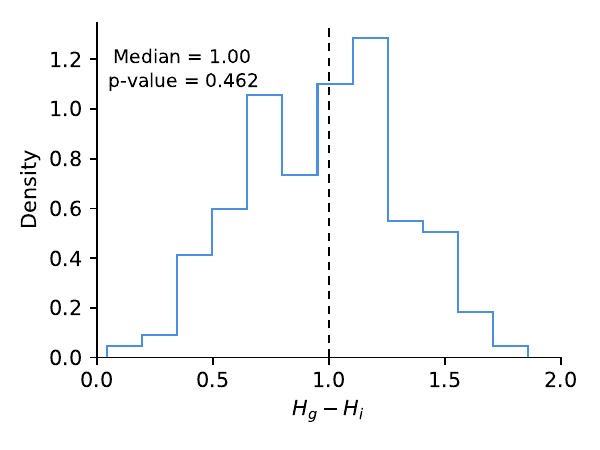}
\includegraphics[width=0.32\textwidth]{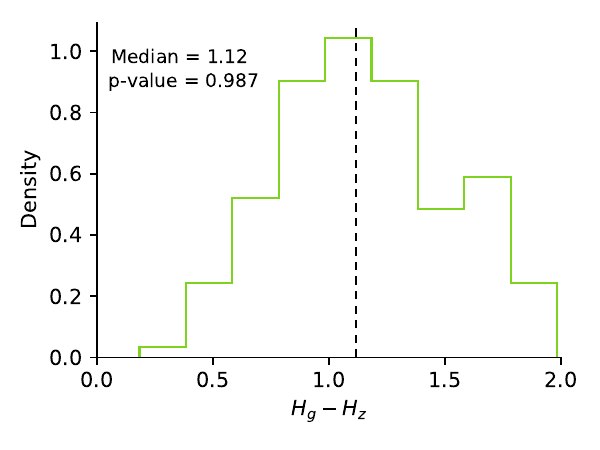}

\caption{Histograms of absolute colors ($H_g-H_i$, $H_g-H_r$, and $H_g-H_z$) from \cite{ofek2012ApJ} [top] and \cite{ferreira2025MNRASabscol} [bottom], shown separately. In each panel, the p-value of the Hartigan dip test is indicated. In all cases, the p-values are well above 0.05, indicating that unimodality cannot be rejected. While $H_g-H_i$ shows the strongest visual hint of bimodality, it remains statistically consistent with a unimodal distribution.
}
\label{appendix:diptest}
\end{figure}

Importantly, the apparent lack of a significant bimodality when using absolute colors is not a new result. Previous studies using them also find no clear evidence for bimodality \citep{ayalaloera2018MNRAS,Alvarez-Candal2019}. Furthermore, using the absolute magnitudes reported in \cite{ofek2012ApJ} and \cite{ferreira2025MNRASabscol}, there is, in principle, no evident bimodal distribution either. To make this comparison quantitative, we performed Hartigan dip tests, which test the null hypothesis of unimodality, on the color distributions reported in \cite{ofek2012ApJ} and \cite{ferreira2025MNRASabscol}. In both cases, the p-values are well above 0.05, indicating that unimodality cannot be rejected (Figure \ref{appendix:diptest}).

However, the Dip Test does not include the uncertainties in the estimation of the p-value. We therefore complemented this analysis by fitting Gaussian Mixture Models (GMMs) and comparing one- and two-component models using the Akaike Information Criterion (AIC) and Bayesian Information Criterion (BIC). To ensure that this comparison is not driven by objects with physically unrealistic behavior, the analysis was restricted to objects with a median spectral slope in the range $-0.1 \leq \beta \leq 0.3$.

\begin{figure}
\centering
\includegraphics[width=\textwidth]{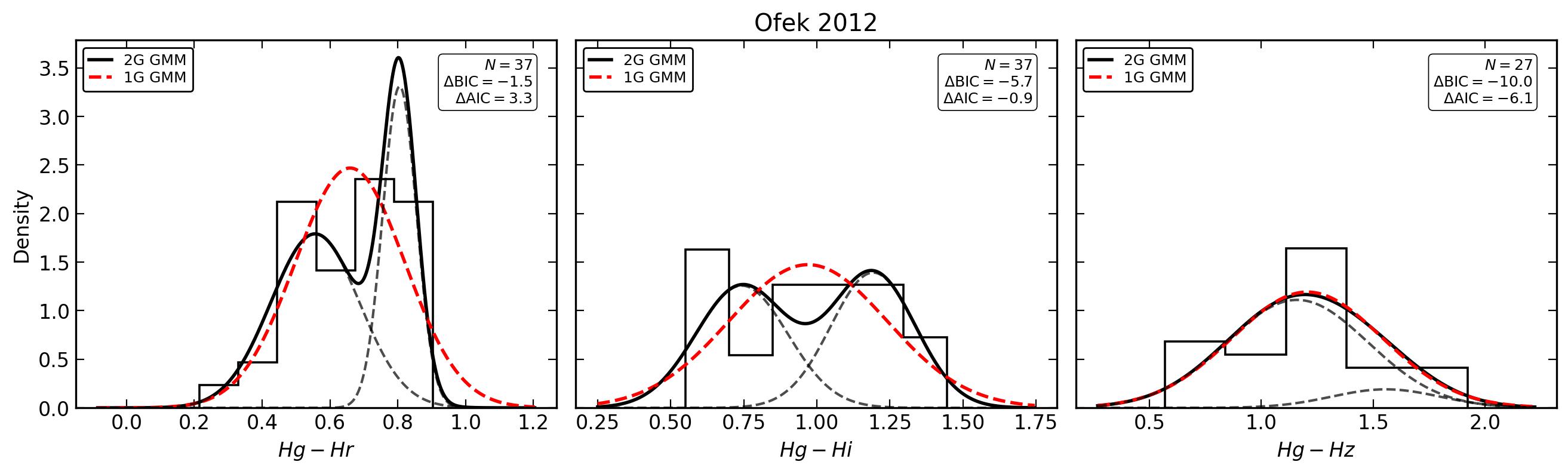}
\vspace{6pt}
\includegraphics[width=\textwidth]{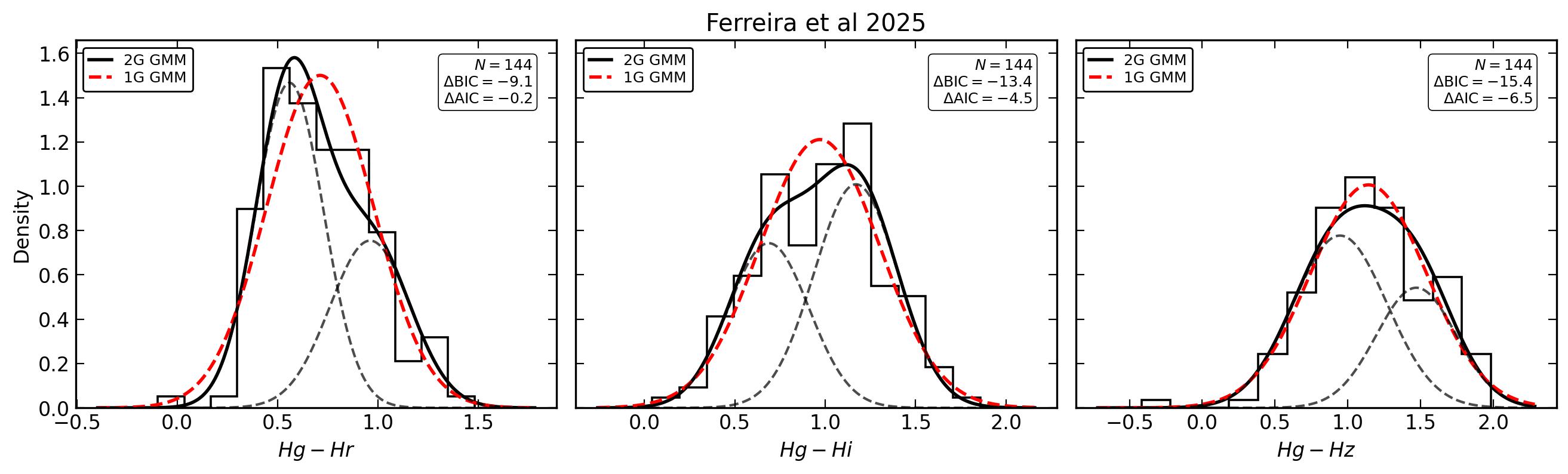}
\vspace{6pt}
\includegraphics[width=\textwidth]{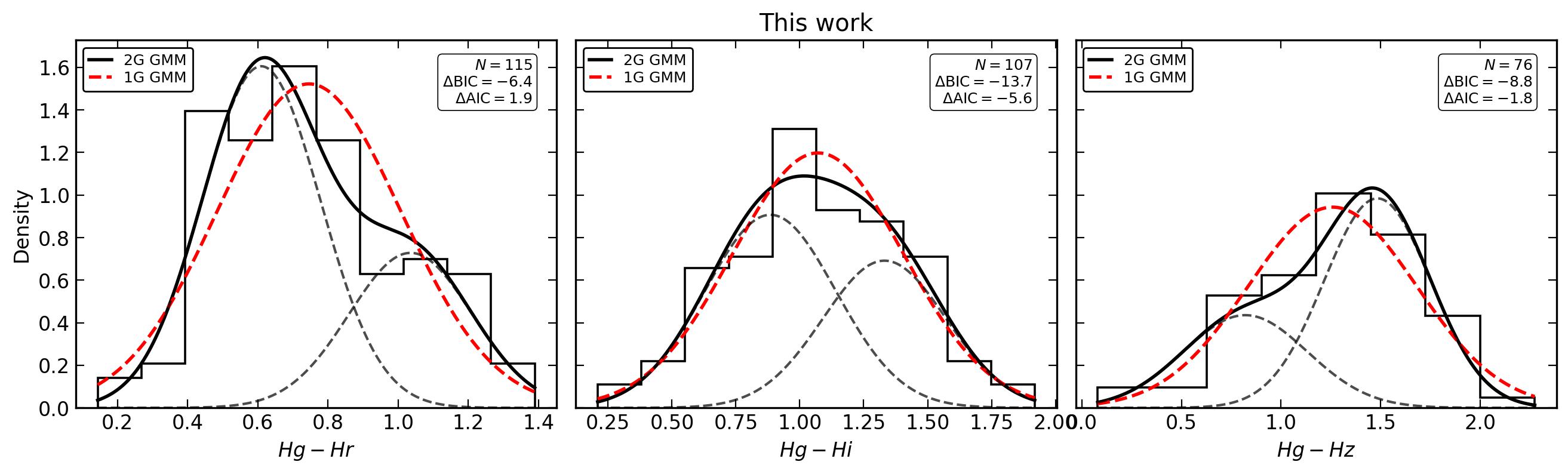}
\caption{Comparison of color distributions and Gaussian Mixture Model (GMM) fits for three datasets: \citet[ top]{ofek2012ApJ}, \citet[ middle]{ferreira2025MNRASabscol}, and this work (bottom). In each panel, the histogram shows the observed distribution, while the dashed red and solid black curves correspond to the one-component (1G) and two-component (2G) GMM fits, respectively. The individual Gaussian components of the 2G model are shown as dashed black lines.
The values of $\Delta$BIC and $\Delta$AIC are reported in each panel, defined as $\Delta$BIC = BIC(1G) - BIC(2G) and $\Delta$AIC = AIC(1G) - AIC(2G), such that negative values indicate a preference for the single-component model. N denotes the number of objects in each sample.}
\label{GMM}
\end{figure}

We computed $\Delta$BIC = BIC(1G) - BIC(2G), where negative values indicate that the single-component model is preferred. In most cases, both criteria favor a unimodal description of the data. The only exception is the $H_g-H_r$ distribution. In the \cite{ofek2012ApJ} sample, the AIC shows a mild preference for a two-component model, while the BIC favors a single component. Given the small size of this dataset and its limited color range, this result is more consistent with possible overfitting than with robust evidence for bimodality. For our sample, the BIC results overall do not support a bimodal distribution. However, the $H_g-H_r$ color distribution shows a slight visual hint of bimodality. In this case, the statistical evidence remains weak, although the AIC, which penalizes model complexity less strongly, yields a marginal preference for a two-component model. In cases where the AIC favors a two-component model, we further examined the fitted distributions by measuring the means and standard deviations of each Gaussian component. If the separation between the component means is smaller than the combined standard deviation, the components are considered to nearly coincide, supporting a unimodal interpretation. Only when the separation exceeds the combined standard deviation, and the AIC favors two components, do we consider bimodal behavior potentially meaningful. In our data and in \cite{ofek2012ApJ}, for the $H_g-H_r$ case, the separation is sufficient to make the statistical evidence for bimodality plausible, though still weak. The final results and comparison between the three works are shown in Figure \ref{GMM}.

While the $H_g-H_r$ distribution shows the strongest visual hint of bimodality, it remains statistically consistent with a unimodal distribution. For completeness, and based on the mineralogical differences found by \cite{pinillaalonso2025NatAs} for the DiSCo sample, we also analyzed their visible spectral slopes using the Hartigan dip test and found no statistically significant evidence for bimodality. To assess the robustness of this result, we performed 1\,000 Monte Carlo realizations, randomly varying the reported slopes within their quoted uncertainties, assuming Gaussian errors. In none of these realizations was unimodality rejected. The objects in common with the spectroscopic sample of \cite{pinillaalonso2025NatAs} are highlighted in Figure~\ref{appendix:bestsample}.  Although the number of objects in common is small, the different spectral types appear to occupy somewhat distinct regions in the color-color diagram, suggesting a tentative separation of surface compositions. 

From all tests and data analyzed, ours and from other absolute magnitudes samples, it is not yet completely clear that the absolute colors distribute in a non-unimodal manner. We discuss possible reasons below.

\begin{table}[h!]
\centering
\caption{Number of objects with determined absolute colors for each filter combination.}
\label{table_colors}
\begin{tabular}{l c | l c}
\hline
\textbf{Color index} & \textbf{N. objects} & \textbf{Color index} & \textbf{N. objects} \\
\hline
$(u-g)$ &  57  & $(r-i)$ &  429 \\
$(u-r)$ &  {59}  & $(r-z)$ &  344 \\
$(u-i)$ &  {59}  &  &  \\
$(u-z)$ &  54  & $(i-z)$ &  341 \\
 &   &  &  \\
$(g-r)$ &  436 &  &  \\
$(g-i)$ &  410 & $(g-z)$ &  353 \\
 &   &         &     \\
\hline
\end{tabular}
\end{table}

\subsubsection{Magnitude-color diagram and relations with orbital parameters}
\begin{figure}
    \centering
    \includegraphics[width=0.6\linewidth]{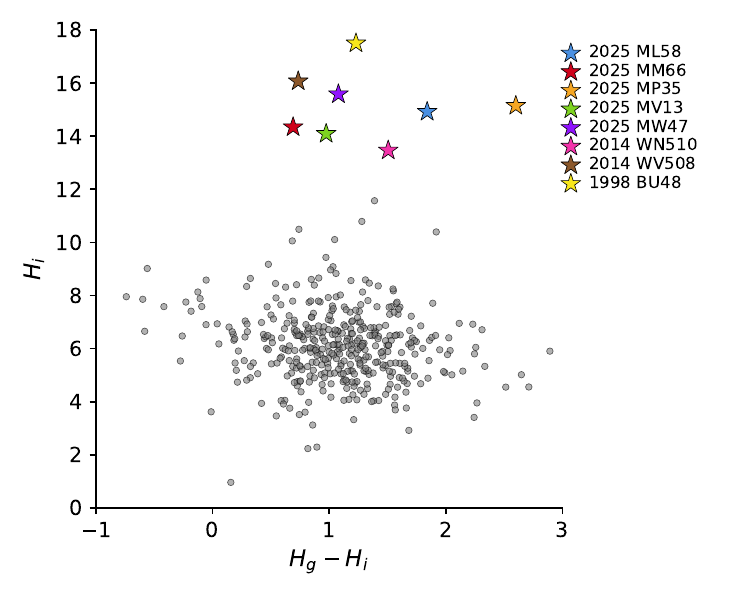}
    \caption{Absolute magnitude ${H_i}$ versus ${H_g - H_i}$. ,Colored stars mark objects observed by Rubin, including both recent discoveries and previously known targets.}
    \label{givsi}
\end{figure}

We also analyzed the absolute colors as a function of absolute magnitude (Figure \ref{givsi}). \citet{Hainaut2002}, in their Figure 3, presented color versus absolute magnitude diagrams and reported that “no striking bimodality appears in the plots.” Similarly, \citet{peixinho2012}, in their Figure 2, found that the full sample—despite showing two apparent peaks—did not provide strong evidence against unimodality, although they found an apparent bimodal distribution of colors when separating the sample into different dynamical classes. On the other hand, we detect a weak negative correlation between absolute color and absolute magnitude (Pearson $r = -0.11$), indicating that smaller or fainter objects tend to be slightly bluer. The Pearson coefficient quantifies the strength and direction of a linear relationship between two variables ($r=1$ for a perfect positive relationship, $r=-1$ for a perfect negative relationship, and $r=0$ for no linear relationship). This trend is also evident in Figure 2 of \citet{peixinho2012}. 

\citet{Hainaut2002} reported that for the full MBOSS population, colors are not correlated with absolute magnitude. But correlations appear when separating by dynamical classes: a positive correlation appears for Classical objects (fainter $M(\alpha)$ are redder), no correlation for Centaurs or Scattered TNOs, and a reversed trend for Plutinos (brighter $M(\alpha)$ are redder). In our dataset, we find evidence for a correlation between color and absolute magnitude when considering all objects. However, this behavior depends on the photometric band considered. When using $H_i$, the correlation is primarily driven by resonant objects, and it is no longer statistically significant once Plutinos are removed. In contrast, when using $H_g$, the correlation remains significant both within the Plutino subsample and among non-resonant objects, indicating that the trend is not solely driven by resonant TNOs.

In Figure \ref{givsi}, all RFL objects display similar color values, including 1998 BU48. While their colors are consistent with the broader TNO population, they occupy a relatively unpopulated region at the faint end of the parameter space, demonstrating that Rubin is probing new areas that were previously poorly sampled ($H$ fainter than 12). 

Previous works suggested possible correlations between the colors and orbital parameters of TNOs. For instance, \citep{Tegler2000} reported that objects with $q>40$~AU tend to be systematically redder  than objects with lower $q$ in $B-R$, a trend also noted by \cite{McBride2003} using $V-J$, although the latter authors cautioned that the apparent absence of blue objects at large perihelia might partly reflect detection biases in the near-infrared bands. Regarding inclination, \citep{Tegler2000} found that red classical objects generally exhibit low inclinations ($i \lesssim 13^\circ$), while \cite{Trujillo2002} and \cite{McBride2003} suggested a weak tendency toward bluer colors at higher inclinations, albeit with limited statistical significance. Recent results by \citep{pinillaalonso2025NatAs} show that when adding infrared information, three different compositional behaviors appear, which may relate to their original locations in the Solar System. In their Fig. 1, it is apparent that the Bowl-type objects have smaller spectral slopes than the other two types. The six objects in common with our data seem to agree with this picture; note that all Bowl-type objects are located in the bottom-left side in Figure \ref{appendix:bestsample}.

\begin{figure}
    \centering
    \includegraphics[width=0.4\textwidth]{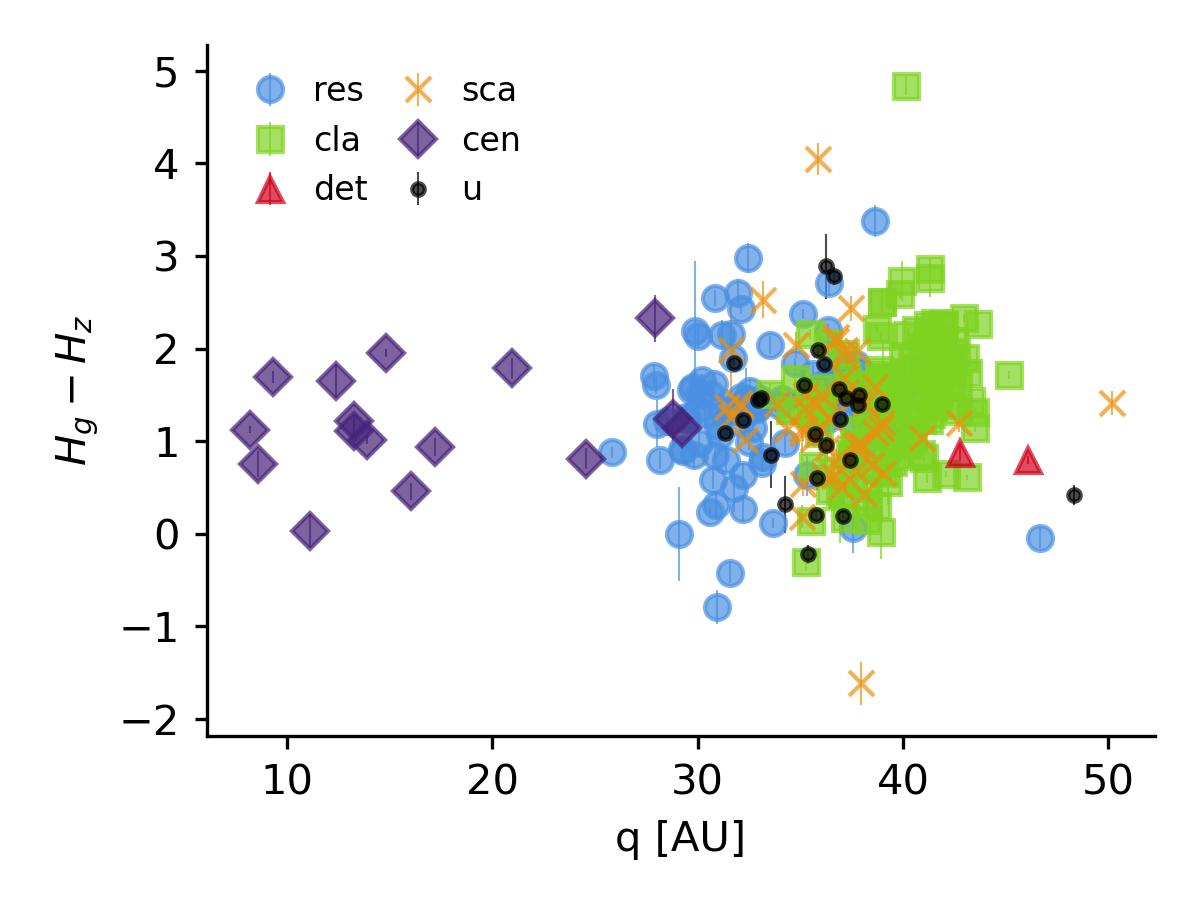}
    \includegraphics[width=0.4\textwidth]{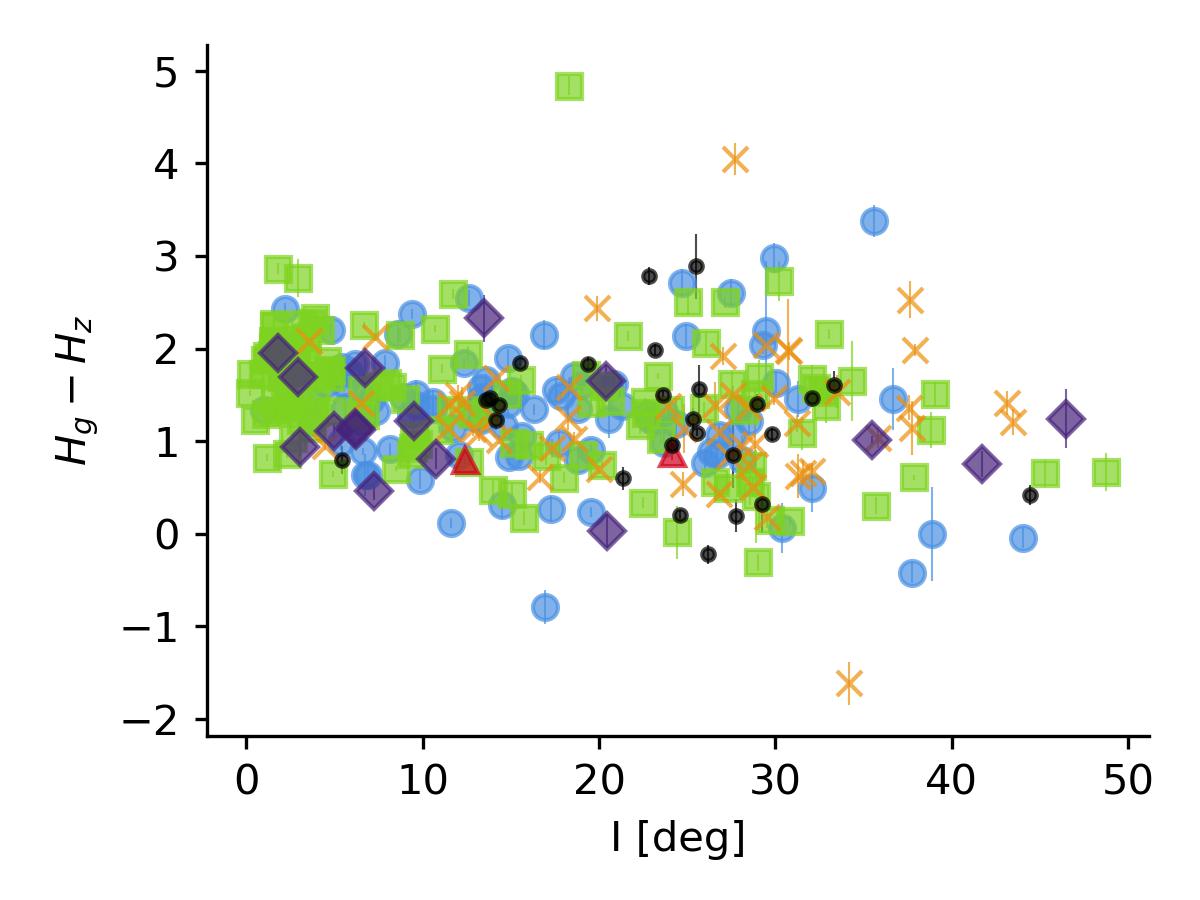}
    \caption{Absolute color $H_g - H_z$ as a function of (left) perihelion distance $q$ and (right) orbital inclination $i$ for trans-Neptunian objects from the filtered dataset. 
Different dynamical classes are shown with distinct symbols and colors: resonant (blue circles), classical (green squares), detached (red triangles), scattered (orange crosses), centaurs (purple diamonds), and unclassified (black points).
Error bars correspond to the standard deviation of the color measurements. Dynamical classes were assigned based on a merge with our dataset and the list from Johnstons Archive (\url{https://www.johnstonsarchive.net/astro/tnoslist.html}).
}
    \label{J-orbital}
\end{figure}

As shown in Figure~\ref{J-orbital}, we also find that objects with $q > 40$~au in our sample are  mostly redder than 0.5 in $H_g-H_z$, consistent with previous findings, except for a handful of bluer objects at $q>45$ au. However, we do not observe a pronounced color discontinuity between objects above and below this threshold. If anything, objects with $q > 35$~AU appear to have a broader color distribution than those with smaller perihelia, which may reflect differences between dynamical classes: classical objects (typically at larger $q$) are seen with a larger distribution of colors, while resonant objects (at smaller $q$) appear bluer, on average. Note, however, that the bluest classical objects also have the larger inclination, therefore being representatives of the hot classical population. Still with the inclinations, no clear global trend is apparent in our data, although there is a possible weak tendency for classical objects to become bluer with increasing inclination, especially for $i>5$ deg.  

\subsection{Phase coloring}
\begin{figure}
    \centering
    \includegraphics[width=\textwidth]{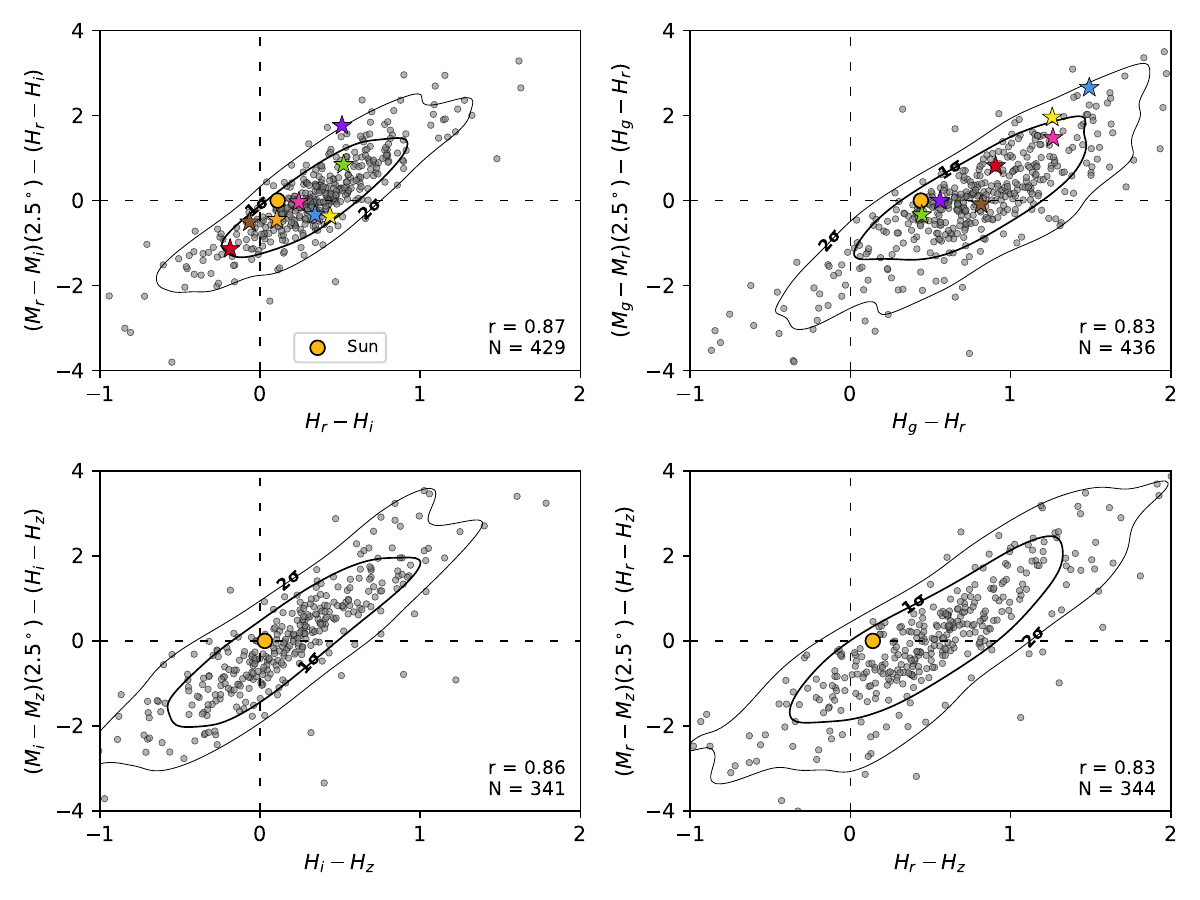}
    \caption{Absolute color ($\alpha = 0^\circ$), versus the color at $\alpha = 2.5^\circ$ minus the absolute color (color at opposition).  Each panel corresponds to one of the four color combinations with the strongest Pearson correlation: $r-i$, $g-r$, $i-z$, and $r-z$. Gray points represent the filtered sample, while black contours indicate the 68\% and 90\% density levels derived from a two-dimensional Gaussian kernel density estimation.  The yellow circle marks the solar color.  Colored stars highlight selected RFL-LSST objects, using the same color scheme adopted in previous figures.  The Pearson correlation coefficient and the number of objects are indicated in each panel.}
    \label{colors_vs_color_lsst}
\end{figure}

To test the effect of phase coloring we computed different colors $H_{\lambda_1}-H_{\lambda_2}$ and $(M_{\lambda_1}-M_{\lambda_2})(\alpha)$, where $M(\alpha)$ is the reduced magnitude al $\alpha>0$ deg. In Figure \ref{colors_vs_color_lsst} we plot the four colors with the largest Pearson correlation coefficients. The figure shows in the x-axis the color at opposition, while in the y-axis the coloring effect (color at $\alpha = 2.5$ deg minus color at opposition). The figure highlights that redder (bluer) objects at opposition become redder (bluer) with increasing $\alpha$, consistent with the findings of \cite{Alvarez-Candal2019,alvarezcandal2022, Alvarez-Candal2024, Colazo2026}.

From the figure, it is clear that there are objects with unrealistic high-values, color differences of $\pm 4$ in the extremes. These values should be considered with care. Nevertheless, it is clear that the relation holds even if the y-axis limit is restricted to a smaller range. These results are limited to the predictive power of a simple linear model fitted to the observational data, but they confirm that colors can become either redder or bluer with increasing phase angles.

\section{Discussion and conclusions}

In this work, we presented a comprehensive photometric analysis of trans-Neptunian objects (TNOs) by combining data from four major surveys: SDSS, Col-OSSOS, DES, and the recent Rubin First Look (RFL) release. Our compiled database includes 43\,677 observations in the $u$, $g$, $r$, $i$, and $z$ filters, from which we derived 2\,193 individual phase curves corresponding to 781 unique objects. The complete database of observations, together with the catalog of phase-curve parameters and the catalog of
colors measurements, is publicly available at OSF\footnote{\url{https://osf.io/nw846}} under the repository of the first author.

From these data, we computed absolute colors for all objects in the sample, obtaining 57 for $(u-g)$, 59 for $(u-r)$, 59 for $(u-i)$, 54 for $(u-z)$, 436 for $(g-r)$, 410 for $(g-i)$, 353 for $(g-z)$, 429 for $(r-i)$, 344 for $(r-z)$, and 341 for $(i-z)$, summing up to a total of 2\,542 absolute-color measurements, an increase by a factor of 25 with respect to \cite{Alvarez-Candal2019}'s $\approx100$ measurements of $(H_V-H_R)$.
We analyzed the existence of more than one mode in our datasets without finding strong evidence to support it. This lack of confirmation of multimodal distributions is a due to several factors: First, our data only include visible colors. Note that the distinction between the brightIR and faintIR populations in \cite{Fraser2023} uses $J$ data, which we do not include; while the NIRB and NIRF populations in \cite{bernardinelli2025AJ} are detected after a thorough statistical analysis, which was out of the scope of this work. Second, the nature of our method increases the uncertainties of our absolute magnitudes which may blur dips or gaps in the distributions. Finally, the anti-correlation between absolute colors and $\Delta\beta$ implies that redder (bluer) objects at opposition tend to have even redder (bluer) colors with increasing $\alpha$, naturally decreasing the width of any dip seen at $\alpha>0$ deg. 

Interestingly, Figure \ref{givsi} shows that the Legacy Survey for Space and Time will fill the space for fainter absolute magnitudes, opening a new realm of color distributions and relations because of its high-quality colors for tens of thousands of objects \citep{murtagh2025AJ,kurlander2025AJ}, of which we see a first glimpse in this work with RFL data.

Therefore, beyond supplying an extensive catalog of TNO observations and derived parameters, our dataset includes the first-ever photometric measurements obtained with the Rubin Observatory during the RFL campaign. We derived phase curves, absolute colors, and spectral slopes for eight RFL objects—five newly discovered TNOs and three previously known ones—and placed them in the context of existing surveys (Col-OSSOS, DES, and SDSS). These early LSST observations demonstrate the tremendous potential of Rubin data: the detected objects occupy sparsely populated regions of parameter space, particularly at faint magnitudes. The inclusion of these data will significantly expand the sampled phase space and shift current observational biases toward smaller and fainter objects. This improved coverage will enhance our ability to assess whether apparent groupings in color or brightness are intrinsic to different dynamical populations, size ranges, or survey selection effects.

We analyzed the phase coloring effect for all objects seen in Figure \ref{colors_vs_color_lsst}, more than a factor of four compared to previous studies.
We confirm previous results showing that colors change with phase angle and may become bluer or redder. Visually, the $H_r-H_z$ seems to be the most affected by the area covered by the data, suggesting that phase color may be better detected in wider wavelength coverage \citep[see also][]{Colazo2026}.

Finally, we investigated potential correlations between photometric parameters and orbital elements. While previous studies have largely focused on relative magnitudes, we extended the analysis to absolute colors. Within our sample, we do not find any strong or statistically significant correlations between color, phase slope, or spectral slope parameters and orbital quantities such as inclination or perihelion distance. This suggests that any such relationships, if present, are subtle and may become more evident as larger and more homogeneous datasets from Rubin Observatory become available in the coming years.

%% Also note that the akcnowlodgment environment does not support long amounts of text. If you have a lot of people and institutions to acknowledge, do not use this command. Instead, create a new \section{Acknowledgments}.

\begin{acknowledgments}
MC was supported by grant No. 2022/45/B/ST9/00267 from the National Science Centre, Poland. 
AAC acknowledges financial support from the Severo Ochoa grant CEX2021-001131-S funded by MCIN/AEI/10.13039/501100011033 and the Spanish project PID2023-153123NB-I00, funded by MCIN/AEI.
MC acknowledges financial support from IDUB 185/07/POB4/0004B.
This work made use of the Rubin Observatory Minor Planet Center upload scripts and data access tools provided by the Asteroid Institute \citep{koumjian2025}. We also thank the Vera C. Rubin Observatory Solar System Science Collaboration (SSSC) for supporting the development of community tools for solar system data analysis.  
We gratefully acknowledge Sean O'Brien for providing the upload scripts used in this work. Sean O'Brien is supported by the UK Science and Technology Facilities Council (STFC) grant ST/X001253/1.
For language editing and translation, we utilized DeepL, ChatGPT, and Grammarly.
\end{acknowledgments}


\vspace{5mm}


\software{astropy \citep{Astropy2013, Astropy2018}; pandas \citep{mckinney2010pandas, reback2020pandas}
          }



\newpage

\appendix

\section{RFL - Phase Curves}\label{LSST-phasecurves}

\begin{figure}[ht]
    \centering
    \includegraphics[width=0.3\linewidth]{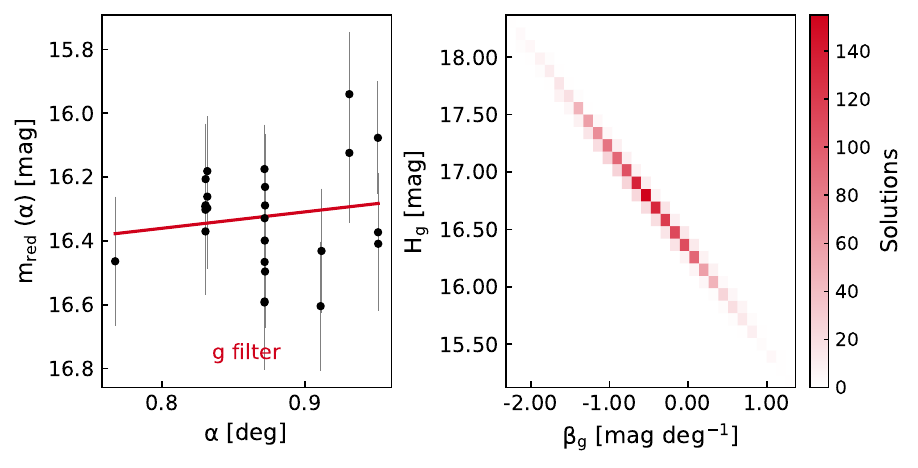}
    \includegraphics[width=0.3\linewidth]{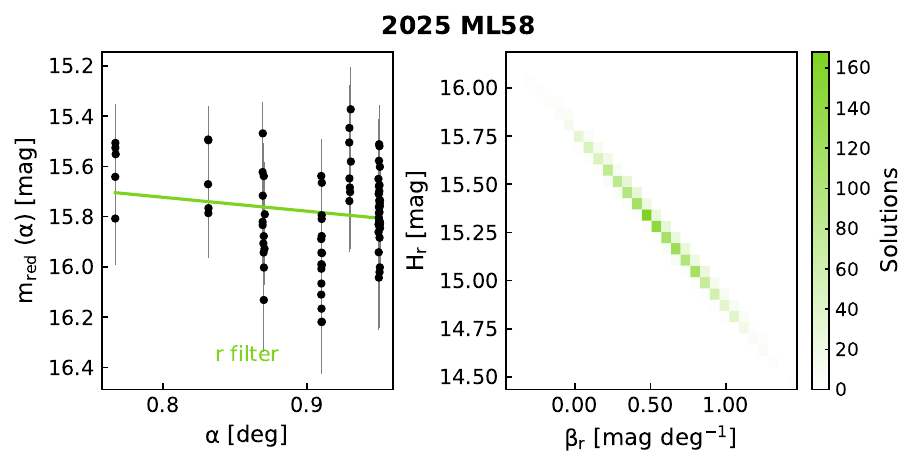}
    \includegraphics[width=0.3\linewidth]{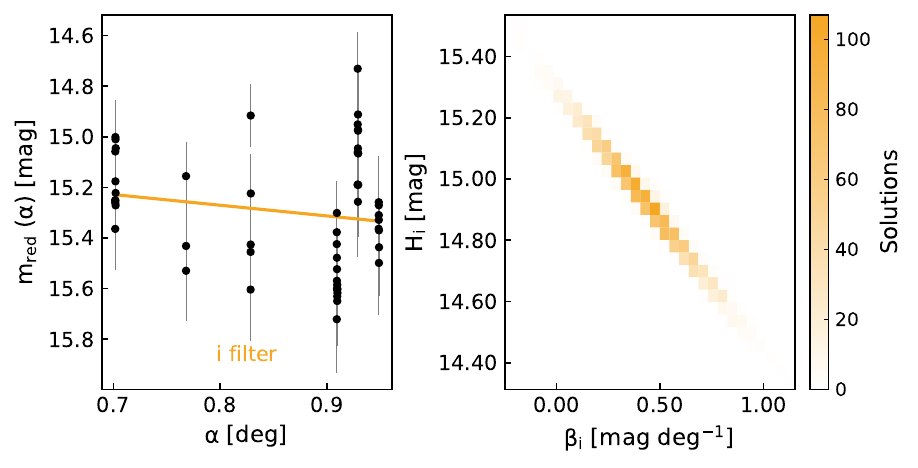}
    \\
    \includegraphics[width=0.3\linewidth]{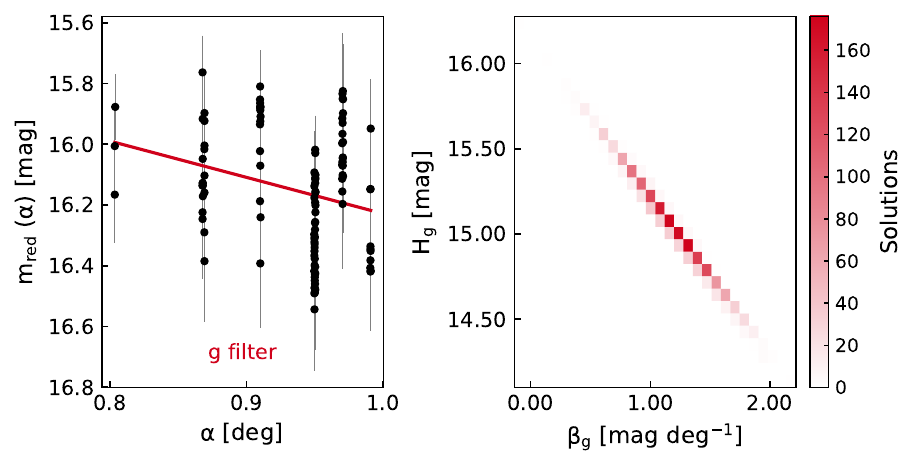}
    \includegraphics[width=0.3\linewidth]{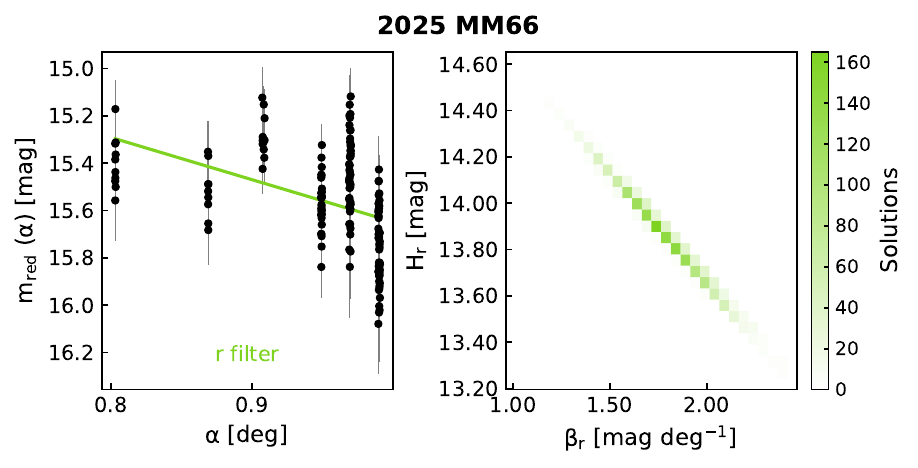}
    \includegraphics[width=0.3\linewidth]{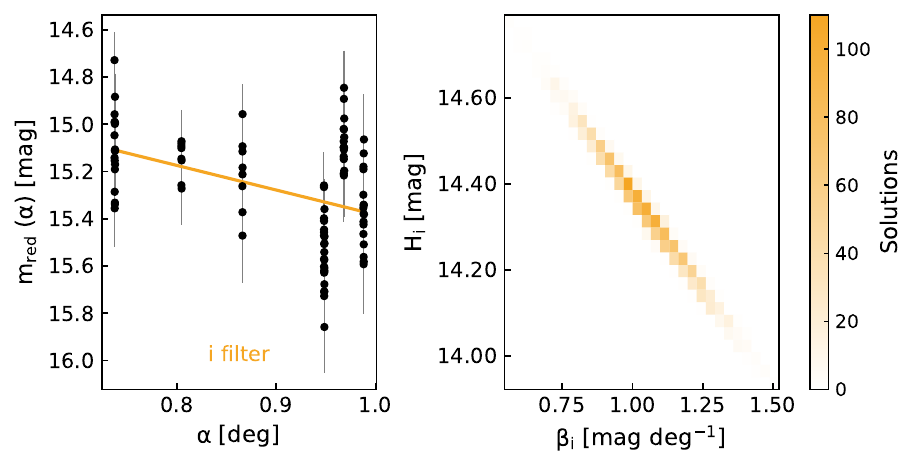}
    \\
    \includegraphics[width=0.3\linewidth]{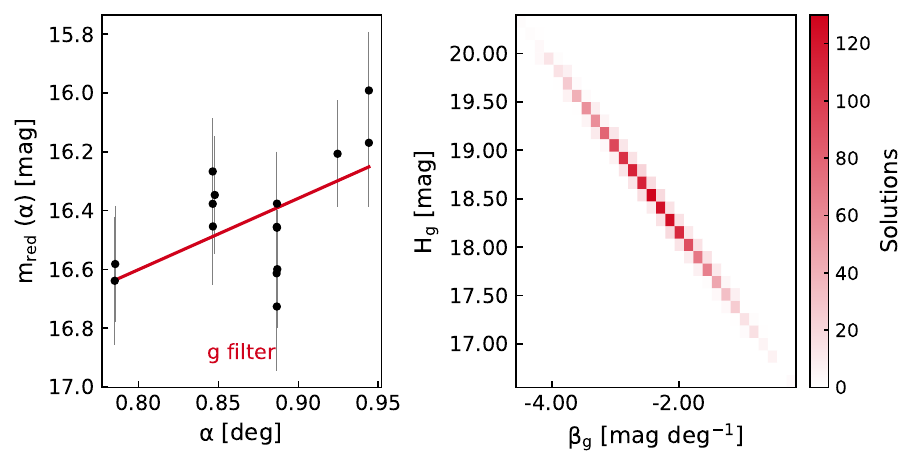}
    \includegraphics[width=0.3\linewidth]{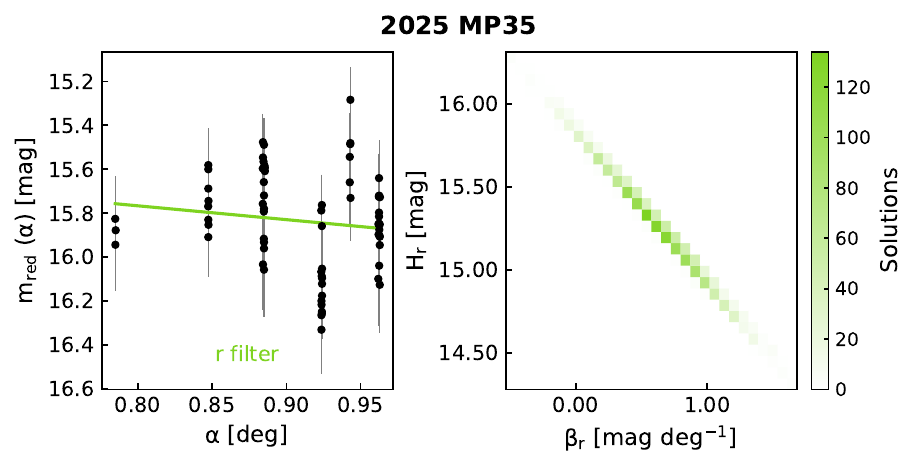}
    \includegraphics[width=0.3\linewidth]{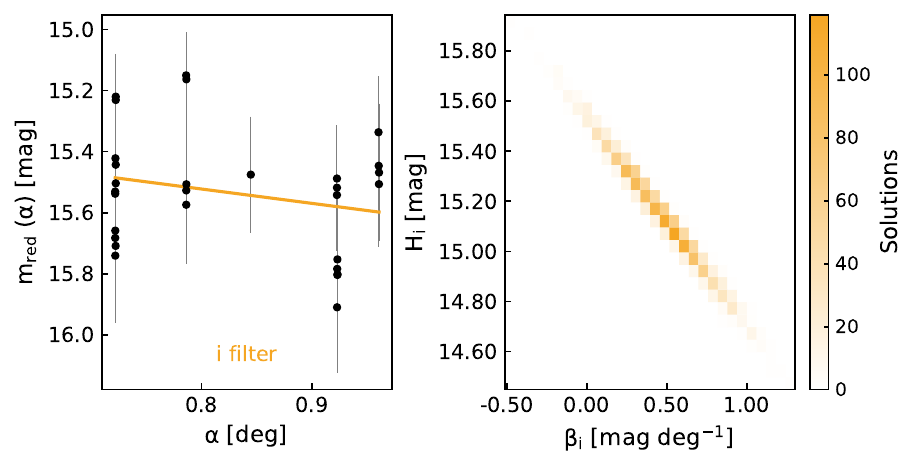}
    \\
    \includegraphics[width=0.3\linewidth]{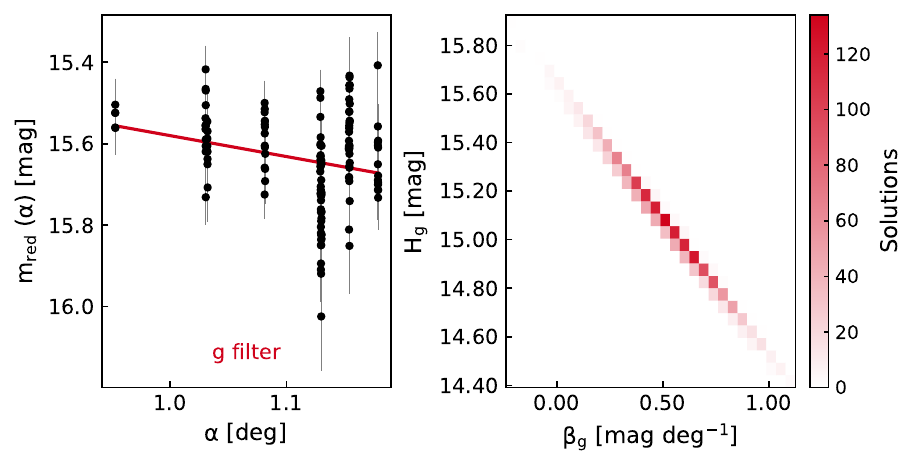}
    \includegraphics[width=0.3\linewidth]{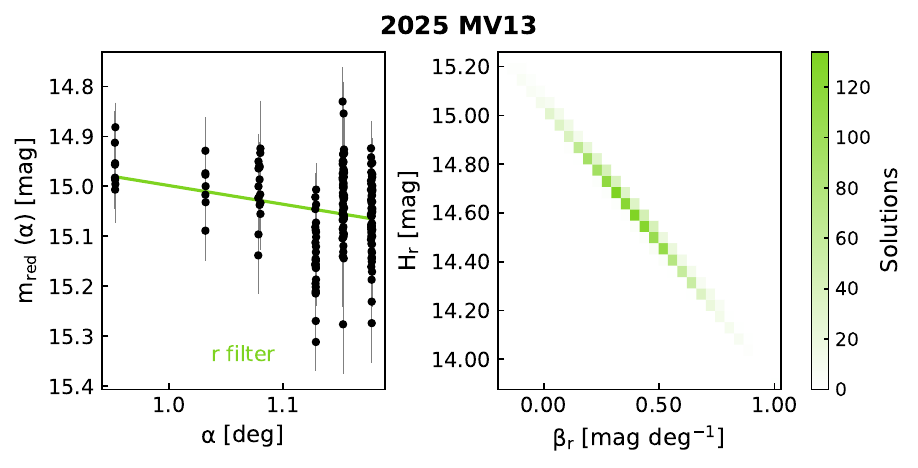}
    \includegraphics[width=0.3\linewidth]{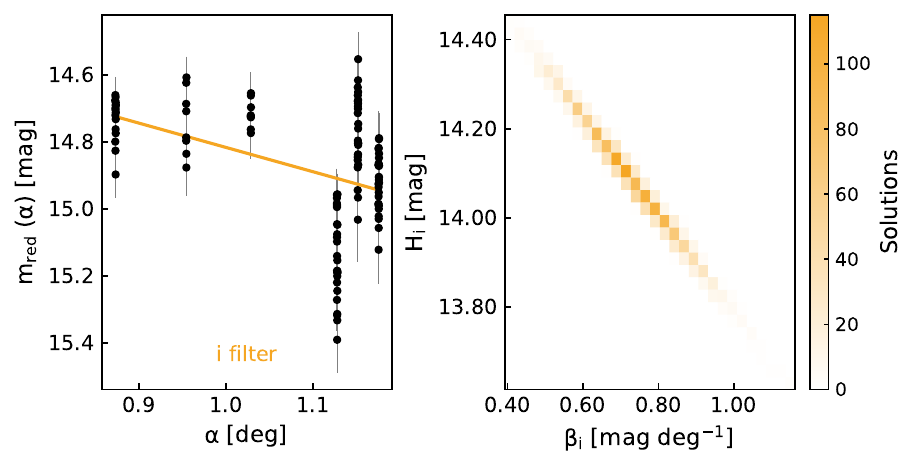}
    \\
    \includegraphics[width=0.3\linewidth]{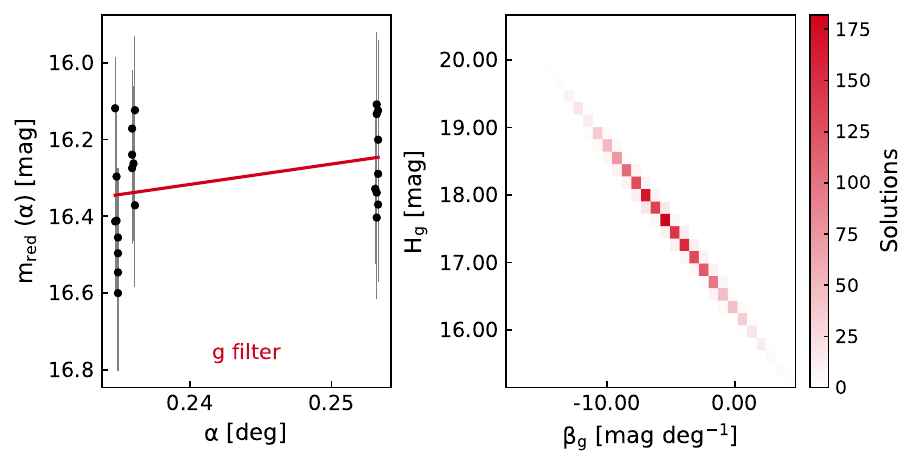}
    \includegraphics[width=0.3\linewidth]{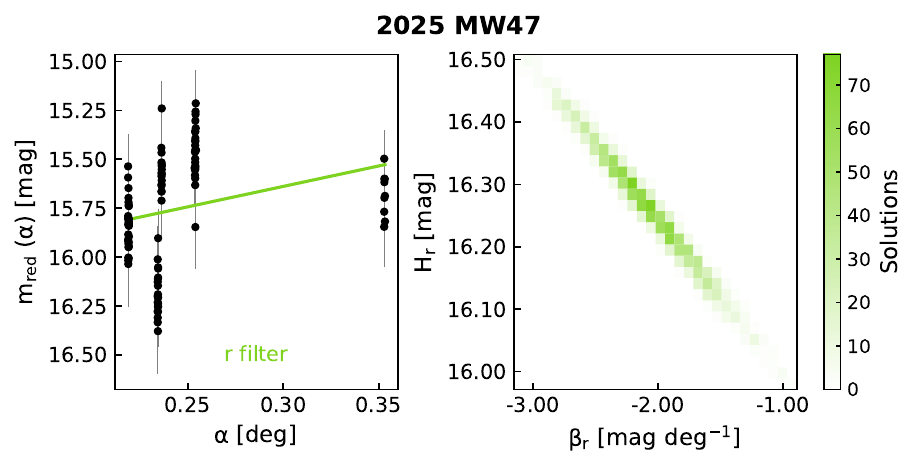}
    \includegraphics[width=0.3\linewidth]{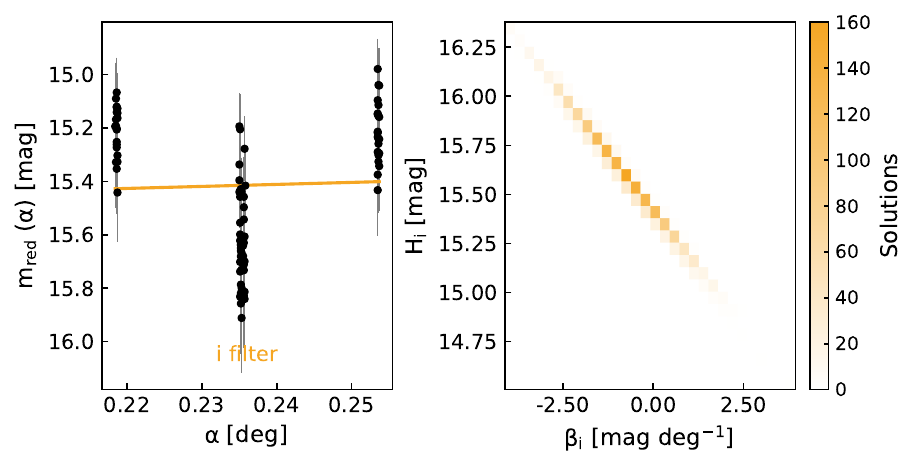}
    \\
    \includegraphics[width=0.3\linewidth]{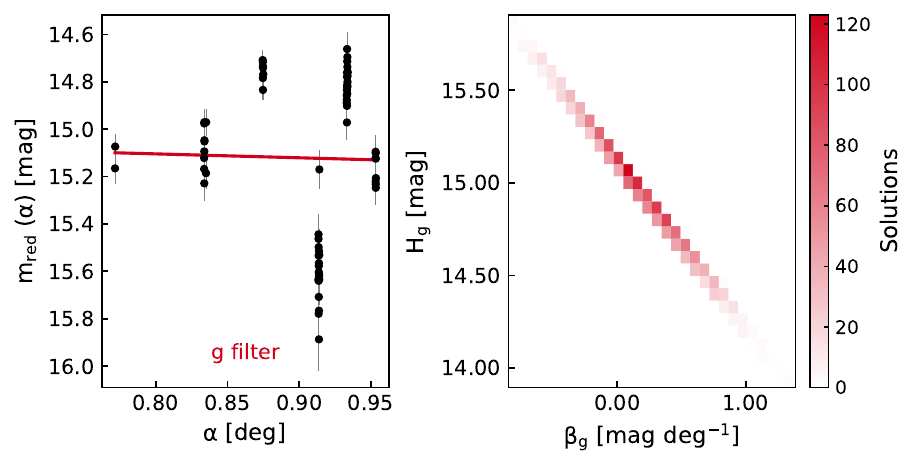}
    \includegraphics[width=0.3\linewidth]{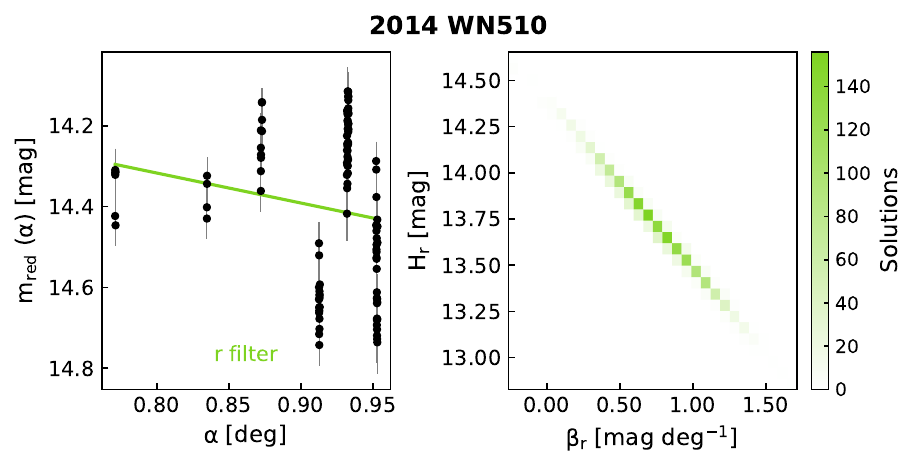}
    \includegraphics[width=0.3\linewidth]{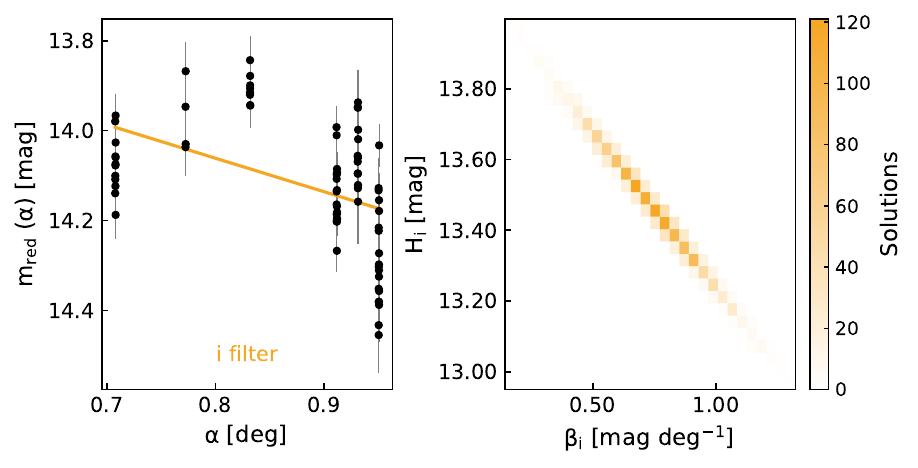}
    \\
    \includegraphics[width=0.3\linewidth]{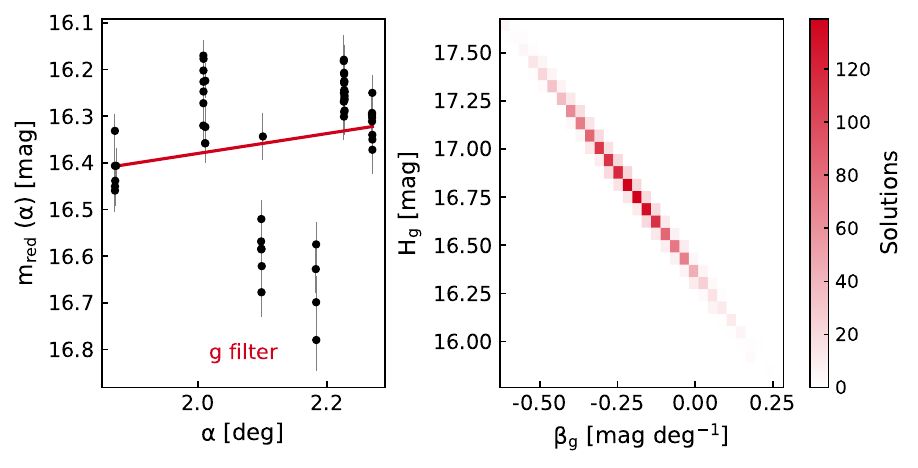}
    \includegraphics[width=0.3\linewidth]{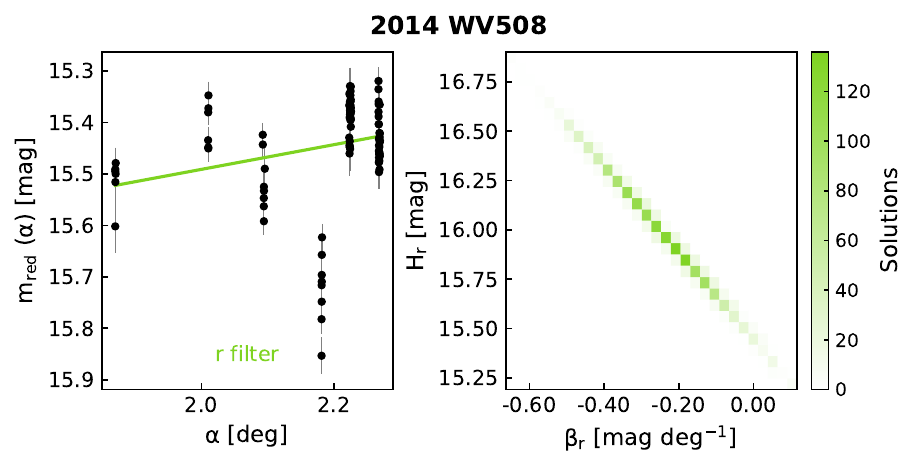}
    \includegraphics[width=0.3\linewidth]{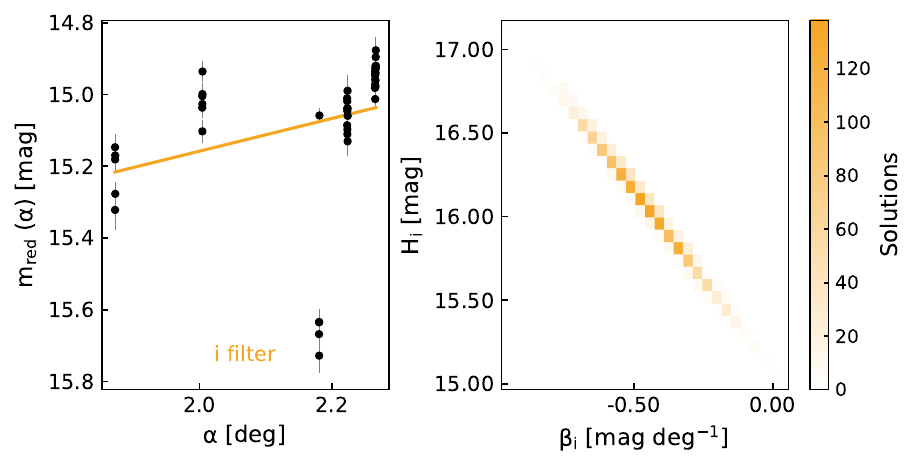}
    \\
    \includegraphics[width=0.3\linewidth]{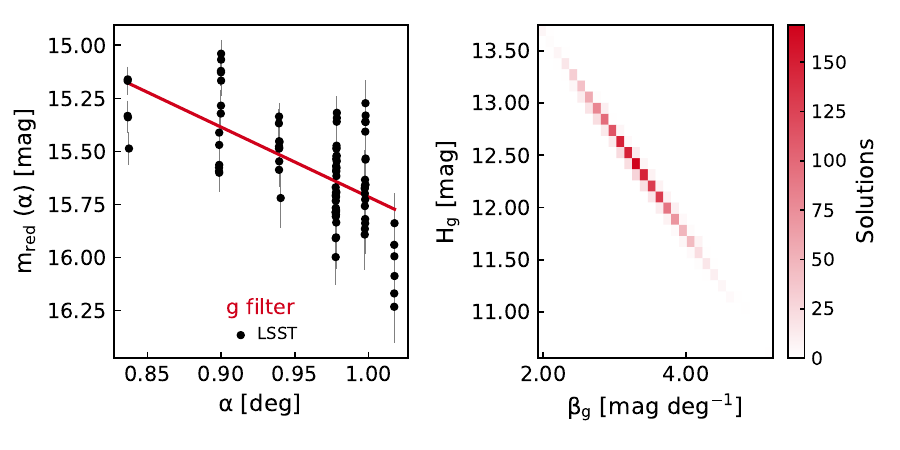}
    \includegraphics[width=0.3\linewidth]{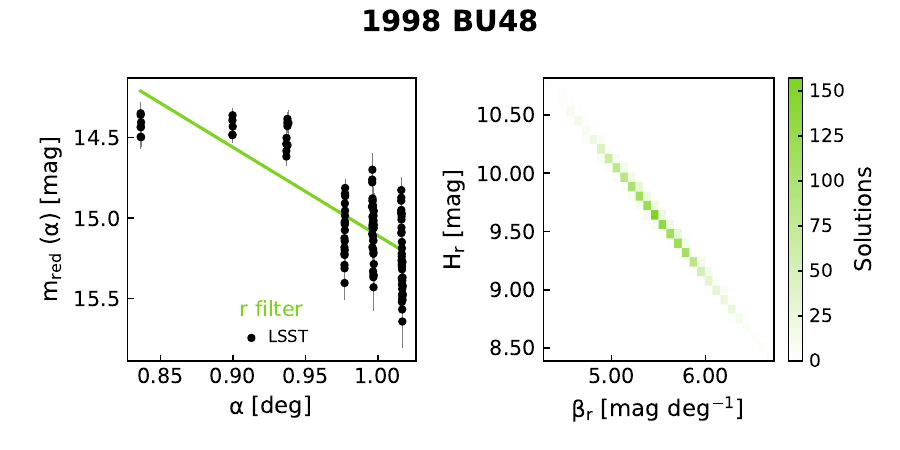}
    \includegraphics[width=0.3\linewidth]{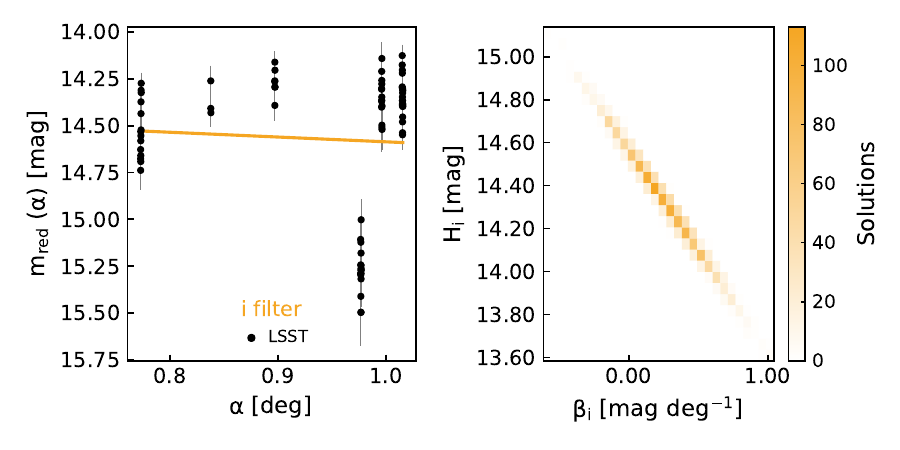}
    \\
    \caption{Phase curve fits for LSST TNOs across different filters. Each pair of panels shows, from left to right, the observed reduced magnitudes versus phase angle with the median linear fit, and the Monte Carlo distribution. The narrow distributions reflect the limited magnitude span of the current LSST data; objects with few observations are therefore only weakly constrained.}
    \label{fig:all_phasecurves}
\end{figure}

\bibliography{sample631}{}

@ARTICLE{Abbott2021,
       author = {{Abbott}, T.~M.~C. and {Adam{\'o}w}, M. and {Aguena}, M. and {Allam}, S. and {Amon}, A. and {Annis}, J. and {Avila}, S. and {Bacon}, D. and {Banerji}, M. and {Bechtol}, K. and {Becker}, M.~R. and {Bernstein}, G.~M. and {Bertin}, E. and {Bhargava}, S. and {Bridle}, S.~L. and {Brooks}, D. and {Burke}, D.~L. and {Carnero Rosell}, A. and {Carrasco Kind}, M. and {Carretero}, J. and {Castander}, F.~J. and {Cawthon}, R. and {Chang}, C. and {Choi}, A. and {Conselice}, C. and {Costanzi}, M. and {Crocce}, M. and {da Costa}, L.~N. and {Davis}, T.~M. and {De Vicente}, J. and {DeRose}, J. and {Desai}, S. and {Diehl}, H.~T. and {Dietrich}, J.~P. and {Drlica-Wagner}, A. and {Eckert}, K. and {Elvin-Poole}, J. and {Everett}, S. and {Evrard}, A.~E. and {Ferrero}, I. and {Fert{\'e}}, A. and {Flaugher}, B. and {Fosalba}, P. and {Friedel}, D. and {Frieman}, J. and {Garc{\'\i}a-Bellido}, J. and {Gaztanaga}, E. and {Gelman}, L. and {Gerdes}, D.~W. and {Giannantonio}, T. and {Gill}, M.~S.~S. and {Gruen}, D. and {Gruendl}, R.~A. and {Gschwend}, J. and {Gutierrez}, G. and {Hartley}, W.~G. and {Hinton}, S.~R. and {Hollowood}, D.~L. and {Honscheid}, K. and {Huterer}, D. and {James}, D.~J. and {Jeltema}, T. and {Johnson}, M.~D. and {Kent}, S. and {Kron}, R. and {Kuehn}, K. and {Kuropatkin}, N. and {Lahav}, O. and {Li}, T.~S. and {Lidman}, C. and {Lin}, H. and {MacCrann}, N. and {Maia}, M.~A.~G. and {Manning}, T.~A. and {Maloney}, J.~D. and {March}, M. and {Marshall}, J.~L. and {Martini}, P. and {Melchior}, P. and {Menanteau}, F. and {Miquel}, R. and {Morgan}, R. and {Myles}, J. and {Neilsen}, E. and {Ogando}, R.~L.~C. and {Palmese}, A. and {Paz-Chinch{\'o}n}, F. and {Petravick}, D. and {Pieres}, A. and {Plazas}, A.~A. and {Pond}, C. and {Rodriguez-Monroy}, M. and {Romer}, A.~K. and {Roodman}, A. and {Rykoff}, E.~S. and {Sako}, M. and {Sanchez}, E. and {Santiago}, B. and {Scarpine}, V. and {Serrano}, S. and {Sevilla-Noarbe}, I. and {Smith}, J. Allyn and {Smith}, M. and {Soares-Santos}, M. and {Suchyta}, E. and {Swanson}, M.~E.~C. and {Tarle}, G. and {Thomas}, D. and {To}, C. and {Tremblay}, P.~E. and {Troxel}, M.~A. and {Tucker}, D.~L. and {Turner}, D.~J. and {Varga}, T.~N. and {Walker}, A.~R. and {Wechsler}, R.~H. and {Weller}, J. and {Wester}, W. and {Wilkinson}, R.~D. and {Yanny}, B. and {Zhang}, Y. and {Nikutta}, R. and {Fitzpatrick}, M. and {Jacques}, A. and {Scott}, A. and {Olsen}, K. and {Huang}, L. and {Herrera}, D. and {Juneau}, S. and {Nidever}, D. and {Weaver}, B.~A. and {Adean}, C. and {Correia}, V. and {de Freitas}, M. and {Freitas}, F.~N. and {Singulani}, C. and {Vila-Verde}, G. and {Linea Science Server}},
        title = "{The Dark Energy Survey Data Release 2}",
      journal = {\apjs},
         year = 2021,
        month = aug,
       volume = {255},
       number = {2},
          eid = {20},
        pages = {20},
          doi = {10.3847/1538-4365/ac00b3},
archivePrefix = {arXiv},
       eprint = {2101.05765},
 primaryClass = {astro-ph.IM},
       adsurl = {https://ui.adsabs.harvard.edu/abs/2021ApJS..255...20A}
}

@ARTICLE{Alvarez-Candal2016,
       author = {{Alvarez-Candal}, A. and {Pinilla-Alonso}, N. and {Ortiz}, J.~L. and {Duffard}, R. and {Morales}, N. and {Santos-Sanz}, P. and {Thirouin}, A. and {Silva}, J.~S.},
        title = "{Absolute magnitudes and phase coefficients of trans-Neptunian objects}",
      journal = {\aap},
         year = 2016,
        month = feb,
       volume = {586},
          eid = {A155},
        pages = {A155},
          doi = {10.1051/0004-6361/201527161},
archivePrefix = {arXiv},
       eprint = {1511.09401},
 primaryClass = {astro-ph.EP},
       adsurl = {https://ui.adsabs.harvard.edu/abs/2016A&A...586A.155A}
}

@ARTICLE{Alvarez-Candal2019,
       author = {{Alvarez-Candal}, Alvaro and {Ayala-Loera}, Carmen and {Gil-Hutton}, Ricardo and {Ortiz}, Jos{\'e} Luis and {Santos-Sanz}, Pablo and {Duffard}, Ren{\'e}},
        title = "{Absolute colours and phase coefficients of trans-Neptunian objects: correlations and populations}",
      journal = {\mnras},
         year = 2019,
        month = sep,
       volume = {488},
       number = {3},
        pages = {3035-3044},
          doi = {10.1093/mnras/stz1880},
archivePrefix = {arXiv},
       eprint = {1907.03036},
 primaryClass = {astro-ph.EP},
       adsurl = {https://ui.adsabs.harvard.edu/abs/2019MNRAS.488.3035A}
}

@ARTICLE{Alvarez-Candal2024,
       author = {{Alvarez-Candal}, Alvaro},
        title = "{The multiwavelength phase curves of small bodies. Phase coloring}",
      journal = {\aap},
         year = 2024,
        month = may,
       volume = {685},
          eid = {A29},
        pages = {A29},
          doi = {10.1051/0004-6361/202348287},
archivePrefix = {arXiv},
       eprint = {2402.11113},
 primaryClass = {astro-ph.EP},
       adsurl = {https://ui.adsabs.harvard.edu/abs/2024A&A...685A..29A}
}

@ARTICLE{Astropy2013,
       author = {{Astropy Collaboration} and {Robitaille}, Thomas P. and {Tollerud}, Erik J. and {Greenfield}, Perry and {Droettboom}, Michael and {Bray}, Erik and {Aldcroft}, Tom and {Davis}, Matt and {Ginsburg}, Adam and {Price-Whelan}, Adrian M. and {Kerzendorf}, Wolfgang E. and {Conley}, Alexander and {Crighton}, Neil and {Barbary}, Kyle and {Muna}, Demitri and {Ferguson}, Henry and {Grollier}, Fr{\'e}d{\'e}ric and {Parikh}, Madhura M. and {Nair}, Prasanth H. and {Unther}, Hans M. and {Deil}, Christoph and {Woillez}, Julien and {Conseil}, Simon and {Kramer}, Roban and {Turner}, James E.~H. and {Singer}, Leo and {Fox}, Ryan and {Weaver}, Benjamin A. and {Zabalza}, Victor and {Edwards}, Zachary I. and {Azalee Bostroem}, K. and {Burke}, D.~J. and {Casey}, Andrew R. and {Crawford}, Steven M. and {Dencheva}, Nadia and {Ely}, Justin and {Jenness}, Tim and {Labrie}, Kathleen and {Lim}, Pey Lian and {Pierfederici}, Francesco and {Pontzen}, Andrew and {Ptak}, Andy and {Refsdal}, Brian and {Servillat}, Mathieu and {Streicher}, Ole},
        title = "{Astropy: A community Python package for astronomy}",
      journal = {\aap},
         year = 2013,
        month = oct,
       volume = {558},
          eid = {A33},
        pages = {A33},
          doi = {10.1051/0004-6361/201322068},
archivePrefix = {arXiv},
       eprint = {1307.6212},
 primaryClass = {astro-ph.IM},
       adsurl = {https://ui.adsabs.harvard.edu/abs/2013A&A...558A..33A}
}

@ARTICLE{Astropy2018,
       author = {{Astropy Collaboration} and {Price-Whelan}, A.~M. and {Sip{\H{o}}cz}, B.~M. and {G{\"u}nther}, H.~M. and {Lim}, P.~L. and {Crawford}, S.~M. and {Conseil}, S. and {Shupe}, D.~L. and {Craig}, M.~W. and {Dencheva}, N. and {Ginsburg}, A. and {VanderPlas}, J.~T. and {Bradley}, L.~D. and {P{\'e}rez-Su{\'a}rez}, D. and {de Val-Borro}, M. and {Aldcroft}, T.~L. and {Cruz}, K.~L. and {Robitaille}, T.~P. and {Tollerud}, E.~J. and {Ardelean}, C. and {Babej}, T. and {Bach}, Y.~P. and {Bachetti}, M. and {Bakanov}, A.~V. and {Bamford}, S.~P. and {Barentsen}, G. and {Barmby}, P. and {Baumbach}, A. and {Berry}, K.~L. and {Biscani}, F. and {Boquien}, M. and {Bostroem}, K.~A. and {Bouma}, L.~G. and {Brammer}, G.~B. and {Bray}, E.~M. and {Breytenbach}, H. and {Buddelmeijer}, H. and {Burke}, D.~J. and {Calderone}, G. and {Cano Rodr{\'\i}guez}, J.~L. and {Cara}, M. and {Cardoso}, J.~V.~M. and {Cheedella}, S. and {Copin}, Y. and {Corrales}, L. and {Crichton}, D. and {D'Avella}, D. and {Deil}, C. and {Depagne}, {\'E}. and {Dietrich}, J.~P. and {Donath}, A. and {Droettboom}, M. and {Earl}, N. and {Erben}, T. and {Fabbro}, S. and {Ferreira}, L.~A. and {Finethy}, T. and {Fox}, R.~T. and {Garrison}, L.~H. and {Gibbons}, S.~L.~J. and {Goldstein}, D.~A. and {Gommers}, R. and {Greco}, J.~P. and {Greenfield}, P. and {Groener}, A.~M. and {Grollier}, F. and {Hagen}, A. and {Hirst}, P. and {Homeier}, D. and {Horton}, A.~J. and {Hosseinzadeh}, G. and {Hu}, L. and {Hunkeler}, J.~S. and {Ivezi{\'c}}, {\v{Z}}. and {Jain}, A. and {Jenness}, T. and {Kanarek}, G. and {Kendrew}, S. and {Kern}, N.~S. and {Kerzendorf}, W.~E. and {Khvalko}, A. and {King}, J. and {Kirkby}, D. and {Kulkarni}, A.~M. and {Kumar}, A. and {Lee}, A. and {Lenz}, D. and {Littlefair}, S.~P. and {Ma}, Z. and {Macleod}, D.~M. and {Mastropietro}, M. and {McCully}, C. and {Montagnac}, S. and {Morris}, B.~M. and {Mueller}, M. and {Mumford}, S.~J. and {Muna}, D. and {Murphy}, N.~A. and {Nelson}, S. and {Nguyen}, G.~H. and {Ninan}, J.~P. and {N{\"o}the}, M. and {Ogaz}, S. and {Oh}, S. and {Parejko}, J.~K. and {Parley}, N. and {Pascual}, S. and {Patil}, R. and {Patil}, A.~A. and {Plunkett}, A.~L. and {Prochaska}, J.~X. and {Rastogi}, T. and {Reddy Janga}, V. and {Sabater}, J. and {Sakurikar}, P. and {Seifert}, M. and {Sherbert}, L.~E. and {Sherwood-Taylor}, H. and {Shih}, A.~Y. and {Sick}, J. and {Silbiger}, M.~T. and {Singanamalla}, S. and {Singer}, L.~P. and {Sladen}, P.~H. and {Sooley}, K.~A. and {Sornarajah}, S. and {Streicher}, O. and {Teuben}, P. and {Thomas}, S.~W. and {Tremblay}, G.~R. and {Turner}, J.~E.~H. and {Terr{\'o}n}, V. and {van Kerkwijk}, M.~H. and {de la Vega}, A. and {Watkins}, L.~L. and {Weaver}, B.~A. and {Whitmore}, J.~B. and {Woillez}, J. and {Zabalza}, V. and {Astropy Contributors}},
        title = "{The Astropy Project: Building an Open-science Project and Status of the v2.0 Core Package}",
      journal = {\aj},
         year = 2018,
        month = sep,
       volume = {156},
       number = {3},
          eid = {123},
        pages = {123},
          doi = {10.3847/1538-3881/aabc4f},
archivePrefix = {arXiv},
       eprint = {1801.02634},
 primaryClass = {astro-ph.IM},
       adsurl = {https://ui.adsabs.harvard.edu/abs/2018AJ....156..123A}
}

@ARTICLE{Bannister2016a,
       author = {{Bannister}, Michele T. and {Alexandersen}, Mike and {Benecchi}, Susan D. and {Chen}, Ying-Tung and {Delsanti}, Audrey and {Fraser}, Wesley C. and {Gladman}, Brett J. and {Granvik}, Mikael and {Grundy}, Will M. and {Guilbert-Lepoutre}, Aur{\'e}lie and {Gwyn}, Stephen D.~J. and {Ip}, Wing-Huen and {Jakubik}, Marian and {Jones}, R. Lynne and {Kaib}, Nathan and {Kavelaars}, J.~J. and {Lacerda}, Pedro and {Lawler}, Samantha and {Lehner}, Matthew J. and {Lin}, Hsing Wen and {Lykawka}, Patryk Sofia and {Marsset}, Michael and {Murray-Clay}, Ruth and {Noll}, Keith S. and {Parker}, Alex and {Petit}, Jean-Marc and {Pike}, Rosemary E. and {Rousselot}, Philippe and {Schwamb}, Megan E. and {Shankman}, Cory and {Veres}, Peter and {Vernazza}, Pierre and {Volk}, Kathryn and {Wang}, Shiang-Yu and {Weryk}, Robert},
        title = "{OSSOS. IV. Discovery of a Dwarf Planet Candidate in the 9:2 Resonance with Neptune}",
      journal = {\aj},
         year = 2016,
        month = dec,
       volume = {152},
       number = {6},
          eid = {212},
        pages = {212},
          doi = {10.3847/0004-6256/152/6/212},
archivePrefix = {arXiv},
       eprint = {1607.06970},
 primaryClass = {astro-ph.EP},
       adsurl = {https://ui.adsabs.harvard.edu/abs/2016AJ....152..212B}
}

@ARTICLE{Bannister2016b,
       author = {{Bannister}, Michele T. and {Kavelaars}, J.~J. and {Petit}, Jean-Marc and {Gladman}, Brett J. and {Gwyn}, Stephen D.~J. and {Chen}, Ying-Tung and {Volk}, Kathryn and {Alexandersen}, Mike and {Benecchi}, Susan D. and {Delsanti}, Audrey and {Fraser}, Wesley C. and {Granvik}, Mikael and {Grundy}, Will M. and {Guilbert-Lepoutre}, Aur{\'e}lie and {Hestroffer}, Daniel and {Ip}, Wing-Huen and {Jakubik}, Marian and {Jones}, R. Lynne and {Kaib}, Nathan and {Kavelaars}, Catherine F. and {Lacerda}, Pedro and {Lawler}, Samantha and {Lehner}, Matthew J. and {Lin}, Hsing Wen and {Lister}, Tim and {Lykawka}, Patryk Sofia and {Monty}, Stephanie and {Marsset}, Michael and {Murray-Clay}, Ruth and {Noll}, Keith S. and {Parker}, Alex and {Pike}, Rosemary E. and {Rousselot}, Philippe and {Rusk}, David and {Schwamb}, Megan E. and {Shankman}, Cory and {Sicardy}, Bruno and {Vernazza}, Pierre and {Wang}, Shiang-Yu},
        title = "{The Outer Solar System Origins Survey. I. Design and First-quarter Discoveries}",
      journal = {\aj},
         year = 2016,
        month = sep,
       volume = {152},
       number = {3},
          eid = {70},
        pages = {70},
          doi = {10.3847/0004-6256/152/3/70},
archivePrefix = {arXiv},
       eprint = {1511.02895},
 primaryClass = {astro-ph.EP},
       adsurl = {https://ui.adsabs.harvard.edu/abs/2016AJ....152...70B}
}

@ARTICLE{Bannister2018,
       author = {{Bannister}, Michele T. and {Gladman}, Brett J. and {Kavelaars}, J.~J. and {Petit}, Jean-Marc and {Volk}, Kathryn and {Chen}, Ying-Tung and {Alexandersen}, Mike and {Gwyn}, Stephen D.~J. and {Schwamb}, Megan E. and {Ashton}, Edward and {Benecchi}, Susan D. and {Cabral}, Nahuel and {Dawson}, Rebekah I. and {Delsanti}, Audrey and {Fraser}, Wesley C. and {Granvik}, Mikael and {Greenstreet}, Sarah and {Guilbert-Lepoutre}, Aur{\'e}lie and {Ip}, Wing-Huen and {Jakubik}, Marian and {Jones}, R. Lynne and {Kaib}, Nathan A. and {Lacerda}, Pedro and {Van Laerhoven}, Christa and {Lawler}, Samantha and {Lehner}, Matthew J. and {Lin}, Hsing Wen and {Lykawka}, Patryk Sofia and {Marsset}, Micha{\"e}l and {Murray-Clay}, Ruth and {Pike}, Rosemary E. and {Rousselot}, Philippe and {Shankman}, Cory and {Thirouin}, Audrey and {Vernazza}, Pierre and {Wang}, Shiang-Yu},
        title = "{OSSOS. VII. 800+ Trans-Neptunian Objects{\textemdash}The Complete Data Release}",
      journal = {\apjs},
         year = 2018,
        month = may,
       volume = {236},
       number = {1},
          eid = {18},
        pages = {18},
          doi = {10.3847/1538-4365/aab77a},
archivePrefix = {arXiv},
       eprint = {1805.11740},
 primaryClass = {astro-ph.EP},
       adsurl = {https://ui.adsabs.harvard.edu/abs/2018ApJS..236...18B}
}

@ARTICLE{Bernardinelli2022,
       author = {{Bernardinelli}, Pedro H. and {Bernstein}, Gary M. and {Sako}, Masao and {Yanny}, Brian and {Aguena}, M. and {Allam}, S. and {Andrade-Oliveira}, F. and {Bertin}, E. and {Brooks}, D. and {Buckley-Geer}, E. and {Burke}, D.~L. and {Carnero Rosell}, A. and {Carrasco Kind}, M. and {Carretero}, J. and {Conselice}, C. and {Costanzi}, M. and {da Costa}, L.~N. and {De Vicente}, J. and {Desai}, S. and {Diehl}, H.~T. and {Dietrich}, J.~P. and {Doel}, P. and {Eckert}, K. and {Everett}, S. and {Ferrero}, I. and {Flaugher}, B. and {Fosalba}, P. and {Frieman}, J. and {Garc{\'\i}a-Bellido}, J. and {Gerdes}, D.~W. and {Gruen}, D. and {Gruendl}, R.~A. and {Gschwend}, J. and {Hinton}, S.~R. and {Hollowood}, D.~L. and {Honscheid}, K. and {James}, D.~J. and {Kent}, S. and {Kuehn}, K. and {Kuropatkin}, N. and {Lahav}, O. and {Maia}, M.~A.~G. and {March}, M. and {Menanteau}, F. and {Miquel}, R. and {Morgan}, R. and {Myles}, J. and {Ogando}, R.~L.~C. and {Palmese}, A. and {Paz-Chinch{\'o}n}, F. and {Pieres}, A. and {Plazas Malag{\'o}n}, A.~A. and {Romer}, A.~K. and {Roodman}, A. and {Sanchez}, E. and {Scarpine}, V. and {Schubnell}, M. and {Serrano}, S. and {Sevilla-Noarbe}, I. and {Smith}, M. and {Soares-Santos}, M. and {Suchyta}, E. and {Swanson}, M.~E.~C. and {Tarle}, G. and {To}, C. and {Varga}, T.~N. and {Walker}, A.~R.},
        title = "{A Search of the Full Six Years of the Dark Energy Survey for Outer Solar System Objects}",
      journal = {\apjs},
         year = 2022,
        month = feb,
       volume = {258},
       number = {2},
          eid = {41},
        pages = {41},
          doi = {10.3847/1538-4365/ac3914},
archivePrefix = {arXiv},
       eprint = {2109.03758},
 primaryClass = {astro-ph.EP},
       adsurl = {https://ui.adsabs.harvard.edu/abs/2022ApJS..258...41B}
}

@ARTICLE{Bernardinelli2020,
       author = {{Bernardinelli}, Pedro H. and {Bernstein}, Gary M. and {Sako}, Masao and {Liu}, Tongtian and {Saunders}, William R. and {Khain}, Tali and {Lin}, Hsing Wen and {Gerdes}, David W. and {Brout}, Dillon and {Adams}, Fred C. and {Belyakov}, Matthew and {Somasundaram}, Aditya Inada and {Sharma}, Lakshay and {Locke}, Jennifer and {Franson}, Kyle and {Becker}, Juliette C. and {Napier}, Kevin and {Markwardt}, Larissa and {Annis}, James and {Abbott}, T.~M.~C. and {Avila}, S. and {Brooks}, D. and {Burke}, D.~L. and {Carnero Rosell}, A. and {Carrasco Kind}, M. and {Castander}, F.~J. and {da Costa}, L.~N. and {De Vicente}, J. and {Desai}, S. and {Diehl}, H.~T. and {Doel}, P. and {Everett}, S. and {Flaugher}, B. and {Garc{\'\i}a-Bellido}, J. and {Gruen}, D. and {Gruendl}, R.~A. and {Gschwend}, J. and {Gutierrez}, G. and {Hollowood}, D.~L. and {James}, D.~J. and {Johnson}, M.~W.~G. and {Johnson}, M.~D. and {Krause}, E. and {Kuropatkin}, N. and {Maia}, M.~A.~G. and {March}, M. and {Miquel}, R. and {Paz-Chinch{\'o}n}, F. and {Plazas}, A.~A. and {Romer}, A.~K. and {Rykoff}, E.~S. and {S{\'a}nchez}, C. and {Sanchez}, E. and {Scarpine}, V. and {Serrano}, S. and {Sevilla-Noarbe}, I. and {Smith}, M. and {Sobreira}, F. and {Suchyta}, E. and {Swanson}, M.~E.~C. and {Tarle}, G. and {Walker}, A.~R. and {Wester}, W. and {Zhang}, Y. and {DES Collaboration}},
        title = "{Trans-Neptunian Objects Found in the First Four Years of the Dark Energy Survey}",
      journal = {\apjs},
         year = 2020,
        month = mar,
       volume = {247},
       number = {1},
          eid = {32},
        pages = {32},
          doi = {10.3847/1538-4365/ab6bd8},
archivePrefix = {arXiv},
       eprint = {1909.01478},
 primaryClass = {astro-ph.EP},
       adsurl = {https://ui.adsabs.harvard.edu/abs/2020ApJS..247...32B}
}

@ARTICLE{Bernardinelli2023,
       author = {{Bernardinelli}, Pedro H. and {Bernstein}, Gary M. and {Jindal}, Nicholas and {Abbott}, T.~M.~C. and {Aguena}, M. and {Alves}, O. and {Andrade-Oliveira}, F. and {Annis}, J. and {Bacon}, D. and {Bertin}, E. and {Brooks}, D. and {Burke}, D.~L. and {Carnero Rosell}, A. and {Carrasco Kind}, M. and {Carretero}, J. and {da Costa}, L.~N. and {Pereira}, M.~E.~S. and {Davis}, T.~M. and {Desai}, S. and {Diehl}, H.~T. and {Doel}, P. and {Everett}, S. and {Ferrero}, I. and {Friedel}, D. and {Frieman}, J. and {Garc{\'\i}a-Bellido}, J. and {Giannini}, G. and {Gruen}, D. and {Herner}, K. and {Hinton}, S.~R. and {Hollowood}, D.~L. and {Honscheid}, K. and {James}, D.~J. and {Kuehn}, K. and {Marshall}, J.~L. and {Mena-Fern{\'a}ndez}, J. and {Menanteau}, F. and {Miquel}, R. and {Ogando}, R.~L.~C. and {Palmese}, A. and {Pieres}, A. and {Plazas Malag{\'o}n}, A.~A. and {Raveri}, M. and {Sanchez}, E. and {Sevilla-Noarbe}, I. and {Smith}, M. and {Suchyta}, E. and {Swanson}, M.~E.~C. and {Tarle}, G. and {To}, C. and {Walker}, A.~R. and {Wiseman}, P. and {Zhang}, Y. and {DES Collaboration}},
        title = "{Photometry of Outer Solar System Objects from the Dark Energy Survey. I. Photometric Methods, Light-curve Distributions, and Trans-Neptunian Binaries}",
      journal = {\apjs},
         year = 2023,
        month = nov,
       volume = {269},
       number = {1},
          eid = {18},
        pages = {18},
          doi = {10.3847/1538-4365/acf6bf},
archivePrefix = {arXiv},
       eprint = {2304.03017},
 primaryClass = {astro-ph.EP},
       adsurl = {https://ui.adsabs.harvard.edu/abs/2023ApJS..269...18B}
}

@INPROCEEDINGS{Boulade2003,
       author = {{Boulade}, Olivier and {Charlot}, Xavier and {Abbon}, P. and {Aune}, Stephan and {Borgeaud}, Pierre and {Carton}, Pierre-Henri and {Carty}, M. and {Da Costa}, J. and {Deschamps}, H. and {Desforge}, D. and {Eppell{\'e}}, Dominique and {Gallais}, Pascal and {Gosset}, L. and {Granelli}, Remy and {Gros}, Michel and {de Kat}, Jean and {Loiseau}, Denis and {Ritou}, J. -. and {Rouss{\'e}}, Jean Y. and {Starzynski}, Pierre and {Vignal}, Nicolas and {Vigroux}, Laurent G.},
        title = "{MegaCam: the new Canada-France-Hawaii Telescope wide-field imaging camera}",
    booktitle = {Instrument Design and Performance for Optical/Infrared Ground-based Telescopes},
         year = 2003,
       editor = {{Iye}, Masanori and {Moorwood}, Alan F.~M.},
       series = {Society of Photo-Optical Instrumentation Engineers (SPIE) Conference Series},
       volume = {4841},
        month = mar,
        pages = {72-81},
          doi = {10.1117/12.459890},
       adsurl = {https://ui.adsabs.harvard.edu/abs/2003SPIE.4841...72B}
}

@ARTICLE{Flaugher2015,
       author = {{Flaugher}, B. and {Diehl}, H.~T. and {Honscheid}, K. and {Abbott}, T.~M.~C. and {Alvarez}, O. and {Angstadt}, R. and {Annis}, J.~T. and {Antonik}, M. and {Ballester}, O. and {Beaufore}, L. and {Bernstein}, G.~M. and {Bernstein}, R.~A. and {Bigelow}, B. and {Bonati}, M. and {Boprie}, D. and {Brooks}, D. and {Buckley-Geer}, E.~J. and {Campa}, J. and {Cardiel-Sas}, L. and {Castander}, F.~J. and {Castilla}, J. and {Cease}, H. and {Cela-Ruiz}, J.~M. and {Chappa}, S. and {Chi}, E. and {Cooper}, C. and {da Costa}, L.~N. and {Dede}, E. and {Derylo}, G. and {DePoy}, D.~L. and {de Vicente}, J. and {Doel}, P. and {Drlica-Wagner}, A. and {Eiting}, J. and {Elliott}, A.~E. and {Emes}, J. and {Estrada}, J. and {Fausti Neto}, A. and {Finley}, D.~A. and {Flores}, R. and {Frieman}, J. and {Gerdes}, D. and {Gladders}, M.~D. and {Gregory}, B. and {Gutierrez}, G.~R. and {Hao}, J. and {Holland}, S.~E. and {Holm}, S. and {Huffman}, D. and {Jackson}, C. and {James}, D.~J. and {Jonas}, M. and {Karcher}, A. and {Karliner}, I. and {Kent}, S. and {Kessler}, R. and {Kozlovsky}, M. and {Kron}, R.~G. and {Kubik}, D. and {Kuehn}, K. and {Kuhlmann}, S. and {Kuk}, K. and {Lahav}, O. and {Lathrop}, A. and {Lee}, J. and {Levi}, M.~E. and {Lewis}, P. and {Li}, T.~S. and {Mandrichenko}, I. and {Marshall}, J.~L. and {Martinez}, G. and {Merritt}, K.~W. and {Miquel}, R. and {Mu{\~n}oz}, F. and {Neilsen}, E.~H. and {Nichol}, R.~C. and {Nord}, B. and {Ogando}, R. and {Olsen}, J. and {Palaio}, N. and {Patton}, K. and {Peoples}, J. and {Plazas}, A.~A. and {Rauch}, J. and {Reil}, K. and {Rheault}, J. -P. and {Roe}, N.~A. and {Rogers}, H. and {Roodman}, A. and {Sanchez}, E. and {Scarpine}, V. and {Schindler}, R.~H. and {Schmidt}, R. and {Schmitt}, R. and {Schubnell}, M. and {Schultz}, K. and {Schurter}, P. and {Scott}, L. and {Serrano}, S. and {Shaw}, T.~M. and {Smith}, R.~C. and {Soares-Santos}, M. and {Stefanik}, A. and {Stuermer}, W. and {Suchyta}, E. and {Sypniewski}, A. and {Tarle}, G. and {Thaler}, J. and {Tighe}, R. and {Tran}, C. and {Tucker}, D. and {Walker}, A.~R. and {Wang}, G. and {Watson}, M. and {Weaverdyck}, C. and {Wester}, W. and {Woods}, R. and {Yanny}, B. and {DES Collaboration}},
        title = "{The Dark Energy Camera}",
      journal = {\aj},
         year = 2015,
        month = nov,
       volume = {150},
       number = {5},
          eid = {150},
        pages = {150},
          doi = {10.1088/0004-6256/150/5/150},
archivePrefix = {arXiv},
       eprint = {1504.02900},
 primaryClass = {astro-ph.IM},
       adsurl = {https://ui.adsabs.harvard.edu/abs/2015AJ....150..150F}
}

@ARTICLE{Fraser2023,
       author = {{Fraser}, Wesley C. and {Pike}, Rosemary E. and {Marsset}, Micha{\"e}l and {Schwamb}, Megan E. and {Bannister}, Michele T. and {Buchanan}, Laura and {Kavelaars}, J.~J. and {Benecchi}, Susan D. and {Tan}, Nicole J. and {Peixinho}, Nuno and {Gwyn}, Stephen D.~J. and {Alexandersen}, Mike and {Chen}, Ying-Tung and {Gladman}, Brett and {Volk}, Kathryn},
        title = "{Col-OSSOS: The Two Types of Kuiper Belt Surfaces}",
      journal = {\psj},
         year = 2023,
        month = may,
       volume = {4},
       number = {5},
          eid = {80},
        pages = {80},
          doi = {10.3847/PSJ/acc844},
archivePrefix = {arXiv},
       eprint = {2206.04068},
 primaryClass = {astro-ph.EP},
       adsurl = {https://ui.adsabs.harvard.edu/abs/2023PSJ.....4...80F}
}

@ARTICLE{Hainaut2002,
       author = {{Hainaut}, O.~R. and {Delsanti}, A.~C.},
        title = "{Colors of Minor Bodies in the Outer Solar System. A statistical analysis}",
      journal = {\aap},
         year = 2002,
        month = jul,
       volume = {389},
        pages = {641-664},
          doi = {10.1051/0004-6361:20020431},
       adsurl = {https://ui.adsabs.harvard.edu/abs/2002A&A...389..641H}
}

@ARTICLE{Hodapp2003,
       author = {{Hodapp}, Klaus W. and {Jensen}, Joseph B. and {Irwin}, Everett M. and {Yamada}, Hubert and {Chung}, Randolph and {Fletcher}, Kent and {Robertson}, Louis and {Hora}, Joseph L. and {Simons}, Douglas A. and {Mays}, Wendy and {Nolan}, Robert and {Bec}, Matthieu and {Merrill}, Michael and {Fowler}, Albert M.},
        title = "{The Gemini Near-Infrared Imager (NIRI)}",
      journal = {\pasp},
         year = 2003,
        month = dec,
       volume = {115},
       number = {814},
        pages = {1388-1406},
          doi = {10.1086/379669},
       adsurl = {https://ui.adsabs.harvard.edu/abs/2003PASP..115.1388H}
}

@ARTICLE{Hook2004,
       author = {{Hook}, I.~M. and {J{\o}rgensen}, Inger and {Allington-Smith}, J.~R. and {Davies}, R.~L. and {Metcalfe}, N. and {Murowinski}, R.~G. and {Crampton}, D.},
        title = "{The Gemini-North Multi-Object Spectrograph: Performance in Imaging, Long-Slit, and Multi-Object Spectroscopic Modes}",
      journal = {\pasp},
         year = 2004,
        month = may,
       volume = {116},
       number = {819},
        pages = {425-440},
          doi = {10.1086/383624},
       adsurl = {https://ui.adsabs.harvard.edu/abs/2004PASP..116..425H}
}

@ARTICLE{Ivezic2001,
       author = {{Ivezi{\'c}}, {\v{Z}}eljko and {Tabachnik}, Serge and {Rafikov}, Roman and {Lupton}, Robert H. and {Quinn}, Tom and {Hammergren}, Mark and {Eyer}, Laurent and {Chu}, Jennifer and {Armstrong}, John C. and {Fan}, Xiaohui and {Finlator}, Kristian and {Geballe}, Tom R. and {Gunn}, James E. and {Hennessy}, Gregory S. and {Knapp}, Gillian R. and {Leggett}, Sandy K. and {Munn}, Jeffrey A. and {Pier}, Jeffrey R. and {Rockosi}, Constance M. and {Schneider}, Donald P. and {Strauss}, Michael A. and {Yanny}, Brian and {Brinkmann}, Jonathan and {Csabai}, Istv{\'a}n and {Hindsley}, Robert B. and {Kent}, Stephen and {Lamb}, Don Q. and {Margon}, Bruce and {McKay}, Timothy A. and {Smith}, J. Allyn and {Waddel}, Patrick and {York}, Donald G. and {SDSS Collaboration}},
        title = "{Solar System Objects Observed in the Sloan Digital Sky Survey Commissioning Data}",
      journal = {\aj},
         year = 2001,
        month = nov,
       volume = {122},
       number = {5},
        pages = {2749-2784},
          doi = {10.1086/323452},
archivePrefix = {arXiv},
       eprint = {astro-ph/0105511},
 primaryClass = {astro-ph},
       adsurl = {https://ui.adsabs.harvard.edu/abs/2001AJ....122.2749I}
}

@ARTICLE{Juric2002,
       author = {{Juri{\'c}}, Mario and {Ivezi{\'c}}, {\v{Z}}eljko and {Lupton}, Robert H. and {Quinn}, Tom and {Tabachnik}, Serge and {Fan}, Xiaohui and {Gunn}, James E. and {Hennessy}, Gregory S. and {Knapp}, Gillian R. and {Munn}, Jeffrey A. and {Pier}, Jeffrey R. and {Rockosi}, Constance M. and {Schneider}, Donald P. and {Brinkmann}, Jonathan and {Csabai}, Istv{\'a}n and {Fukugita}, Masataka},
        title = "{Comparison of Positions and Magnitudes of Asteroids Observed in the Sloan Digital Sky Survey with Those Predicted for Known Asteroids}",
      journal = {\aj},
         year = 2002,
        month = sep,
       volume = {124},
       number = {3},
        pages = {1776-1787},
          doi = {10.1086/341950},
archivePrefix = {arXiv},
       eprint = {astro-ph/0202468},
 primaryClass = {astro-ph},
       adsurl = {https://ui.adsabs.harvard.edu/abs/2002AJ....124.1776J}
}

@ARTICLE{Schwamb2019,
       author = {{Schwamb}, Megan E. and {Fraser}, Wesley C. and {Bannister}, Michele T. and {Marsset}, Micha{\"e}l and {Pike}, Rosemary E. and {Kavelaars}, J.~J. and {Benecchi}, Susan D. and {Lehner}, Matthew J. and {Wang}, Shiang-Yu and {Thirouin}, Audrey and {Delsanti}, Audrey and {Peixinho}, Nuno and {Volk}, Kathryn and {Alexandersen}, Mike and {Chen}, Ying-Tung and {Gladman}, Brett and {Gwyn}, Stephen D.~J. and {Petit}, Jean-Marc},
        title = "{Col-OSSOS: The Colors of the Outer Solar System Origins Survey}",
      journal = {\apjs},
         year = 2019,
        month = jul,
       volume = {243},
       number = {1},
          eid = {12},
        pages = {12},
          doi = {10.3847/1538-4365/ab2194},
archivePrefix = {arXiv},
       eprint = {1809.08501},
 primaryClass = {astro-ph.EP},
       adsurl = {https://ui.adsabs.harvard.edu/abs/2019ApJS..243...12S}
}

@ARTICLE{DES2016,
       author = {{Dark Energy Survey Collaboration} and {Abbott}, T. and {Abdalla}, F.~B. and {Aleksi{\'c}}, J. and {Allam}, S. and {Amara}, A. and {Bacon}, D. and {Balbinot}, E. and {Banerji}, M. and {Bechtol}, K. and {Benoit-L{\'e}vy}, A. and {Bernstein}, G.~M. and {Bertin}, E. and {Blazek}, J. and {Bonnett}, C. and {Bridle}, S. and {Brooks}, D. and {Brunner}, R.~J. and {Buckley-Geer}, E. and {Burke}, D.~L. and {Caminha}, G.~B. and {Capozzi}, D. and {Carlsen}, J. and {Carnero-Rosell}, A. and {Carollo}, M. and {Carrasco-Kind}, M. and {Carretero}, J. and {Castander}, F.~J. and {Clerkin}, L. and {Collett}, T. and {Conselice}, C. and {Crocce}, M. and {Cunha}, C.~E. and {D'Andrea}, C.~B. and {da Costa}, L.~N. and {Davis}, T.~M. and {Desai}, S. and {Diehl}, H.~T. and {Dietrich}, J.~P. and {Dodelson}, S. and {Doel}, P. and {Drlica-Wagner}, A. and {Estrada}, J. and {Etherington}, J. and {Evrard}, A.~E. and {Fabbri}, J. and {Finley}, D.~A. and {Flaugher}, B. and {Foley}, R.~J. and {Fosalba}, P. and {Frieman}, J. and {Garc{\'\i}a-Bellido}, J. and {Gaztanaga}, E. and {Gerdes}, D.~W. and {Giannantonio}, T. and {Goldstein}, D.~A. and {Gruen}, D. and {Gruendl}, R.~A. and {Guarnieri}, P. and {Gutierrez}, G. and {Hartley}, W. and {Honscheid}, K. and {Jain}, B. and {James}, D.~J. and {Jeltema}, T. and {Jouvel}, S. and {Kessler}, R. and {King}, A. and {Kirk}, D. and {Kron}, R. and {Kuehn}, K. and {Kuropatkin}, N. and {Lahav}, O. and {Li}, T.~S. and {Lima}, M. and {Lin}, H. and {Maia}, M.~A.~G. and {Makler}, M. and {Manera}, M. and {Maraston}, C. and {Marshall}, J.~L. and {Martini}, P. and {McMahon}, R.~G. and {Melchior}, P. and {Merson}, A. and {Miller}, C.~J. and {Miquel}, R. and {Mohr}, J.~J. and {Morice-Atkinson}, X. and {Naidoo}, K. and {Neilsen}, E. and {Nichol}, R.~C. and {Nord}, B. and {Ogando}, R. and {Ostrovski}, F. and {Palmese}, A. and {Papadopoulos}, A. and {Peiris}, H.~V. and {Peoples}, J. and {Percival}, W.~J. and {Plazas}, A.~A. and {Reed}, S.~L. and {Refregier}, A. and {Romer}, A.~K. and {Roodman}, A. and {Ross}, A. and {Rozo}, E. and {Rykoff}, E.~S. and {Sadeh}, I. and {Sako}, M. and {S{\'a}nchez}, C. and {Sanchez}, E. and {Santiago}, B. and {Scarpine}, V. and {Schubnell}, M. and {Sevilla-Noarbe}, I. and {Sheldon}, E. and {Smith}, M. and {Smith}, R.~C. and {Soares-Santos}, M. and {Sobreira}, F. and {Soumagnac}, M. and {Suchyta}, E. and {Sullivan}, M. and {Swanson}, M. and {Tarle}, G. and {Thaler}, J. and {Thomas}, D. and {Thomas}, R.~C. and {Tucker}, D. and {Vieira}, J.~D. and {Vikram}, V. and {Walker}, A.~R. and {Wechsler}, R.~H. and {Weller}, J. and {Wester}, W. and {Whiteway}, L. and {Wilcox}, H. and {Yanny}, B. and {Zhang}, Y. and {Zuntz}, J.},
        title = "{The Dark Energy Survey: more than dark energy - an overview}",
      journal = {\mnras},
         year = 2016,
        month = aug,
       volume = {460},
       number = {2},
        pages = {1270-1299},
          doi = {10.1093/mnras/stw641},
archivePrefix = {arXiv},
       eprint = {1601.00329},
 primaryClass = {astro-ph.CO},
       adsurl = {https://ui.adsabs.harvard.edu/abs/2016MNRAS.460.1270D}
}

@InProceedings{mckinney2010pandas,
  author    = { {W}es {M}c{K}inney },
  title     = { {D}ata {S}tructures for {S}tatistical {C}omputing in {P}ython },
  booktitle = { {P}roceedings of the 9th {P}ython in {S}cience {C}onference },
  pages     = { 56 - 61 },
  year      = { 2010 },
  editor    = { {S}t\'efan van der {W}alt and {J}arrod {M}illman },
  doi       = { 10.25080/Majora-92bf1922-00a }
}

@ARTICLE{McBride2003,
       author = {{McBride}, Neil and {Green}, Simon F. and {Davies}, John K. and {Tholen}, David J. and {Sheppard}, Scott S. and {Whiteley}, Robert J. and {Hillier}, Jon K.},
        title = "{Visible and infrared photometry of Kuiper Belt objects: searching for evidence of trends}",
      journal = {\icarus},
         year = 2003,
        month = feb,
       volume = {161},
       number = {2},
        pages = {501-510},
          doi = {10.1016/S0019-1035(02)00041-6},
       adsurl = {https://ui.adsabs.harvard.edu/abs/2003Icar..161..501M}
}

@software{reback2020pandas,
    author       = {The pandas development team},
    title        = {pandas-dev/pandas: Pandas},
    month        = feb,
    year         = 2020,
    publisher    = {Zenodo},
    version      = {latest},
    doi          = {10.5281/zenodo.3509134},
    url          = {https://doi.org/10.5281/zenodo.3509134}
}

@ARTICLE{Sergeyev2021,
       author = {{Sergeyev}, Alexey V. and {Carry}, Benoit},
        title = "{A million asteroid observations in the Sloan Digital Sky Survey}",
      journal = {\aap},
         year = 2021,
        month = aug,
       volume = {652},
          eid = {A59},
        pages = {A59},
          doi = {10.1051/0004-6361/202140430},
archivePrefix = {arXiv},
       eprint = {2108.05749},
 primaryClass = {astro-ph.EP},
       adsurl = {https://ui.adsabs.harvard.edu/abs/2021A&A...652A..59S}
}

@ARTICLE{Warner2009,
       author = {{Warner}, Brian D. and {Harris}, Alan W. and {Pravec}, Petr},
        title = "{The asteroid lightcurve database}",
      journal = {\icarus},
         year = 2009,
        month = jul,
       volume = {202},
       number = {1},
        pages = {134-146},
          doi = {10.1016/j.icarus.2009.02.003},
       adsurl = {https://ui.adsabs.harvard.edu/abs/2009Icar..202..134W}
}

@ARTICLE{tsiganis2005NiceModel,
       author = {{Tsiganis}, K. and {Gomes}, R. and {Morbidelli}, A. and {Levison}, H.~F.},
        title = "{Origin of the orbital architecture of the giant planets of the Solar System}",
      journal = {\nat},
         year = 2005,
        month = may,
       volume = {435},
       number = {7041},
        pages = {459-461},
          doi = {10.1038/nature03539},
       adsurl = {https://ui.adsabs.harvard.edu/abs/2005Natur.435..459T}
}

@ARTICLE{nesvornymorbidelli2012,
       author = {{Nesvorn{\'y}}, David and {Morbidelli}, Alessandro},
        title = "{Statistical Study of the Early Solar System's Instability with Four, Five, and Six Giant Planets}",
      journal = {\aj},
         year = 2012,
        month = oct,
       volume = {144},
       number = {4},
          eid = {117},
        pages = {117},
          doi = {10.1088/0004-6256/144/4/117},
archivePrefix = {arXiv},
       eprint = {1208.2957},
 primaryClass = {astro-ph.EP},
       adsurl = {https://ui.adsabs.harvard.edu/abs/2012AJ....144..117N}
}

@ARTICLE{Trujillo2002,
       author = {{Trujillo}, Chadwick A. and {Brown}, Michael E.},
        title = "{A Correlation between Inclination and Color in the Classical Kuiper Belt}",
      journal = {\apjl},
         year = 2002,
        month = feb,
       volume = {566},
       number = {2},
        pages = {L125-L128},
          doi = {10.1086/339437},
archivePrefix = {arXiv},
       eprint = {astro-ph/0201040},
 primaryClass = {astro-ph},
       adsurl = {https://ui.adsabs.harvard.edu/abs/2002ApJ...566L.125T}
}

@misc{koumjian2025,
  author       = {Koumjian, A. and B612 Foundation | Asteroid Institute},
  year         = {2025},
  title        = {Vera C. Rubin Observatory Observations in the Minor Planet Center},
  publisher    = {Zenodo},
  doi          = {10.5281/zenodo.17047589},
  url          = {https://doi.org/10.5281/zenodo.17047589},
  note         = {[Data set]}
}

@ARTICLE{luujewitt1996AJ,
       author = {{Luu}, Jane and {Jewitt}, David},
        title = "{Color Diversity Among the Centaurs and Kuiper Belt Objects}",
      journal = {\aj},
         year = 1996,
        month = nov,
       volume = {112},
        pages = {2310},
          doi = {10.1086/118184},
       adsurl = {https://ui.adsabs.harvard.edu/abs/1996AJ....112.2310L}
}

@ARTICLE{perna2010,
       author = {{Perna}, D. and {Barucci}, M.~A. and {Fornasier}, S. and {DeMeo}, F.~E. and {Alvarez-Candal}, A. and {Merlin}, F. and {Dotto}, E. and {Doressoundiram}, A. and {de Bergh}, C.},
        title = "{Colors and taxonomy of Centaurs and trans-Neptunian objects}",
      journal = {\aap},
         year = 2010,
        month = feb,
       volume = {510},
          eid = {A53},
        pages = {A53},
          doi = {10.1051/0004-6361/200913654},
archivePrefix = {arXiv},
       eprint = {0912.2621},
 primaryClass = {astro-ph.EP},
       adsurl = {https://ui.adsabs.harvard.edu/abs/2010A&A...510A..53P}
}

@ARTICLE{ferreira2025MNRASabscol,
       author = {{Ferreira}, F.~S. and {Camargo}, J.~I.~B. and {Boufleur}, R. and {Banda-Huarca}, M.~V. and {Pieres}, A. and {Peixoto}, V.~F. and {Assafin}, M. and {Bernardinelli}, P.~H. and {Lin}, H.~W. and {Braga-Ribas}, F. and {Gomes-Junior}, A. and {Vieira-Martins}, R. and {da Costa}, L.~N. and {Abbott}, T.~M.~C. and {Aguena}, M. and {Allam}, Sahar S. and {Alves}, O. and {Annis}, J. and {Bacon}, D. and {Brooks}, D. and {Burke}, D.~L. and {Carnero Rosell}, A. and {Carretero}, J. and {Desai}, S. and {Doel}, P. and {Everett}, S. and {Ferrero}, I. and {Frieman}, J. and {Garc{\'\i}a-Bellido}, J. and {Gatti}, M. and {Gazta{\~n}aga}, E. and {Giannini}, G. and {Gruen}, D. and {Gruendl}, R.~A. and {Herner}, K. and {Hinton}, S.~R. and {Hollowood}, D.~L. and {Honscheid}, K. and {James}, D.~J. and {Kuehn}, K. and {Lee}, S. and {Marshall}, J.~L. and {Mena-Fern{\'a}ndez}, J. and {Miquel}, R. and {Myles}, J. and {Palmese}, A. and {Plazas Malag{\'o}n}, A.~A. and {Pereira}, M.~E.~S. and {Samuroff}, S. and {Sanchez}, E. and {Sanchez Cid}, D. and {Servila-Noarbe}, I. and {Smith}, M. and {Suchyta}, E. and {Swanson}, M.~E.~C. and {Tarle}, G. and {To}, C. and {Tucker}, D.~L. and {de Vicente}, J. and {Vikram}, V. and {Walker}, A.~R. and {Weaverdyck}, N.},
        title = "{Year six photometric measurements of known Trans-Neptunian Objects and Centaurs by the Dark Energy Survey}",
      journal = {\mnras},
         year = 2025,
        month = jun,
       volume = {540},
       number = {1},
        pages = {460-482},
          doi = {10.1093/mnras/staf650},
archivePrefix = {arXiv},
       eprint = {2504.16927},
 primaryClass = {astro-ph.EP},
       adsurl = {https://ui.adsabs.harvard.edu/abs/2025MNRAS.540..460F}
}

@ARTICLE{bernardinelli2025AJ,
       author = {{Bernardinelli}, Pedro H. and {Bernstein}, Gary M. and {Abbott}, T.~M.~C. and {Aguena}, M. and {Allam}, S.~S. and {Brooks}, D. and {Carnero Rosell}, A. and {Carretero}, J. and {da Costa}, L.~N. and {Pereira}, M.~E.~S. and {Davis}, T.~M. and {De Vicente}, J. and {Desai}, S. and {Diehl}, H.~T. and {Doel}, P. and {Everett}, S. and {Flaugher}, B. and {Frieman}, J. and {Garc{\'\i}a-Bellido}, J. and {Gaztanaga}, E. and {Gruendl}, R.~A. and {Gutierrez}, G. and {Herner}, K. and {Hinton}, S.~R. and {Hollowood}, D.~L. and {Honscheid}, K. and {James}, D.~J. and {Kuehn}, K. and {Lahav}, O. and {Lee}, S. and {Marshall}, J.~L. and {Mena-Fern{\'a}ndez}, J. and {Miquel}, R. and {Myles}, J. and {Plazas Malag{\'o}n}, A.~A. and {Samuroff}, S. and {Sanchez}, E. and {Santiago}, B. and {Sevilla-Noarbe}, I. and {Smith}, M. and {Suchyta}, E. and {Tarle}, G. and {Tucker}, D.~L. and {Vikram}, V. and {Walker}, A.~R. and {Weaverdyck}, N. and {DES Collaboration}},
        title = "{Photometry of Outer Solar System Objects from the Dark Energy Survey. II. A Joint Analysis of Trans-Neptunian Absolute Magnitudes, Colors, Light Curves and Dynamics}",
      journal = {\aj},
         year = 2025,
        month = jun,
       volume = {169},
       number = {6},
          eid = {305},
        pages = {305},
          doi = {10.3847/1538-3881/adc459},
archivePrefix = {arXiv},
       eprint = {2501.01551},
 primaryClass = {astro-ph.EP},
       adsurl = {https://ui.adsabs.harvard.edu/abs/2025AJ....169..305B}
}

@ARTICLE{telgerroma1998Natur,
       author = {{Tegler}, S.~C. and {Romanishin}, W.},
        title = "{Two distinct populations of Kuiper-belt objects}",
      journal = {\nat},
         year = 1998,
        month = mar,
       volume = {392},
       number = {6671},
        pages = {49-51},
          doi = {10.1038/32108},
       adsurl = {https://ui.adsabs.harvard.edu/abs/1998Natur.392...49T}
}

@ARTICLE{Tegler2000,
       author = {{Tegler}, S.~C. and {Romanishin}, W.},
        title = "{Extremely red Kuiper-belt objects in near-circular orbits beyond 40 AU}",
      journal = {\nat},
         year = 2000,
        month = oct,
       volume = {407},
       number = {6807},
        pages = {979-981},
          doi = {10.1038/35039572},
       adsurl = {https://ui.adsabs.harvard.edu/abs/2000Natur.407..979T}
}

@ARTICLE{barucci2000AJ,
       author = {{Barucci}, M.~A. and {Romon}, J. and {Doressoundiram}, A. and {Tholen}, D.~J.},
        title = "{Compositional Surface Diversity in the Trans-Neptunian Objects}",
      journal = {\aj},
         year = 2000,
        month = jul,
       volume = {120},
       number = {1},
        pages = {496-500},
          doi = {10.1086/301416},
       adsurl = {https://ui.adsabs.harvard.edu/abs/2000AJ....120..496B}
}

@ARTICLE{peixinho2012,
       author = {{Peixinho}, N. and {Delsanti}, A. and {Guilbert-Lepoutre}, A. and {Gafeira}, R. and {Lacerda}, P.},
        title = "{The bimodal colors of Centaurs and small Kuiper belt objects}",
      journal = {\aap},
         year = 2012,
        month = oct,
       volume = {546},
          eid = {A86},
        pages = {A86},
          doi = {10.1051/0004-6361/201219057},
archivePrefix = {arXiv},
       eprint = {1206.3153},
 primaryClass = {astro-ph.EP},
       adsurl = {https://ui.adsabs.harvard.edu/abs/2012A&A...546A..86P}
}

@INCOLLECTION{peixinho2025cent,
       author = {{Peixinho}, Nuno and {Licandro}, Javier and {Lilly}, Eva and {Alvarez-Candal}, Alvaro and {Souza-Feliciano}, A.~C. and {Seccull}, Tom},
        title = "{Surface Properties and Composition}",
    booktitle = {Centaurs},
         year = 2025,
       editor = {{Volk}, Kathryn and {Womack}, Maria and {Steckloff}, Jordan},
        pages = {5-1},
          doi = {10.1088/978-0-7503-5588-9ch5},
       adsurl = {https://ui.adsabs.harvard.edu/abs/2025cent.book....5P}
}

@ARTICLE{marsset2023PSJ,
       author = {{Marsset}, Micha{\"e}l and {Fraser}, Wesley C. and {Schwamb}, Megan E. and {Buchanan}, Laura E. and {Pike}, Rosemary E. and {Volk}, Kathryn and {Peixinho}, Nuno and {Benecchi}, Susan and {Bannister}, Michele T. and {Tan}, Nicole J. and {Kavelaars}, J.~J.},
        title = "{Col-OSSOS: Evidence for a Compositional Gradient Inherited from the Protoplanetary Disk?}",
      journal = {\psj},
         year = 2023,
        month = sep,
       volume = {4},
       number = {9},
          eid = {160},
        pages = {160},
          doi = {10.3847/PSJ/ace7d0},
archivePrefix = {arXiv},
       eprint = {2206.04096},
 primaryClass = {astro-ph.EP},
       adsurl = {https://ui.adsabs.harvard.edu/abs/2023PSJ.....4..160M}
}

@ARTICLE{verbiscer2022PSJ,
       author = {{Verbiscer}, Anne J. and {Helfenstein}, Paul and {Porter}, Simon B. and {Benecchi}, Susan D. and {Kavelaars}, J.~J. and {Lauer}, Tod R. and {Peng}, Jinghan and {Protopapa}, Silvia and {Spencer}, John R. and {Stern}, S. Alan and {Weaver}, Harold A. and {Buie}, Marc W. and {Buratti}, Bonnie J. and {Olkin}, Catherine B. and {Parker}, Joel and {Singer}, Kelsi N. and {Young}, Leslie A. and {New Horizons Science Team}},
        title = "{The Diverse Shapes of Dwarf Planet and Large KBO Phase Curves Observed from New Horizons}",
      journal = {\psj},
         year = 2022,
        month = apr,
       volume = {3},
       number = {4},
          eid = {95},
        pages = {95},
          doi = {10.3847/PSJ/ac63a6},
       adsurl = {https://ui.adsabs.harvard.edu/abs/2022PSJ.....3...95V}
}

@ARTICLE{rabino2007AJ,
       author = {{Rabinowitz}, David L. and {Schaefer}, Bradley E. and {Tourtellotte}, Suzanne W.},
        title = "{The Diverse Solar Phase Curves of Distant Icy Bodies. I. Photometric Observations of 18 Trans-Neptunian Objects, 7 Centaurs, and Nereid}",
      journal = {\aj},
         year = 2007,
        month = jan,
       volume = {133},
       number = {1},
        pages = {26-43},
          doi = {10.1086/508931},
archivePrefix = {arXiv},
       eprint = {astro-ph/0605745},
 primaryClass = {astro-ph},
       adsurl = {https://ui.adsabs.harvard.edu/abs/2007AJ....133...26R}
}

@ARTICLE{ofek2012ApJ,
       author = {{Ofek}, Eran O.},
        title = "{Sloan Digital Sky Survey Observations of Kuiper Belt Objects: Colors and Variability}",
      journal = {\apj},
         year = 2012,
        month = apr,
       volume = {749},
       number = {1},
          eid = {10},
        pages = {10},
          doi = {10.1088/0004-637X/749/1/10},
archivePrefix = {arXiv},
       eprint = {1202.1538},
 primaryClass = {astro-ph.EP},
       adsurl = {https://ui.adsabs.harvard.edu/abs/2012ApJ...749...10O}
}

@ARTICLE{ayalaloera2018MNRAS,
       author = {{Ayala-Loera}, C. and {Alvarez-Candal}, A. and {Ortiz}, J.~L. and {Duffard}, R. and {Fern{\'a}ndez-Valenzuela}, E. and {Santos-Sanz}, P. and {Morales}, N.},
        title = "{Absolute colours and phase coefficients of trans-Neptunian objects: H$_{V}$ - H$_{R}$ and relative phase coefficients}",
      journal = {\mnras},
         year = 2018,
        month = dec,
       volume = {481},
       number = {2},
        pages = {1848-1857},
          doi = {10.1093/mnras/sty2363},
archivePrefix = {arXiv},
       eprint = {1808.09938},
 primaryClass = {astro-ph.EP},
       adsurl = {https://ui.adsabs.harvard.edu/abs/2018MNRAS.481.1848A}
}

@ARTICLE{alvarezcandal2022,
       author = {{Alvarez-Candal}, A. and {Jimenez Corral}, S. and {Colazo}, M.},
        title = "{Absolute colors and phase coefficients of asteroids}",
      journal = {\aap},
         year = 2022,
        month = nov,
       volume = {667},
          eid = {A81},
        pages = {A81},
          doi = {10.1051/0004-6361/202243479},
archivePrefix = {arXiv},
       eprint = {2209.13246},
 primaryClass = {astro-ph.EP},
       adsurl = {https://ui.adsabs.harvard.edu/abs/2022A&A...667A..81A}
}

@ARTICLE{alvarezcandal2025,
       author = {{Alvarez-Candal}, Alvaro and {Rizos}, Juan Luis and {Colazo}, Milagros and {Duffard}, Ren{\'e} and {Morate}, David and {Carruba}, Valerio and {Camargo}, Julio I.~B. and {G{\'o}mez-Toribio}, Andr{\'e}s},
        title = "{A catalog of near-IR absolute magnitudes of Solar System small bodies}",
      journal = {\aap},
         year = 2025,
        month = sep,
       volume = {701},
          eid = {A231},
        pages = {A231},
          doi = {10.1051/0004-6361/202554269},
archivePrefix = {arXiv},
       eprint = {2507.22259},
 primaryClass = {astro-ph.EP},
       adsurl = {https://ui.adsabs.harvard.edu/abs/2025A&A...701A.231A}
}

@ARTICLE{murtagh2025AJ,
       author = {{Murtagh}, Joseph and {Schwamb}, Megan E. and {Merritt}, Stephanie R. and {Bernardinelli}, Pedro H. and {Kurlander}, Jacob A. and {Cornwall}, Samuel and {Juri{\'c}}, Mario and {Fedorets}, Grigori and {Holman}, Matthew J. and {Eggl}, Siegfried and {Nesvorn{\'y}}, David and {Volk}, Kathryn and {Jones}, R. Lynne and {Yoachim}, Peter and {Moeyens}, Joachim and {Kubica}, Jeremy and {Oldag}, Drew and {West}, Maxine and {Chandler}, Colin Orion},
        title = "{Predictions of the LSST Solar System Yield: Discovery Rates and Characterizations of Centaurs}",
      journal = {\aj},
         year = 2025,
        month = aug,
       volume = {170},
       number = {2},
          eid = {98},
        pages = {98},
          doi = {10.3847/1538-3881/ade1db},
archivePrefix = {arXiv},
       eprint = {2506.02779},
 primaryClass = {astro-ph.EP},
       adsurl = {https://ui.adsabs.harvard.edu/abs/2025AJ....170...98M}
}

@ARTICLE{ivezic2019ApJ-LSST,
       author = {{Ivezi{\'c}}, {\v{Z}}eljko and {Kahn}, Steven M. and {Tyson}, J. Anthony and {Abel}, Bob and {Acosta}, Emily and {Allsman}, Robyn and {Alonso}, David and {AlSayyad}, Yusra and {Anderson}, Scott F. and {Andrew}, John and {Angel}, James Roger P. and {Angeli}, George Z. and {Ansari}, Reza and {Antilogus}, Pierre and {Araujo}, Constanza and {Armstrong}, Robert and {Arndt}, Kirk T. and {Astier}, Pierre and {Aubourg}, {\'E}ric and {Auza}, Nicole and {Axelrod}, Tim S. and {Bard}, Deborah J. and {Barr}, Jeff D. and {Barrau}, Aurelian and {Bartlett}, James G. and {Bauer}, Amanda E. and {Bauman}, Brian J. and {Baumont}, Sylvain and {Bechtol}, Ellen and {Bechtol}, Keith and {Becker}, Andrew C. and {Becla}, Jacek and {Beldica}, Cristina and {Bellavia}, Steve and {Bianco}, Federica B. and {Biswas}, Rahul and {Blanc}, Guillaume and {Blazek}, Jonathan and {Blandford}, Roger D. and {Bloom}, Josh S. and {Bogart}, Joanne and {Bond}, Tim W. and {Booth}, Michael T. and {Borgland}, Anders W. and {Borne}, Kirk and {Bosch}, James F. and {Boutigny}, Dominique and {Brackett}, Craig A. and {Bradshaw}, Andrew and {Brandt}, William Nielsen and {Brown}, Michael E. and {Bullock}, James S. and {Burchat}, Patricia and {Burke}, David L. and {Cagnoli}, Gianpietro and {Calabrese}, Daniel and {Callahan}, Shawn and {Callen}, Alice L. and {Carlin}, Jeffrey L. and {Carlson}, Erin L. and {Chandrasekharan}, Srinivasan and {Charles-Emerson}, Glenaver and {Chesley}, Steve and {Cheu}, Elliott C. and {Chiang}, Hsin-Fang and {Chiang}, James and {Chirino}, Carol and {Chow}, Derek and {Ciardi}, David R. and {Claver}, Charles F. and {Cohen-Tanugi}, Johann and {Cockrum}, Joseph J. and {Coles}, Rebecca and {Connolly}, Andrew J. and {Cook}, Kem H. and {Cooray}, Asantha and {Covey}, Kevin R. and {Cribbs}, Chris and {Cui}, Wei and {Cutri}, Roc and {Daly}, Philip N. and {Daniel}, Scott F. and {Daruich}, Felipe and {Daubard}, Guillaume and {Daues}, Greg and {Dawson}, William and {Delgado}, Francisco and {Dellapenna}, Alfred and {de Peyster}, Robert and {de Val-Borro}, Miguel and {Digel}, Seth W. and {Doherty}, Peter and {Dubois}, Richard and {Dubois-Felsmann}, Gregory P. and {Durech}, Josef and {Economou}, Frossie and {Eifler}, Tim and {Eracleous}, Michael and {Emmons}, Benjamin L. and {Fausti Neto}, Angelo and {Ferguson}, Henry and {Figueroa}, Enrique and {Fisher-Levine}, Merlin and {Focke}, Warren and {Foss}, Michael D. and {Frank}, James and {Freemon}, Michael D. and {Gangler}, Emmanuel and {Gawiser}, Eric and {Geary}, John C. and {Gee}, Perry and {Geha}, Marla and {Gessner}, Charles J.~B. and {Gibson}, Robert R. and {Gilmore}, D. Kirk and {Glanzman}, Thomas and {Glick}, William and {Goldina}, Tatiana and {Goldstein}, Daniel A. and {Goodenow}, Iain and {Graham}, Melissa L. and {Gressler}, William J. and {Gris}, Philippe and {Guy}, Leanne P. and {Guyonnet}, Augustin and {Haller}, Gunther and {Harris}, Ron and {Hascall}, Patrick A. and {Haupt}, Justine and {Hernandez}, Fabio and {Herrmann}, Sven and {Hileman}, Edward and {Hoblitt}, Joshua and {Hodgson}, John A. and {Hogan}, Craig and {Howard}, James D. and {Huang}, Dajun and {Huffer}, Michael E. and {Ingraham}, Patrick and {Innes}, Walter R. and {Jacoby}, Suzanne H. and {Jain}, Bhuvnesh and {Jammes}, Fabrice and {Jee}, M. James and {Jenness}, Tim and {Jernigan}, Garrett and {Jevremovi{\'c}}, Darko and {Johns}, Kenneth and {Johnson}, Anthony S. and {Johnson}, Margaret W.~G. and {Jones}, R. Lynne and {Juramy-Gilles}, Claire and {Juri{\'c}}, Mario and {Kalirai}, Jason S. and {Kallivayalil}, Nitya J. and {Kalmbach}, Bryce and {Kantor}, Jeffrey P. and {Karst}, Pierre and {Kasliwal}, Mansi M. and {Kelly}, Heather and {Kessler}, Richard and {Kinnison}, Veronica and {Kirkby}, David and {Knox}, Lloyd and {Kotov}, Ivan V. and {Krabbendam}, Victor L. and {Krughoff}, K. Simon and {Kub{\'a}nek}, Petr and {Kuczewski}, John and {Kulkarni}, Shri and {Ku}, John and {Kurita}, Nadine R. and {Lage}, Craig S. and {Lambert}, Ron and {Lange}, Travis and {Langton}, J. Brian and {Le Guillou}, Laurent and {Levine}, Deborah and {Liang}, Ming and {Lim}, Kian-Tat and {Lintott}, Chris J. and {Long}, Kevin E. and {Lopez}, Margaux and {Lotz}, Paul J. and {Lupton}, Robert H. and {Lust}, Nate B. and {MacArthur}, Lauren A. and {Mahabal}, Ashish and {Mandelbaum}, Rachel and {Markiewicz}, Thomas W. and {Marsh}, Darren S. and {Marshall}, Philip J. and {Marshall}, Stuart and {May}, Morgan and {McKercher}, Robert and {McQueen}, Michelle and {Meyers}, Joshua and {Migliore}, Myriam and {Miller}, Michelle and {Mills}, David J.},
        title = "{LSST: From Science Drivers to Reference Design and Anticipated Data Products}",
      journal = {\apj},
         year = 2019,
        month = mar,
       volume = {873},
       number = {2},
          eid = {111},
        pages = {111},
          doi = {10.3847/1538-4357/ab042c},
archivePrefix = {arXiv},
       eprint = {0805.2366},
 primaryClass = {astro-ph},
       adsurl = {https://ui.adsabs.harvard.edu/abs/2019ApJ...873..111I}
}

@ARTICLE{kurlander2025AJ,
       author = {{Kurlander}, Jacob A. and {Bernardinelli}, Pedro H. and {Schwamb}, Megan E. and {Juri{\'c}}, Mario and {Murtagh}, Joseph and {Chandler}, Colin Orion and {Merritt}, Stephanie R. and {Nesvorn{\'y}}, David and {Vokrouhlick{\'y}}, David and {Jones}, R. Lynne and {Fedorets}, Grigori and {Cornwall}, Samuel and {Holman}, Matthew J. and {Eggl}, Siegfried and {Oldag}, Drew and {West}, Maxine and {Kubica}, Jeremy and {Yoachim}, Peter and {Moeyens}, Joachim and {Kiker}, Kathleen and {Buchanan}, Laura E.},
        title = "{Predictions of the LSST Solar System Yield: Near-Earth Objects, Main Belt Asteroids, Jupiter Trojans, and Trans-Neptunian Objects}",
      journal = {\aj},
         year = 2025,
        month = aug,
       volume = {170},
       number = {2},
          eid = {99},
        pages = {99},
          doi = {10.3847/1538-3881/add685},
archivePrefix = {arXiv},
       eprint = {2506.02487},
 primaryClass = {astro-ph.EP},
       adsurl = {https://ui.adsabs.harvard.edu/abs/2025AJ....170...99K}
}

@ARTICLE{pinillaalonso2025NatAs,
       author = {{Pinilla-Alonso}, Noem{\'\i} and {Brunetto}, Rosario and {De Pr{\'a}}, M{\'a}rio N. and {Holler}, Bryan J. and {H{\'e}nault}, Elsa and {Feliciano}, Ana Carolina de Souza and {Lorenzi}, Vania and {Pendleton}, Yvonne J. and {Cruikshank}, Dale P. and {M{\"u}ller}, Thomas G. and {Stansberry}, John A. and {Emery}, Joshua P. and {Schambeau}, Charles A. and {Licandro}, Javier and {Harvison}, Brittany and {McClure}, Lucas and {Guilbert-Lepoutre}, Aur{\'e}lie and {Peixinho}, Nuno and {Bannister}, Michele T. and {Wong}, Ian},
        title = "{A JWST/DiSCo-TNOs portrait of the primordial Solar System through its trans-Neptunian objects}",
      journal = {Nature Astronomy},
         year = 2025,
        month = feb,
       volume = {9},
        pages = {230-244},
          doi = {10.1038/s41550-024-02433-2},
       adsurl = {https://ui.adsabs.harvard.edu/abs/2025NatAs...9..230P}
}

@article{Greenstreet2026,
doi = {10.3847/2041-8213/ae2a30},
url = {https://doi.org/10.3847/2041-8213/ae2a30},
year = {2026},
month = {jan},
publisher = {The American Astronomical Society},
volume = {996},
number = {2},
pages = {L33},
author = {Greenstreet, Sarah and Li, Zhuofu (Chester) and Vavilov, Dmitrii E. and Singh, Devanshi and Jurić, Mario and Ivezić, Željko and Eggl, Siegfried and Koumjian, Alec and Moeyens, Joachim and Carruba, Valerio and Womack, Maria and Granvik, Mikael and Alexov, Anastasia and Antilogus, Pierre and Baumanć, B̌rian J. and Bellm, Eric C. and Boucaud, Alexandre and Bradshaw, Andrew and Carlin, Jeffrey L. and Chiang, Hsin-Fang and Daly, Philip N. and Daruich, Felipe and Daubard, Guillaume and Dennihy, Erik and Deppe, Stephanie JH and Drass, Holger and Gangler, Emmanuel and Le Guillou, Laurent and Guy, Leanne P. and Hascall, Patrick A. and Ingraham, Patrick and Jee, M. James and Jenness, Tim and Kahn, Steven M. and Kannawadi, Arun and Kelvin, Lee S. and Kurlander, Jacob A. and Laporte, Didier and Lust, Nate B. and Lutfi, Mostafa and MacArthur, Lauren A. and Mainetti, Gabriele and Marc, Moniez and Plazas Malagón, Andrés A. and Mejías, David Jiménez and Menanteau, F̌elipe and Mills, David J. and O’Mullane, William and Neto, Angelo Fausti and Neveu, Jeremy and Nourbakhsh, Erfan and Park, HyeYun and Patterson, Maria T and Peterson, John R. and Quint, Bruno C. and Ribeiro, Tiago and Ridgway, Stephen T. and van Reeven, Wouter and Sebag, Jacques and Sedaghat, Nima and Shaw, Richard A. and Strauss, Alan L. and Suberlak, Krzysztof and Sullivan, Ian S. and Swinbank, John D. and Thomas, Sandrine and Thornton, Adam and Wood-Vasey, W. M. and Walter, Christopher W. and Ward, Charlotte and Willman, Beth},
title = {Lightcurves, Rotation Periods, and Colors for Vera C. Rubin Observatory’s First Asteroid Discoveries},
journal = {The Astrophysical Journal Letters}
}

@ARTICLE{Colazo2026,
       author = {{Colazo}, Milagros and {Oszkiewicz}, Dagmara and {Po{\'z}niak}, Patrycja and {Alvarez-Candal}, Alvaro and {Carry}, Benoit and {Wenda}, Ola and {Stefanowska}, Wiktoria},
        title = "{Detectability of asteroid phase coloring based on phase curves obtained from large spectro-photometric surveys}",
      journal = {\icarus},
         year = 2026,
        month = mar,
       volume = {447},
          eid = {116891},
        pages = {116891},
          doi = {10.1016/j.icarus.2025.116891},
       adsurl = {https://ui.adsabs.harvard.edu/abs/2026Icar..44716891C}
}
\bibliographystyle{aasjournal}

\allauthors


\end{document}